\documentclass{article} 

\usepackage[T1]{fontenc}
\usepackage{microtype}
\usepackage[english]{babel}
\usepackage{float}
\usepackage[export]{adjustbox}
\usepackage[utf8]{inputenx}
\usepackage{amssymb}
\usepackage[nottoc,numbib]{tocbibind}
\usepackage{tabularx}
\usepackage{setspace}
\usepackage{hyperref}

\usepackage[backend=bibtex]{biblatex}
\usepackage{graphicx}
\usepackage{subfig}
\usepackage{amssymb}

\usepackage[noend]{algpseudocode}
\usepackage[section]{algorithm}

\usepackage{tikz}
  \usetikzlibrary{arrows,shadows,positioning}
\usepackage{pgf-umlsd}
\usetikzlibrary{positioning}
\usepackage{ctable}

\renewcommand{\mess}[4][0]{
  \stepcounter{seqlevel}   
  \path
  (#2)+(0,-\theseqlevel*\unitfactor-0.7*\unitfactor) node (mess from) {};
  \addtocounter{seqlevel}{#1}
  \path
  (#4)+(0,-\theseqlevel*\unitfactor-0.7*\unitfactor) node (mess to) {};
  \draw[->,>=angle 60] (mess from) -- (mess to) node[midway, above]
  {#3};

  \node (\detokenize{#3} from) at (mess from) {};
  \node (\detokenize{#3} to) at (mess to) {};
}
\usepackage{varwidth}

\usepackage{titlesec}
\titleformat{\paragraph}
{\normalfont\normalsize\bfseries}{\theparagraph}{1em}{}
\titlespacing*{\paragraph}{0pt}{3.25ex plus 1ex minus .2ex}{1.5ex plus .2ex}

\usepackage{cancel}
\usepackage{enumerate}
\usepackage{enumitem}

\usepackage{hyperref}
\usepackage{amsthm}
\usepackage{amsmath}

\theoremstyle{definition}
\newtheorem{definition}{Definition}[section]
\newtheorem{theorem}{Theorem}[section]

\usepackage[backend=bibtex]{biblatex}
\title{Authentication as a service: Shamir Secret Sharing with byzantine components}

\makeatletter
\renewcommand\@date{{%
  \vspace{-\baselineskip}%
  \large\centering
  \begin{tabular}{@{}c@{}}
    Andrea Bissoli\textsuperscript{} \\
    \normalsize bissoli.1543640@studenti.uniroma1.it
  \end{tabular}%
  \quad and\quad
  \begin{tabular}{@{}c@{}}
    Fabrizio d'Amore\textsuperscript{} \\
    \normalsize damore@diag.uniroma1.it
  \end{tabular}

  \bigskip

  \textsuperscript{}\textit{Dipartimento di Ingegneria informatica automatica e gestionale Antonio Ruberti (DIAG), Sapienza University of Rome}
  \bigskip

  \date{May 18, 2018}     
}}
\makeatother

\begin{document}

    \maketitle

  \begin{abstract}
  	We present a practical methodology for securing the classical password-based authentication scheme, since in many cases this type of authentication is given as a requirement. We propose a solution based on the well-known $(k,n)$ threshold scheme of Shamir for sharing a secret, where in our case the secret is the password itself and $(k,n)$ threshold scheme means that $n$ password-derived secrets (shares) are created and $k\leq n$ shares are necessary and sufficient for reconstructing the password, while $k-1$ are not sufficient. The scheme is information-theoretic secure.
	\\\\\
	Since each of the $n$ shares is stored on a different host (shareholder), an attacker will need to compromise $k$ different shareholders for obtaining an amount of data sufficient for reconstructing the secret. Furthermore, in order to be resistant to the compromising of the server (dealer) coordinating the shareholders we define a variant of the classic Shamir, where the Shamir's abscissae are unknown to dealer and shareholders, making the reconstruction impossible even in the case of dealer and shareholders compromised. In addition we use the Pedersen technique for allowing the verification of shares.
	\\\\
	For the described scenario we have designed two protocols allowing the communication between application, dealer and shareholders, so that the relevant players can participate to the phases of registration (user’s sign-up, to be carried out once),  and of authentication (user’s login).
	\\\
	We analyse several scenarios where dealer and/or shareholders are partially/totally compromised and confirm that none of them is enabling the attacker to break the authentication. Furthermore we focus on cases where one or more byzantine servers are presented, analysing the impact on the framework and show the adopted mechanisms to be safe against these kinds of attacks.
	\\\\
	We have developed a prototype demonstrating that our framework works correctly, effectively and efficiently. It provides a first feasibility study that will provide a base for structured and engineered cloud-based implementations aiming at providing an authentication-as-a-service.

  \end{abstract}

  %%%%%%%%%%%%%%%%%%%%%%%%%%%%%%%%%
  %      INTRODUCTION       %
  %%%%%%%%%%%%%%%%%%%%%%%%%%%%%%%%%
  \newpage
  \section{Introduction} \label{chap:Introduction}
    Nowadays whoever wants to use online resources that provide user custom services has to sign up in order to later log in. In this way the online service can understand who the consumer of the service is and provide specific information related to that user.
    \\\\
    The majority, if not the totality, of online services use the password-based authentication model. The password-based authentication consists in using a secret password in order to recognize the user. The secret is set in a previous step and the scheme relies on human cognitive ability to remember the shared secret. 
    \\\\
    The growth of the number of the online services lead the users to keep in mind a lot of passwords for each service. The challenge for the user is remembering a lot of passwords for numerous services such as: emails, social, banking and financial accounts. The difficulty of memorizing many passwords lead the user to use the \textbf{same password} for each service. This is a huge issue because not all online systems have a good security standard so if an attacker manages to steal the password in one service, they can try to use the same information for other online systems signed by the user. 
    \\\\
    The last big event happened in December, 2017 when 1.4 billion \cite{1_4BillionClearTextCredentials} cleartext credentials were discovered. The source of the data are from 133 new breaches apart from 252 previous breaches. It is an aggregated, interactive database that allows fast (one second response) searches and new breach imports. This database makes finding passwords faster and easier than ever before. It is a very powerful tool and anyone with no programming language knowledge can use it in order to try to discover the password of somebody else. 
    Until today, in addition to this case, there have been a lot of leaks, the most famous are \cite{MostDataBranches}:
    \begin{itemize}
      \item 2012 - LinkedIn, 65 million accounts compromised
      \item 2012 - Dropbox,  68 million accounts compromised
      \item 2014 - eBay, 145 million users compromised
      \item 2014 - Yahoo, 3 billion user accounts
      \item 2016 - Adult Friend Finder, More than 412.2 million accounts
      \item 2016 - Uber, Personal information of 57 million Uber users and 600,000 drivers exposed.
      \item 2017 - Equifax,  145 million accounts compromised
    \end{itemize}

  %%%%%%%%%%%%%%%%%%%%%%%%%%%%%%%%%
  %      STATE OF THE ART     %
  %%%%%%%%%%%%%%%%%%%%%%%%%%%%%%%%%

  \section{State of the art}\label{chap:State of the art}
    In this chapter, the state of the art of the two main concepts will be examined. The two arguments are the password-based authentication and the Shamir secret sharing scheme idea. First of all the scenario and related problems about password-based authentication paradigm will be studied, then, in the second part the state of the art of the secret sharing scheme will be discussed. Furthermore, in the second part, the Shamir secret sharing scheme, its related limits and how to overcome them will be presented. In the middle of the two topics the basic idea of how the secret sharing scheme is used in the password-based authentication scheme will be explained.

    \subsection{Password-based authentication}
      \subsubsection{Overview}
      % qua spiego un pò come funziona. Prendo pezzi dal paper e dal sito internet
      In agreement with \cite{PasswordBaseAuth}, the user-id and password information, is a cost effective and efficient method of maintaining a shared secret between a user and a computer system in order to provide a mechanism of user online authentication. 
      The identification and authorization step is needed for online services that provide per-user operation.
      \\\\
      The aim of the authentication is to verify the user's identity. 
      There are many different implementations of user authentication, and every method belongs of in one of the three widely acknowledged classes. These classes are based on the relationship between a user and information that user uses in order to authenticate themself \cite{PasswordBasedAuthenticationMasterThesis}: 

      \begin{itemize}
        \item \textbf{Something the user knows.} \textit{Password based authentication} belongs to this class. The verification is done on the credentials, which are the user-id and password, in order to prove that the incoming request is from a legitimate user-id. 

        \item \textbf{Something the user has.} Every method requiring some kind of authentication token falls into this class.

        \item \textbf{Something the user is.} Mostly biometrics such as fingerprints, retinal scans pupil images or face ID. The idea is to find unique information that confirms a user's identity.
      \end{itemize}
      Furthermore, it is possible to combine different categories in order to create the so called "\textit{multi factor authentication}" extra security layer that is considered much stronger than methods only from one category. For example the two factor authentication is an extra layer of security that requires, in addition to a password and username, something that only user possesses. This is achieved with a physical token that generates random numbers or through utilising mobile phone SMS technology, by which the code is sent to the user in an SMS. 
      \\\\
      In any case the message scheme is the following: the user presents the data, if the data match with the data previously stored on the service, the service returns a value that represents authorization; otherwise it does not.
      \\\\
      Everybody agrees that the password-based authentication model is easiest and the cheapest when compared to the other two. In this model the user is not annoyed by additional accessories and administrators can easily implement the password-based authentication with equipment to both store and retrieve the password. Hence with the advent of the Internet, and therefore with growth of web-sites, everyone has adopted the first scheme. Yet these features are dependent on user behaviour and users like to have weak passwords and tend to lose them. 
      \\\\
      Moreover nowadays, another problem is that each user has to remember many credentials for different accounts. The difficulty of memorizing many passwords leads the user to use the same password for each service. This is a huge issue because not all online systems have a good security standard so if an attacker succeeds in stealing the password in one service, they can try to use the same information for other online systems frequented by a user.
      \\\\
      Since the secret is present in both actors, service and user, it has to manage both sides securely. In the former case no one can force a user to follow a secure procedure. In the latter case there are much better prospects. The designer of the system can install defence elements like firewalls and IPSs in order to protect databases of user passwords from active attack and use a \textbf{secure passwords store} method against passive attacks. That is, even if an attacker manages to penetrate the service and steal the database they will be unable to retrieve any sensitive information.
      \\\\
      The following are some secure password-storing methods:
      \begin{itemize}
        \item \textbf{Plaintext passwords.} The password is memorised with no secure store techniques. Storing passwords in plaintext must be avoided at all costs.

        \item \textbf{Encrypted passwords.} Cryptographic tools are necessary for storing passwords in a secure manner. Thus, if an attacker is able to break into the service and steal the databases of user passwords they must decrypt each password but without the encryption key it is a computationally intensive task. Passwords stored in this way are immune to attacks.

        \item \textbf{Hashed passwords.} Hash functions are the most prevalent way of storing passwords securely. If an attacker gets access to the system's database, they can not retrieve the plaintext passwords because the hash is not a reversible function. The following are modern hashing algorithms:
          \begin{itemize}
            \item MD-5
            \item SHA-1
            \item SHA-2
            \item SHA-3
          \end{itemize} 
        Observe that both SHA-1 \cite{SHA1_Broken} and MD5 \cite{MD5_broken} are “cryptographically broken”, therefore it is better use SHA-2 or SHA-3 solutions. At the bottom of this page a table lists the major hashing functions with the related broken status.
        \\\\ 
        Hashing passwords algorithms are based on hashing algorithms in order to achieve their purposes. Password hashing requires the following properties\cite{AboutSecurePasswordHashing}: 
        \begin{itemize}
            \item Have a unique salt \footnote{A salt is a non-secret \textit{unique value} random data that is used as an additional input to a one-way function that "hashes" data, a password or passphrase. A new salt is randomly generated for each password. Salts are closely related to the concept of nonce. The primary function of salts is to defend against dictionary attacks or against its hashed equivalent, a pre-computed rainbow table attack \cite{Salt_definition}.} per password (salt may only be used once across the database containing all password hashes) to prevent a bruteforce attack.
            \item Fast on software (executing the function once must be relatively fast).
            \item Low on hardware (executing the function concurrently should be slow, this is to prevent brute forcing on distributed systems).
        \end{itemize}

        The currently safe algorithms are Argon2, PBKDF2, bcrypt and scrypt.
        
      \end{itemize}  

      \begin{figure}[h]
        \centering
        \includegraphics[width=1.0\textwidth]{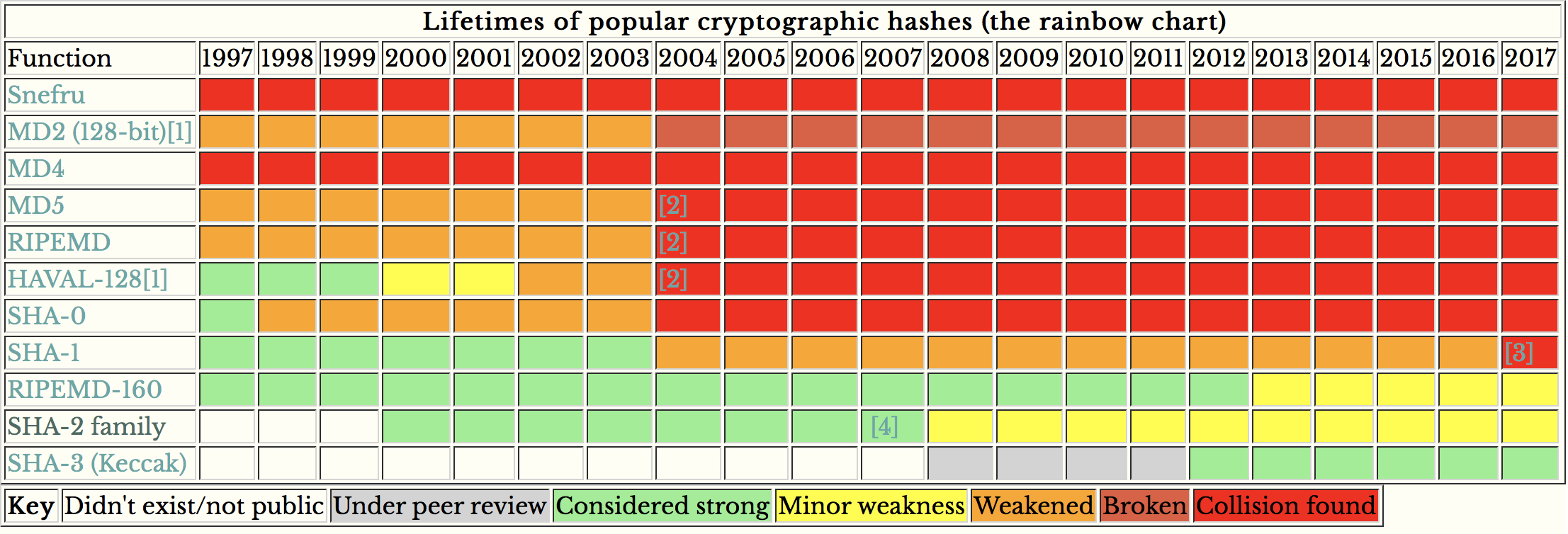}
        \caption{Lifetimes of popular cryptographic hashes \cite{LifetimesCryptographicHashes} }
        \label{fig:LifetimesCryptographicHashes }
      \end{figure}

      \subsubsection{Password-based risk and attacks}
      % spiego i vari tipo di attacco
        Password-based authentication has many risks and real problems, even if a strong and secure passwords store implementation is used. This is due to the fact that if a user chooses a weak password all security countermeasures become useless because the password could be vulnerable to various kinds of attacks. The user has to find a trade-off between a strong password that is safe against an attacker who tries to guess it and a password that the user is able to remember without writing it down or using the same password for many services.
        The well-known possible attacks are the following:
        \begin{itemize}
          \item \textbf{Dictionary attack.} A dictionary attack is a method based on a predetermined list (called a dictionary) of possible password values where the attacker systematically tries every word in the list. 

          \item \textbf{Brute force.} A brute force attack is similar to the above one, but the attacker systematically attempts every possible alphanumeric password combination in the keyspace.

          \item \textbf{Lookup tables.} Lookup tables use the following general idea: they pre-calculate the hashes of the possible passwords of a password dictionary and store them with their corresponding password in a lookup table data structure. The lookup tables are useful because they save computation time and hash for each string in the stolen database is no longer needed. The search is done in the following manner: the automatic program searches for the hash, if it is found, it displays the corresponding value, that is the password \cite{reverse-lookup-table-attack_definition}.

          \item \textbf{Reverse lookup tables.} Reverse lookup tables work in the following manner: 
            \begin{itemize}
              \item The attacker builds a lookup table that maps each password hash (from the stolen database)to a list of users who have that hash.

              \item The attacker, uses the dictionary or brute-force idea and hashes each password guess. They then use the lookup table to retrieve a list of users who have the guessed password. 
            \end{itemize} 
          This kind of attack is especially effective when many users use the same common password.
          Observe that this method works without having to pre-compute a lookup table and the attacker can systematically try many hashes at the same time \cite{reverse-lookup-table-attack_definition}.
            
          \item \textbf{Birthday Attack.} A birthday attack exploits weaknesses in the mathematical algorithms used to generate hashes.

          \item \textbf{Hybrid Password Attack.} A hybrid password attack is a combination of different password attacks.
      \end{itemize}

      \subsubsection{Real cases}
      % Ripendo i casi introdotti nell'intro ed ne elenco altri -->
        As we have seen in the introduction there have been many cases of data leakage until today. The following is a list of ten biggest known breaches known in the 21st century \cite{17_BiggestDataBreaches21stCentury}:  

        \begin{enumerate}
          \item Yahoo, 2013/14 - 3 billion user accounts.
          \item Adult Friend Finder, 2016 - More than 412.2 million accounts.
          \item eBay, 2014 - 145 million users.
          \item Equifax, 2017 - Personal information of 143 million consumers; 209.000 consumers also had their credit card data.
          \item Heartland Payment Systems, 2008 - 134 million credit cards through SQL injection to install spyware on Heartland's data systems.
          \item Target Stores, 2013 - Credit/debit card information and/or contact information of up to 110 million people.
          \item TJX Companies, 2006 - 94 million credit cards.
          \item Uber, 2016 - Personal information of 57 million Uber users and 600,000 drivers.
          \item JP Morgan Chase, 2014 - 76 million households and 7 million small businesses accounts.
          \item US Office of Personnel Management (OPM), 2012 - Personal information of 22 million current and former federal employees.
        \end{enumerate}
        There have been many cases of data leakage until today thus it is necessary to find and study a new way to improve the security of the password-based authentication paradigm.
      
    \subsection{A new way of thinking}\label{NewWayOfThinking}
        Nowadays in the password-based authentication paradigm there are two main actors: a client and a server that exchange information with each other. As we have seen in the previous section the client authentication is based on a password so when the client wants to log in to online services on a server, they send their personal information (username and password). The server handles this information in order to understand whether or not the request is valid. Once the server has done the appropriate checks, it allows the client to enter.
        \begin{figure}[h]
          \centering
          \includegraphics[width=0.7\textwidth]{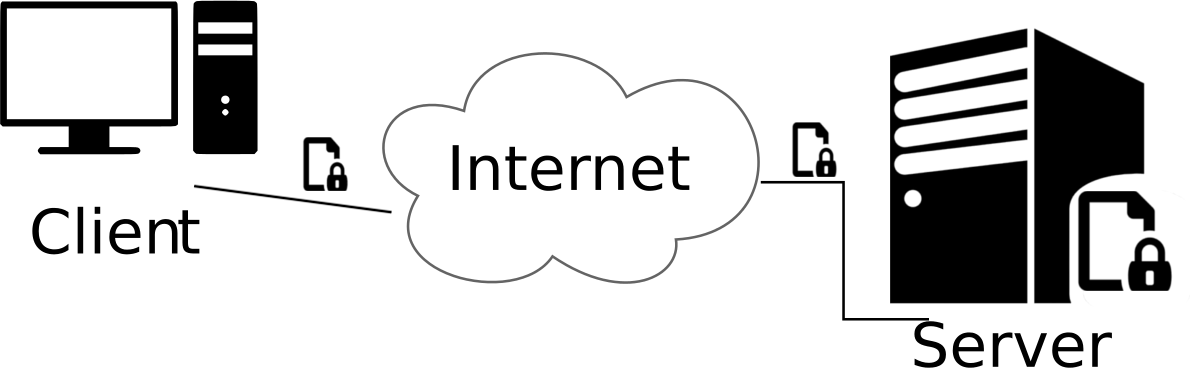}
          \caption{Classical Client-Server model}
          \label{fig:classical_cs_model}
        \end{figure}
        \\
        In the scheme, represented in fig \ref{fig:classical_cs_model}, both the client and the server are vulnerable to an attack, which means  that an attacker can decide which side to hit or to attack both at the same time. Obviously if an attacker decides to attack the client they will have specific information about the user, but if an attacker decides to penetrate the server, they will get much more information because they will get information about all users registered on that server. To make a comparison with the real world it is as if a thief has to choose between stealing a wallet or robbing a bank. 
        \\\\
        The second type of attack is more difficult than the first one, but if it is successful, it will cause much more damage. Moreover for a defender it is easier to defend a server rather than a client,  because latter is subjected to the behaviour of its user. Defence gets more difficult if users are less sensitive about security and unaware of risks.  Since we have two attack schemes we will have two defence schemes and for a deeper analysis we must choose whether to defend each single user or defend the server that stores huge amounts of information.
        \\\\
        The basic problem is that servers are vulnerable by definition. There are many bugs that may be used by attackers. 
        Thanks to these vulnerabilities attackers can retrieve all the sensitive information they want about users.
        \\\\
        The problem is that all information is saved on a single server, so if it is attacked, it will lose all its sensitive information. The new way of thinking could be to break the secret into $n$ pieces and share them with $n$ servers. Only the union of all small secrets can rebuild the original information. In this way attackers have to gather all the $n$-pieces if they want to rebuild the original secret. The family name of these schemes is \textit{secret sharing scheme} and there are many implementations. 
        \\\\
        In order to achieve our goal the better implementation is Shamir.  With \textit{Shamir secret sharing} it is possible to implement an authentication service that works in the following manner: usernames and passwords are not saved on a single server but when users sign up, their personal data, i.e. username and password, are broken down and distributed on a set of servers. At authentication time, the set of servers, with an adequate minimum message exchanges, will recompose the necessary information to evaluate if an incoming request is from a legitimate user or not. 
        \begin{figure}[h]
          \centering
          \includegraphics[width=0.7\textwidth]{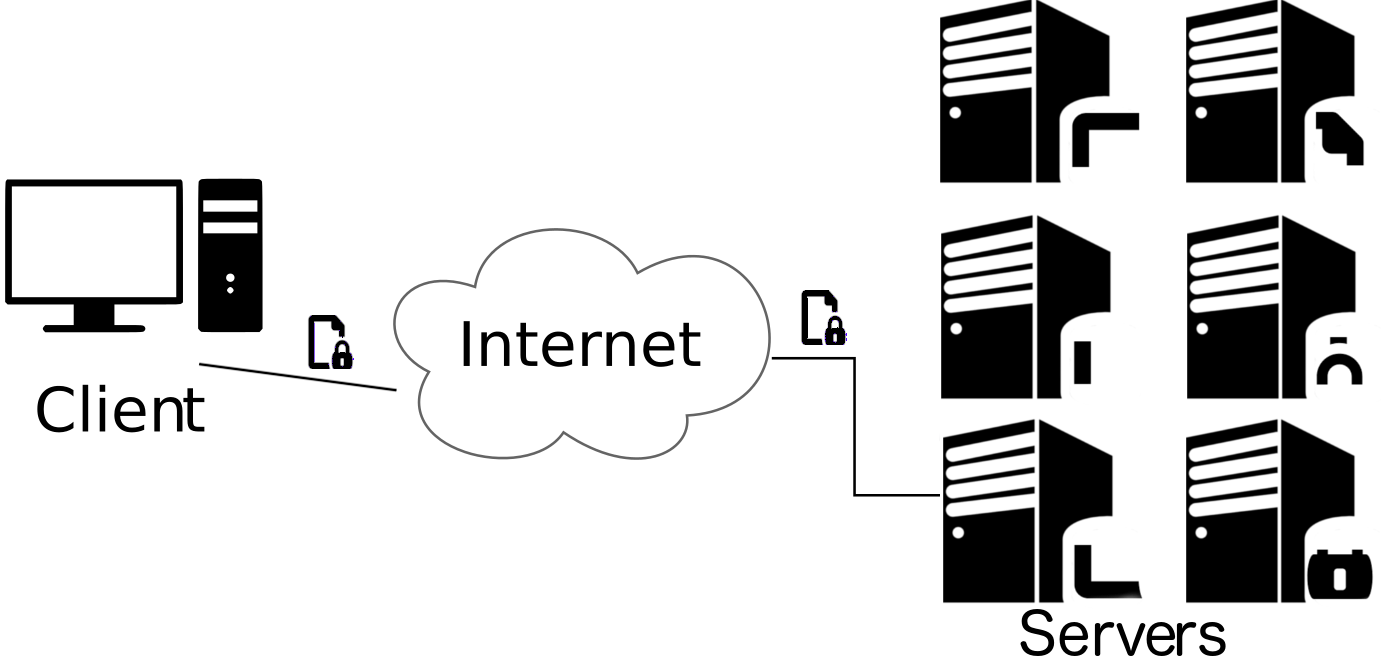}
          \label{fig:multi_cs_model}
          \caption{Multi Client-Server model}
        \end{figure}
    
    \subsection{Secret Sharing Schemes}
      In this section a secret sharing scheme idea is presented. A real working solution is provided to solve the problem of breaking the secret into $n$-pieces. In the introduction a general concept of the issue was explained, but now we must  introduce it in a structured way. We are getting to the heart of the matter. We know that the idea is to break the secret $S$ into many pieces. We must find a way to break the secret $S$ into $n$-pieces, and understand how it is possible to rebuild the original secret $S$.
      
      \subsubsection{Overview}
        In order to understand how secret sharing schemes work we will start from the base case, where there is a secret $S$, represented by an integer we want to split between two people $P_1$ and $P_2$. The easiest solution is:
        \begin{itemize}
          \item $P_1$ receives a random number $s_1= r$
          \item $P_2$ receives $s_2= S-r$
        \end{itemize}  
        To reconstruct the original message $P_1$ and $P_2$ must sum the shares received. Thus:
        \begin{equation*}
          \begin{split}
            S &= s_1+s_2\\
            & = r + S-r\\
            & = S
          \end{split} 
        \end{equation*}
        If we want to generalize this concept it will be possible to break the secret $S$ among $n$ people. Therefore:
        \begin{itemize}
          \item $\forall\ i \in \{1,\dots,n-1\}: P_i$ receives a random number $s_i= r_i\mod k$, where $k$ is an integer greater than all possible S.
          \item $P_n$ receives $s_n = S -\big(\sum_{i=1}^{n-1} r_i \mod k\big)$
        \end{itemize} 
        To reconstruct the original message all $P_i$ must sum the shares received. Thus:
        \begin{equation*}
          \begin{split}
            S &= \sum_{i=1}^{n} s_i\\
            & = s_1+s_2+\dots+s_n\\
            & = r_1\mod k+r_2\mod k+\dots+S -\big(\sum_{i=1}^{n-1} r_i\mod k\big)\\
            & = r_1 +r_2 +\dots+r_{n-1}+S-r_1 +r_2 +\dots+r_{n-1}\\
            & = \cancel{r_1} +\cancel{r_2} +\dots+\cancel{r_{n-1}}+S-\cancel{r_1} +\cancel{r_2} +\dots+\cancel{r_{n-1}}\\
            & = S
          \end{split} 
        \end{equation*}
        As we can see from this simple scheme if we want to reconstruct the original message $S$ all parties must agree and put their shares together. This model presents many difficulties if someone disagrees on the reconstruction or someone is unable to reply with their share.
        \\\\
        We are looking for a scheme where also with $t<n$ shares it is possible to reconstruct the original secret $S$. This kind of scheme are called \textbf{threshold schemes}.

        \begin{definition}{\textbf{(t,n)-Threshold scheme}}
          Let $t$ and $n$ be two positive integers with $t<n$ a threshold scheme is one way to share a secret $S$ among $n$ participants. The peculiarity of this model is that any subset of $t$ participants can reconstruct the secret S, while no subset of cardinality less than $t$ can do so.
        \end{definition}
        \leavevmode
        The implementation of this scheme can be done in different ways. The first two schemes were invented by Adi Shamir\cite{[2]SSSHow_to_ShareSecret.pdf} and George Blakley\cite{[2]BLAKLEY-Safeguarding-cryptographic-keys.pdf} independently in 1979. Shamir's idea is different from Blakley's. The former is based on the fact that 2 points are sufficient to define a line, 3 points are sufficient to define a parabola, and so forth. That is taking $t$ points it is possible to define a polynomial of degree t-1 with unique solution. The latter is based on the geometry of hyperplanes, in general taking $n$ non-parallel (n-1)-dimensional hyperplanes is sufficient to locate a specific point.
        \\\\
        In 1983 Mignotte\cite{[2]Mignotte1983} and Asmuth-Bloom\cite{[2]Asmuth-Bloom1983} proposed another kind of solution based on Chinese remainder theorem.

      \subsubsection{Notation}
        We will now attempt to clarify some notations that we will be using in the whole text. We are dealing with the case in which $t$ and $n$ are two positive integers s.t. $t<n$. A threshold scheme (t,n) is a method of sharing a secret $S$ among a set of $P$ participants, thus:
        \begin{itemize}
          \item Every group of participants with cardinality $\geq t$ can reconstruct the original secret
          \item Every group of participants with cardinality $\leq t$ cannot reconstruct the original secret
        \end{itemize}
        The value S is chosen by a special participant called \textbf{Dealer} and it will be indicated with the letter D. The set of participants is indicated with the letter P, 
        \[P = \{ P_i: 1\leq\ i\leq n\}\] 
        We have two possibilities:
        \begin{itemize}
          \item $D\in P$
          \item $D\not\in P$
        \end{itemize}
        When D wants to share the secret S among the participants, he assigns to each participant a partial information, called \textit{share}. Each \textit{share} is kept secret by each participant, and the dealer D distributes shares in a secret way. 
        \\\\
        Given a subset of participants $Q \subseteq P$ that join their shares in order to recompose the original secret S, if $|Q| \geq t$ they are able to obtain the original message, otherwise with $|Q|< t$ that is not possible. Recapping we have:
        \begin{itemize}
          \item $P = \{ P_i: 1\leq\ i\leq n\}$, the set of participant,players or shareholders.
          \item $D$, that is the dealer.
          \item $S$, that is the original secret
          \item $s_i$, that is i-th share given to the $P_i$ player.
        \end{itemize}
        In the next section we will see how it is possible to break the secret S and how it is possible to reconstruct it, thanks to Shamir's work. Then we will show that with less than $t$ shares it is not possible to obtain S.
    
    \subsection{Shamir Secret Sharing}
      In this section Shamir's scheme is introduced and it is deeply analysed in order to understand how it works. Shamir's idea works under the assumption that everyone is honest and everyone follows the protocol.
  
      \subsubsection{Sharing phase}
        % presentazione dello schema di shamir, sharing phase e reconstruction phase.
        Shamir's scheme works under the following assumptions in order to prove the security of the scheme:

        \begin{enumerate}[label=(\alph*)]
            \item the model works with $Z_p$, with $p$ prime.
          \item $p$ has to be greater than n, that is $p\ge n+1$, or we can say in a different way, $n<|Z_p|$
          \item the secret S has to be an element of $Z_p$, that is: $S \in Z_p$
        \end{enumerate}

        \begin{itemize}
          \item \textbf{Start phase}
            \begin{enumerate}
              \item D chooses n \textbf{distinct elements}. We have: \[ x_i\ with\ 1\leq i \leq n\]
              \item D distributes each $x_i$ to each $P_i$. The $x_i$ and $p$ values are public.
            \end{enumerate}

            \item \textbf{Sharing phase}
            \begin{enumerate}
            \item D wants to share S$\in Z_p$. 
            
            \item D secretly and randomly (uniformly and independently) chooses $t-1$ elements of $Z_p$, thus:
            \[a_1,a_2,\dots,a_{t-1} \in Z_p\]

            \item D builds a polynomial function s.t.:
            \begin{equation*}
              \begin{split}
                P(x) &= (a_0+a_1x+a_2x^2+\dots+a_{t-1}x^{t-1})\mod{p}\\
                &= a_0 + \Big(\sum_{j=1}^{t-1}a_jx^j \mod{p} \Big)\\
                assign\ a_0 =S\\
                &= S +\Big(\sum_{j=1}^{t-1}a_jx^j\mod{p} \Big)
              \end{split} 
            \end{equation*}
            D makes a random polynomial $P(x)$ of maximum degree $t-1$ and the constant term is the secret S. 
            \item For $1\leq i \leq n$, D computes:\[s_i = P(x_i)\]
            
            \item For $1\leq i \leq n$, D secretly distributes the share $s_i$ to each shareholder $P_i$. In the end each player has a point $(x_i,s_i)$ of the polynomial chart.
            \end{enumerate}

          \item \textbf{Reconstruction phase}\\
          If it is necessary to rebuild the secret S, this is possible only if the cardinality of the subset of $P$ is greater than $t$, otherwise the reconstruction phase fails. There are two ways to obtain S:
          \begin{itemize}
            \item through the system of linear equations
            \item through the polynomial interpolation concept
          \end{itemize}

          In the next section we will consider further how it is possible to reconstruct the original message and how it works and why it is secure. 
        \end{itemize}

      \subsubsection{Reconstruction phase}
        % Qui presentiamo i due modi per ricostruire il segreto, con le dimostrazioni sul perche funziona a lv teorico.
        \paragraph{System of linear equations}
          Given
          \begin{itemize}
            \item $Q = \{P_1,P_2,\dots,P_t\}$ s.t. $Q\subseteq P$ that wants to reconstruct S
            \item $P(x) \in Z_p$ the secret polynomial chosen by D with maximum degree $t-1$, thus:
    \[P(x) = (a_0+a_1x+a_2x^2+\dots+a_{t-1}x^{t-1})\mod{p}\]
            \item $s_i  = P(x_i)$ for $1\leq i \leq t$
            \item $a_1,\dots,a_{t-1} \in Z_p$ chosen by D in a random way.
            \item $a_0 = S \in Z_p$
          \end{itemize}
          Shareholders keep together the t-shares and they will obtain a system of t equations in unknowns $a_0,\dots,a_{t-1} \in Z_p$ since the $s_i$ are linear in the coefficients of $P(x)$ for $1\leq i \leq t$. The system of t equations is the following:
          \[
          \begin{cases} 
            a_0+a_1x_1+a_2x_1^2+\dots+a_{t-1}x_1^{t-1} = s_1\\ 
            a_0+a_1x_2+a_2x_2^2+\dots+a_{t-1}x_2^{t-1} = s_2\\ 
             \vdots\\
            a_0+a_1x_t+a_2x_t^2+\dots+a_{t-1}x_t^{t-1} = s_t\\ 
          \end{cases}\]
          In the matrix form the equations system $Aa= s $ is:
          \[\begin{pmatrix}
           1 & x_1 & x_1^2 & \dots & x_1^{t-1}\\ 
           1 & x_2 & x_2^2 & \dots & x_2^{t-1}\\ 
           \vdots & \vdots & \vdots& \vdots & \vdots \\
            1 & x_t & x_t^2 & \dots & x_t^{t-1}
           \end{pmatrix}
           \begin{pmatrix}
           a_0 \\ 
           a_1 \\
           \vdots\\
           a_{t-1}
           \end{pmatrix}
           =
           \begin{pmatrix}
           s_1 \\ 
           s_2 \\
           \vdots\\
           s_{t}
           \end{pmatrix} \]
           If the equations are linearly independent, there will be a \textbf{unique solution} and $a_0$ will reveal the original secret S.
           \\\\
           In order to prove the uniqueness of the solution we will use a Vandermonde matrix. The only three things to know are:

          \begin{definition}{\textbf{(Vandermonde matrix)}} A Vandermonde's matrix is a matrix whose rows (or columns) have elements starting from 1 and each item follows a geometric progression. 
          \[V = \begin{pmatrix}
           1 & \alpha_1^2 & \alpha_1^2 & \dots & \alpha_1^{n-1}\\ 
           1 & \alpha_2^2 & \alpha_2^2 & \dots & \alpha_2^{n-1}\\ 
           1 & \alpha_3^2 & \alpha_3^2 & \dots & \alpha_3^{n-1}\\ 
           \vdots & \vdots & \vdots& \vdots & \vdots \\
           1 & \alpha_m^2 & \alpha_m^2 & \dots & \alpha_m^{n-1}
           \end{pmatrix}\]
          \end{definition}
           
          \begin{theorem}\label{V}   A square Vandermonde matrix of order n has determinant:
          \[ det(V) = \prod_{1\leq i \leq j \leq n} (\alpha_j - \alpha_i)\]
       
          \end{theorem}

          \begin{theorem}{\textbf{Rouché–Capelli}}\label{RC}  
          The system of $n$ equations and $n$ unknowns $Ax =b$ has a unique solution $ \Leftrightarrow$ det(A)$\neq 0$. 
          \end{theorem}
           In order to prove the uniqueness of the solution of the system we have to demonstrate that the determinant is $\neq 0$, thanks to the Rouché–Capelli theorem.
           \\\\
            Given a $t$ x $t$ A matrix admits a unique solution because:
            \begin{itemize}
              \item thanks to \ref{V} we have:
               \[ det(A) = \prod_{1\leq j \leq k \leq n} (x_k - x_j)\mod{p}\]

              \item thanks to \ref{RC} and since all $x_i$ values are distinct, then each term $(x_k - x_j)\neq 0$, then $det(A)\neq 0$, so the system always admits a unique solution.

              \item This proves that in a threshold scheme $t$-shareholders will always be able to determine the original secret $S$, if they put their $t$-shares together.

            \end{itemize}
        Now we want to present the case where $t-1$ participants try to reconstruct S. When $P_1,P_2,\dots,P_{t-1}$ share each $s_i$ they obtain a $t-1$ equation system in t unknowns $a_0,\dots,a_{t-1} \in Z_p$. The system of linear equations is the following:
        \[
          \begin{cases} 
            a_0+a_1x_1+a_2x_1^2+\dots+a_{t-1}x_1^{t-1} = s_1\\ 
            a_0+a_1x_2+a_2x_2^2+\dots+a_{t-1}x_2^{t-1} = s_2\\ 
             \vdots\\
            a_0+a_1x_{t-1}+a_2x_{t-1}^2+\dots+a_{t-1}x_{t-1}^{t-1} = s_{t-1}\\ 
          \end{cases}\]
        Since for each polynomial the unknown terms are chosen by D, the $t-1$ shareholders are unable to find $a_0$ because in $Z_p$ the system will always have a non-unique solution.
        \\\\
        \textbf{Example}
          
        Given that:
        \begin{itemize}
          \item $p=17$, $n=5$, $t=3$
          \item public coordinates $x_i =i\ for\ 1\leq i \leq 5$
          \item subset of participants $Q\subseteq P$ s.t. $Q =\{P_1, P_3, P_5\}$
          \item the polynomial $P(x) = a_0+a_1x+a_2x^2$
          \item Dealer chooses $a_0 = 13$, $a_1 = 10$, $a_2 = 2$
        \end{itemize}
        The shares given to the shareholders in $Z_{17}$ will be:
        \begin{equation*}
          \begin{split}
        s_1= P(x_1=1)&= (a_0+a_1+a_2)\mod{17} = (13+10+2) = 25\mod{17} = 8 \\
        s_3=P(x_3=3)&=(a_0+3a_1+9a_2)\mod{17}=(13+30+18) = 61\mod{17} = 10 \\
        s_5=P(x_5=5)&=(a_0+5a_1+25a_2)\mod{17}=(13+50+50) = 113\mod{17} = 11 \\
        \end{split} 
      \end{equation*}
        Finally the system of linear equations will be the
        following:
        \[\begin{cases} 
          P(1)= a_0+a_1+a_2 \\
          P(3)= a_0+3a_1+9a_2\\
          P(5)= a_0+5a_1+8a_2
        \end{cases}\]
        It admits a unique solution in $Z_{17}$ that is: $a_0 = 13$, $a_1 = 10$, $a_2 = 2$, so the original secret is $a_0= S= 13$.

        \paragraph{Polynomial interpolation resolutions}
        The polynomial interpolation allows us to resolve the problem of function approximation that have a non-linear trend. We will use the Lagrange's interpolation in order to achieve our purpose. The Lagrange's interpolation is based on the Lagrange polynomials:
        \begin{definition}{\textbf{(Lagrange polynomials)}} For a given set of distinct points $x_{j}$ and numbers $y_{j}$, the Lagrange polynomial is the polynomial of lowest degree that assumes at each point $x_{j}$ the corresponding value $y_{j}$ (i.e. the functions coincide at each point). The polynomial has the following form:
        \[P_{n-1}(x) = a_0+a_1x+a_2x^2+a_3x^3+\dots = \sum_{i=0}^{n-1} a_ix^i\]
        $P_{n-1}(x)$ is the polynomial of (n-1)-degree with $n$ coefficients $a_i$.
        \end{definition}

        We want to interpolate $n$ points $(x_i, y_i)$ with $i=0,1,2,\dots,n$. Interpolating a polynomial, given a set of distinct points, means that the polynomial has to pass through all points  $(x_i, y_i)$. Hence each point has the following relationship:
        \[P_{n-1}(x_i) = y_i\]  
        For $n$ points we have:
        \begin{equation*}
          \begin{split}
            P_{n-1}(x_0) &= y_0\\
            P_{n-1}(x_1) &= y_1\\
            P_{n-1}(x_2) &= y_2\\
            \dots\\
            P_{n-1}(x_{n-1}) &= y_{n-1}\\
          \end{split} 
        \end{equation*}

        The solution will be $\vec{a}$ the vector of coefficients that allows us to determinate $P_{n-1}(x)$.
        The matrix form of the equations system $Aa= s $ is:

        \[\begin{pmatrix}
           1 & x_0 & x_0^2 & \dots & x_0^{n-1}\\ 
           1 & x_1 & x_1^2 & \dots & x_1^{n-1}\\ 
           \vdots & \vdots & \vdots& \vdots & \vdots \\
            1 & x_{n-1} & x_{n-1}^2 & \dots & x_{n-1}^{n-1}
           \end{pmatrix}
           \begin{pmatrix}
           a_0 \\ 
           a_1 \\
           \vdots\\
           a_{n-1}
           \end{pmatrix}
           =
           \begin{pmatrix}
           y_0 \\ 
           y_1 \\
           \vdots\\
           y_{n-1}
           \end{pmatrix} \]
      In order to prove the uniqueness of the solution of the system we must demonstrate that the determinant is $\neq 0$. As mentioned in the previous section the A-matrix is in the Vandermond form and the determinant is:  
      \[ det(A) = \prod_{1\leq i \leq j \leq n-1} (x_i - x_j)\]
      All values $x_i$ are distinct, then each term $(x_k - x_j)\neq 0$, then $det(A)\neq 0$, so the system always admits a unique solution.
      \\
      Once it has been proven that there is a unique solution, the real question is how it is possible to calculate it. The Lagrange's method presents an alternative way to resolve the system since the application of known methods could be difficult to apply.

      \begin{theorem}{\textbf{Lagrange polynomial interpolation}}
        \\Let $(x_0, y_0),\dots\ (x_{n-1}, y_{n-1})$ be $n$ distinct points, s.t $y_0,\dots y_{n-1} \in Z_p$ and values $x_0,\dots x_{n-1}$ are all different. There exists exactly one polynomial P of degree at most $n-1$, such that $P(x_i) = y_i$ for all $i \in {0, \dots , n-1}$.\\
        This polynomial P is the following:
        \[ P(x) = \sum_{j=0}^{n-1}\Big( y_j \prod_{0 \leq k\leq n-1, k\neq j} \frac{x-x_k}{x_j-x_k}\Big)\]
      \end{theorem}

      The above theorem claims that the polynomial with of degree at most $n-1$ is unique and gives us a formula that can be used to compute the polynomial $P(x)$ given the $(x_i,s_i)$ points.
      Hence, given that
          \begin{itemize}
            \item $P(x) \in Z_p$ the secret polynomial chosen by D with of degree at most $t-1$, such that:
    \[P(x) = (a_0+a_1x+a_2x^2+\dots+a_{t-1}x^{t-1})\mod{p}\]
            \item $s_i  = P(x_i)$ for $1\leq i \leq t$
            \item $a_1,\dots,a_{t-1} \in Z_p$ chosen by D in a random way.
            \item $a_0 = S \in Z_p$
          \end{itemize}
      The reconstruction formula to be used is:
      \begin{equation}\label{lagrandePoly}
        P(x) = \sum_{j=0}^{t-1}\Big( s_j \prod_{0 \leq k\leq t-1, k\neq j} \frac{x-x_k}{x_j-x_k}\Big)\mod{p}    
      \end{equation}
      \\
      The Q subgroup of shareholders that want to reconstruct S can calculate $P(x)$ through the \ref{lagrandePoly} equation. Since the solution is unique the result is the same as if we calculated with the known methods. 
      \\\\
      Since the Q subgroup does not need to know all the $P(x)$ polynomial it is sufficient that it calculate only the $P(0) = a_0 = S$ constant term. Hence, the \ref{lagrandePoly} equation can be rewritten setting $x=0$:
      \begin{equation}\label{eq:2}
        \begin{split}
        S = a_0&=P(0)= \sum_{j=0}^{t-1}\Big( s_j \prod_{0 \leq k\leq t-1, k\neq j} \frac{x_k}{x_k-x_j}\Big)\mod{p}\\
        with\ v_j &= \Big(\prod_{0 \leq k\leq t-1, k\neq j} \frac{x_k}{x_k-x_j}\Big)\mod{p}\\
        S &= \sum_{j=0}^{t-1}\big(s_j v_j\big)\mod{p}
        \end{split} 
      \end{equation}
      The original secret S is a modulo p linear combination of t-shares. 
      \\\\
      In the case of 
      \begin{itemize}
        \item $Q = \{P_1,P_2,\dots,P_{t-1}\}$ and
        \item $Q\subseteq P$ 
      \end{itemize}
      If the $t$-participants try to reconstruct $S$ joining together the $s_1,\dots,s_{t-1}$ shares, they will fail. This is because for each possible value of $S$, there exists exactly one polynomial P of degree at most t, so that:
      \[P(x_0)=S,\ P(x_1)=s_1,\ P(x_2)=s_2,\dots P(x_{t-1})=s_{t_1}\]
      Hence all values of S are equally possible. 
      \\\\
      \textbf{Example}
      \\  
      Given:
      \begin{itemize}
        \item $p=17$, $n=5$, $t=3$
        \item public coordinates $x_i =i\ for\ 1\leq i \leq 5$
        \item subset of participants $Q\subseteq P$ s.t. $Q =\{P_1, P_3, P_5\}$
        \item the polynomial $P(x) = a_0+a_1x+a_2x^2$
        \item Dealer chooses $a_0 = 13$, $a_1 = 10$, $a_2 = 2$
      \end{itemize}
      The shares given to the shareholders in $Z_{17}$ will be:
      \begin{equation*}
          \begin{split}
        s_1=P(x_1=1)&= (a_0+a_1+a_2)\mod{17} = (13+10+2) = 25\mod{17} = 8 \\
        s_3=P(x_3=3)&=(a_0+3a_1+9a_2)\mod{17}=(13+30+18) = 61\mod{17} = 10 \\
        s_5=P(x_5=5)&=(a_0+5a_1+25a_2)\mod{17}=(13+50+50) = 113\mod{17} = 11 \\
        \end{split} 
      \end{equation*}

      The shareholders can compute $v_j$
      \begin{equation*}
          \begin{split}
            v_1 &= \Big(\prod_{0 \leq k\leq t-1, k\neq j} \frac{x_k}{x_k-x_1}\Big)\mod{17}\\
            &= \Big(\frac{x_3}{x_3-x_1} \frac{x_5}{x_5-x_1}\Big)\mod{17} = \frac{x_3 x_5}{(x_3-x_1)(x_5-x_1)}\\
            &= \frac{3*5}{(3-1)(5-1)} = \frac{15}{(2)(4)} = \frac{15}{(-2)(-4)} = 15*(-2)^{-1}(-4)^{-1}\\
            &= 15\mod{17}*(-2\mod{17})^{-1}\mod{17}*(-4\mod{17})^{-1}\mod{17}\\
            &= 15*(15)^{-1}\mod{17} *(13)^{-1}\mod{17} = (15*8*4)\mod{17} = 480\mod{17} = 4\\
            v_3 &= \Big(\prod_{0 \leq k\leq t-1, k\neq j} \frac{x_k}{x_k-x_3}\Big)\mod{p}\\
            &= \Big(\frac{x_1}{x_1-x_3} \frac{x_5}{x_5-x_3}\Big)\mod{p} = \frac{x_1 x_5}{(x_1-x_3)(x_5-x_3)}\mod{p}\\
            &= \frac{1*5}{(1-3)(5-3)}\mod{17} = 5*(-2)^{-1}*(2)^{-1}\mod{17} =3
      \end{split} 
        \end{equation*}
      \begin{equation*}
          \begin{split}
            v_5 &= \Big(\prod_{0 \leq k\leq t-1, k\neq j} \frac{x_k}{x_k-x_5}\Big)\mod{p}\\
            &= \Big(\frac{x_1}{x_1-x_5} \frac{x_3}{x_3-x_5}\Big)\mod{p} = \frac{x_1 x_3}{(x_1-x_5)(x_3-x_5)}\mod{p}\\
            &=\frac{1*3}{(1-5)(3-5)}\mod{17}= 3*(-4)^{-1}(-2)^{-1}\mod{17} = 11
          \end{split} 
        \end{equation*}

        Finally the participants can compute S in the following way:
        \begin{equation*}
          \begin{split}
            S&= \sum_{j=0}^{t-1}\big(s_j v_j\big)\mod{p}\\
            &= (s_1v_1+s_3v_3+s_5v_5)\mod{p}\\
            &= (8*4+10*3+11*11)=183\mod{17} = 13
          \end{split} 
        \end{equation*}

    %\newpage
    %\subfile{Chapters/2-State_ofthe_art/Shamir_limits/ShamirLimits}  

    \subsection{Shamir's limits and how to overcome them}
      %non è stato ideato contro gli attacchi attivi (sia del dealer che dei shareholders), introduzione di cosa andremo a parlare. Perché non funziona, ipotetici attacchi (quelli del attack model 1.0), cioè dealer can cheat. Players can cheat. In che modo. 
      Shamir's secret sharing presented in the previous chapter is based on the idea that both the shareholders and the dealer are honest. Indeed someone could cheat. The problem of cheating has to be seen from both sides, that is what happens if the dealer D cheats and what happens if some participants cheat. First of all we will analyse the cases related to the cheating dealer, then those concerning the cheating players and finally the two cases together.
      \\\\
      There are different categorizations: the attacker could be \textbf{passive} or \textbf{active} and in the distribution systems there are \textbf{honest} or \textbf{dishonest}(\textbf{malicious}) actors. The difference is subtle because an honest or dishonest player may or may not modify the protocol. The passive attacker cannot interact with any of the parties involved in the system but he could for example steal the actor's database. The active attacker could be seen as a dishonest player because he changes the communication protocol. Hence an honest player could also be passive, or a dishonest player could also be passive one. However it is not possible to have an honest and active actor. 
      \\\\
      The dealer could be:
      \begin{itemize}
        \item \textbf{honest} - A dealer is honest only if the secret reconstructed by any combination of k shareholder is the same. In this case, we say that the players' shares are consistent.
        \item \textbf{dishonest} - The dishonest dealer can act in different ways according to the phase he is treating. In the Sharing phase the misleading dealer may be distributing shares $s_1, ... , s_n$ to the players so that when players $i_1, ... , i_k$ put their shares together, they get the secret $S$, but when players $j_1, ... , j_k$ put their shares together, they get the secret $S' \neq S$.
        \\\\
        In the reconstruction phase a malicious dealer can force shareholders to send other shares $s_i$ of another original secret $S_1$ to it in order to rebuild another original secret. By repeating this procedure, the attacker can rebuild, the whole dataset.
          
        \item \textbf{passive} - The passive attacker takes control of the dealer and can only take some information. The dealer deals (in the two phases) only with user specific information. In the Sharing and in the Reconstruction phase the attacked dealer can retrieve S. 
      \end{itemize}

      A shareholder could be:
      \begin{itemize}
        \item \textbf{honest} - A shareholder is honest only if in the Reconstruction phase he sends the dealer D the share $s_i$ given to him at the Sharing phase.

        \item \textbf{dishonest} - The dishonest shareholder could send altered shares to the dealer in order to compromise the Reconstruction phase.
          
        \item \textbf{passive} - The passive attacker takes control of the shareholder and can only take information about the share from infected players.
      \end{itemize}

      \subsubsection{Verifiable Secret Sharing}
        A basic secret sharing scheme is based on the assumption that all the scheme's actors run the protocol without modifying anything. In many real scenarios a secret sharing scheme is also required to be safe against malicious parties. The concept of Verifiable Secret Sharing was introduced for the first time by Chor, Goldwasser, Micali and Awerbuch \cite{[3]BChor-SGoldwasser-SMicali-BAwerbuch-VerifiableSecretSharing-and-achieving-simultaneity-in-the-presence-of-faults} in 1985. They introduced the concept that every user can verify if they have received a valid share. 
        \\\\
        A VSS scheme has to be safe against the following two types of active attacks:
        \begin{itemize}
          \item the misleading dealer sends incorrect shares $s_i$ to some shareholder in order to sabotage the reconstruction phase.

          \item participants send wrong shares during the reconstruction protocol.
        \end{itemize}
        Shamir’s scheme is not a VSS scheme, since it is not safe against these types of attacks.
        \\\\
        In order to have a strong VSS it has to be secure against a malicious dealer \textbf{and} shareholders. In literature there are some papers that discuss only one case, for example Rabin and Ben-Or \cite{[3]Tal_Rabin_BenOr_VSS_and_multiparty_protocols_with_honest_majority} in 1989 hypothesize a trusted dealer and thanks to the Check Vectors concept they achieve their purpose. The proposed solution is secure if more than half of the players are honest and there are appropriate communication channels.
        \\\\
        Generally in the VSS scheme the shareholders are allowed to send messages to each other and to the dealer in order to verify the correctness of their shares. Furthermore, in the reconstruction phase each player must be able to prove the correctness of their share.
        \\\\
        In general a VSS scheme has to implement the following communication model:
        \begin{itemize}
        \item dealer and shareholder can send \textbf{secret messages} to each other; and
        \item  assume the existence of a \textbf{broadcast channel} between dealer and players.
        \end{itemize}
        In 1987 Feldman \cite{[3]Feldman-A-PracticalScheme-for-Non-interactiveVerifiableSecretSharing} proposed another type of VSS scheme based on publicly known committed values. Feldman's scheme uses the computationally secure hard problem known as discrete logarithm problem in order to preserve the security of S. Later, in 1992 Pedersen \cite{[8]Pedersen_Non-Interactive_and_Information-TheoreticSecure_VerifiableSecretSharing} used a commitment scheme to overcome Feldman’s VSS scheme. He proposed a theoretically secure scheme with one exception: if a malicious dealer is able to solve the discrete logarithm problem they could distribute incorrect shares.
        \\\\
        Hence, we can categorize the VSS schemes in two classes: 
        \begin{itemize}
          \item \textbf{Interactive.} Interactive VSS schemes require interactions between dealer and shareholder, this means that the dealer has to be active for the whole duration of the protocol. The problem of this kind of scheme is the number of interactions required in order to achieve the aim. These kinds of schemes use authentication tags, checking vectors, or similar ideas.

          \item \textbf{Non-interactive.} Feldman and Pedersen's schemes are called non-interactive because the distribution protocol does not require any interaction between the dealer and participants, or among participants (themselves).
        \end{itemize}
        The non-interactive schemes are better than interactive ones because the latter require some interaction among users. Moreover, in the interactive scheme the communication complexity is exponential.
        \\\\
        In literature, VSS schemes are also categorized according to the adversarial computational power: \textbf{computational VSS} schemes and \textbf{unconditional VSS schemes}. The computational VSS is safe against adversaries who are computationally bounded, while in the second one the adversary may possess unbounded computational power. The computational VSS schemes are more practical and efficient in terms of message and communication complexities compared to the unconditional schemes.
        \\\\
        Another special case is a Publicly verifiable secret sharing (PVSS) scheme, introduced by Stadler \cite{[11]MarkusStadler-PubliclyVerifiableSecretSharing}, it ensures that anyone (not just the participants) is able to verify the correctnesses of shares distributed by the dealer 
        \\\\
        In the end, VSS is important because it allows a dealer and shareholders to work also with the presence of faults.
          
      \subsubsection{Pedersen's VSS}
        % Riprenrimao da fieldamn, dicendo che studieremo meglio questo schema perchè è quello che ci serve.

        % Partendo da Fieldman ed arrivando a Pedersen. Vedendo formule e dimostrazione
        We will examine Pedersen's VSS in-depth in order to understand how it works. We will do so because the proposed solutions will be based upon Pedersen's work. Pedersen's scheme is \textbf{non-interactive}, it requires unidirectional private links from the dealer to the shareholder and players speak only via the broadcast channel. Pedersen uses Fieldman's work as a basis and tries to improve it. In Feldman’s VSS scheme the committed values are publicly known and the privacy of secret $s$ depends in the difficulty of solving the discrete logarithm problem so it is computationally secure. Pedersen built a non-interactive and information-theoretically secure VSS scheme based on Feldman’s VSS scheme. There is an exception: when the malicious dealer is able to solve the discrete logarithm problem they can distribute incorrect shares.
        \\\\
        The studied scenario by Pedersen in his paper \cite{[8]Pedersen_Non-Interactive_and_Information-TheoreticSecure_VerifiableSecretSharing} is the following: let the dealer be the person who has a secret and distributes it to $n$ shareholders, where n$>$2, if the dealer completely trusts the shareholder, it could give the secret to this person and then avoid the difficulties of having a secret sharing scheme. Thus in many applications the dealer does not trust the shareholder completely, and therefore it should be expected that (some of) the shareholders will not trust the dealer either. 
        \\\\
        Pedersen' s VSS  scheme  is constructed by combining:
        \begin{enumerate}[label=\alph*.,start=1]
          \item Shamir's scheme
          \item Commitment scheme
        \end{enumerate}
        Indeed, this VSS could be seen as a \textbf{distributed commitment scheme}.

        \paragraph{The Commitment Scheme}
          In the simple commitment scheme there are the committer and the verifier. The \textbf{committer}  electronically locks the message in a box and sends the box to the verifier, the box is called \textbf{commitment}. The \textbf{verifier} receives the locked message and verifies that a certain message is contained in the box through extra information sent by the committer.
          \\\\
          The committer sends the commitment to an $s \in \mathbb{Z}_q$ by choosing  $t \in \mathbb{Z}_q$ at random and computing:
          \begin{center}
            $c = g^s h^t\mod{q}$
          \end{center}
          The commitment can later be opened by revealing s and t and this proves that c reveals no information about s, and that the committer cannot open a commitment to s as $s' \neq s$ unless it can find $\log_g (h)$.
          \\\\
          Before beginning the verifier must choose the following parameters: 
          \begin{itemize}
            \item Large primes $p$ and $q$ such that $q$ divides $p-1$
            \item Generator $g$ of the order-q subgroup of $\mathbb{Z}_p^*$
            \item Random secret $a \in$ $\mathbb{Z}_q$ 
            \item Let h be element of $G_q$, where $G_q$ is the unique subgroup of $Z_p^*$. In order to choose $h$, as shown  in the original paper, it can be calculated in the following way:
              \begin{center}
                $h = g^a\mod{q}$
              \end{center}
            $a$ has to keep secret.
          \end{itemize}
          The elements $g$ and $h$ are in $G_q$ so \textbf{nobody} knows $\log_g h$. We have to notice that in a secret sharing scheme these elements must be taken in a safe way because the correctness of the scheme is based on these two values. Hence they can either be chosen by a  \textbf{trusted centre}, when the system is initialized, or by (some of) the participants using a coin-flipping protocol.
          For example in a real scenario these two elements could be taken in set up phase where we hypothesize that no one can cheat. 
          \\\\
          The security of Pedersen Commitments is the following \cite{[8]Slide-Pedersen-Non-Interactive_and_Information-Theoretic-Secure-Verifiable-Secret-Sharing} :
          
          \begin{itemize}
          \item \textbf{Perfectly hiding commitment scheme}
            \begin{itemize}
                \item Given commitment $c$, every value $s$ is equally likely to be the value committed in $c$
                \item Given s, t and any $s^\prime$, there \textbf{exists} $t^\prime$ such that 
                \begin{equation}
                  \begin{split}
                    g^s h^t &= g^{s^\prime} h^{t^\prime}\\
                    \text{let } h&=g^a\mod{q}\\
                    g^s g^{at} &= g^{s^\prime} g^{at^\prime}\\
                    s+at &= s^\prime +at^\prime\\
                    at^\prime &= s-s^\prime+at\\
                    t^\prime &= (s-s^\prime)a^{-1}+t\mod{q}
                  \end{split} 
                \end{equation}
                To compute $t^\prime$ the committer must know $a$, but this it is not possible because $a$ is chosen by the verifier and it is kept secure.
                
            \end{itemize}
          
          \item \textbf{Computationally binding commitment scheme}:\\
          If the sender can find different s and $s^\prime$ both of which open commitment $c = g^s h^t\mod{q}$, then it can solve discrete log:
              \begin{itemize}
              
                \item Suppose sender knows s,t,$s^\prime$,$r^\prime$ s.t. $g^s h^t = g^{s^\prime} h^{t^\prime}\mod{q}$
                \item Since $h = g^a\mod{q}$, this means $s+at = s^\prime+at^\prime \mod{q}$
                \item Hence, committer can compute $a= \frac{(s^\prime-s)}{(t-t^\prime)}\mod{q}$
                \item This implies the committer is able to compute discrete logarithm of h, but it is unfeasible.      
              \end{itemize}
          \end{itemize}

        \paragraph{Non-interactive Verifiable Secret Sharing}
          Now we are going to see how to merge the commitment scheme seen above with Shamir's scheme.
          \\
          If we pick up $p$ and $q$ so that $q$ divides $p-1$, and let $g,h$ be elements of $G_q$, where $G_q$ is the unique subgroup of $Z_p^*$. There are n participants $P_1, P_2, ... , P_n$ and a dealer D who will divide a secret $S \in \mathbb{Z}_q$.

          \begin{enumerate}
            \item \textbf{Sharing Phase.} Dealer D executes the following steps:
            
              \begin{itemize}
                \item Choose F $\in Z_q[x]$ of degree at most $(t-1)$ s.t. $F(0)=a_0 =S$ and computes $s_i = F(i)$ for $1\leq i \leq n$. 

                \item Let $F(x) = a_0+a_1x+...+a_{t-1}x^{t-1}\mod{p}$, where all coefficients $a_i$ are chosen at random for $1\leq i \leq \text{t-1}$ are in $G_q$.
                
                \item Choose $b_0,..., b_{t-1} \in G_q$ at random. Let $K(x) = b_0+b_1x+...+b_{t-1}x^{t-1}\mod{p}$
              
                \item Computes the pair: $(s_i, t_i) = (F(x_i), K(x_i)) \ for \ i = 1, . . . ,n$

                \item Computes  $c_j = g^{a_j} h^{b_j}\mod{p} \ for \ j = 0,.., t-1$

                \item Outputs a list of n shares $(s_i, t_i) $ and distributes each share to corresponding participants $P_i$ \textbf{privately}. D also \textbf{broadcasts} $c_j$.
              \end{itemize}

            \item \textbf{Share verification}
            Each participant $P_i$, who has received the share $(s_i, t_i) $  and all broadcast information, can verify that the share defines a secret by testing:
            \begin{equation}\label{perdersenProof}
               \ g^{s_i} h^{t_i} = \prod_{j=0}^{t-1}{c_j^{i^j}} (\mod{p}) 
            \end{equation}
          The proof is the following:

          \begin{equation}\label{eq:commitment}
            \begin{split}
              g^{s_i} h^{t_i} &= g^{F(x_i)} h^{K(x_i)} = g^{(a_0+a_1x_i+\dots+a_{t-1}x_i^{t-1})} h^{(b_0+a_1x_i+\dots+b_{t-1}x_i^{t-1})}\\
              &\text{knowing: } a^{m+n} = a^m a^n\\
              &=(g^{a_0}g^{a_1x_i}\dots g^{a_{t-1}x_i^{t-1}}) (h^{b_0} h^{b_1x_i}\dots h^{b_{t-1}x_i^{t-1}})\\
              &= \prod_{j=0}^{t-1}g^{a_j x_i^j}\ \ \prod_{j=0}^{t-1}h^{b_j x_i^j}\\
              &= \prod_{j=0}^{t-1}g^{a_j x_i^j}\ h^{b_j x_i^j}\\
              &\text{knowing: } (ab)^n= a^n b^n \text{ and: }(a^{n^m}) = a^{nm}\\
              &= \prod_{j=0}^{t-1}g^{a_j^{x_i^j}}\ h^{b_j^{x_i^j}} = \prod_{j=0}^{t-1}(g^{a_j}\ h^{b_j})^{x_i^j}\\
              &\text{knowing: } c_j=(g^{a_j}\ h^{b_j})\\
              &=\prod_{j=0}^{t-1} c_j^{x_i^j} \qed
            \end{split} 
          \end{equation}
          \textit{Observation.} \ref{eq:commitment} can also test if the given coordinate $x_i$ is correct. Indeed, if the shareholder receives a wrong coordinate the check fails.
          % to do.. mettere anche il caso in cui la coordinata x_i non e quella vera
          
            \item \textbf{Reconstruction phase}\\
            It is the same as of Shamir’s scheme. Since this phase is done in a second moment, the dealer who does not trust the shareholders, can check if the shares $(s_i, t_i)$ sent by the participants are correct thanks to the \ref{perdersenProof} formula.
          \end{enumerate}
          In Pedersen’s scheme, the value $g^{a_0}$ is not made publicly known, that is, the secret S is embedded in the commitment $c_0 = g^{a_0} h^{b_0}$. Thus, no information about the secret S is revealed directly and even if an attacker with unlimited computing power can solve $log_g h$ , the attacker still gets no information about the secret S.
          \\
          Perdersen's VSS scheme is secure against cheating by the dealer and cheating by at most $t$-1 of the participants as long as $2(t-1)< n$.

  %%%%%%%%%%%%%%%%%%%%%%%%%%%%%%%%%
  %      LOGICAL ARCHITECTURE   %
  %%%%%%%%%%%%%%%%%%%%%%%%%%%%%%%%%
  \newpage
  \section{Logical architecture}\label{chap:Logical architecture}
    In this chapter, we will examine logical architecture. In the first part the actors and their roles are presented. In the second part three different solutions are dealt with and for each scheme an attack model and attack model check are discussed. The first solution is made under strong assumptions; then the different levels of impairment are studied supposing that the strong assumptions are no longer true. Finally two different protocols are explained and they are compared so as to understand which is the better one.

    \subsection{Actors and roles presentation}
      % presentazione dei vari attori e ruoli. A questo punto il chiamato fino ad ora External Server sarà rinominato Service. Cioè non si vuole per il momento introdurre l’idea di authentication-as-a-service quindi è il server dove è situato il servizio al quel l’utente vuole accedere.
      In this chapter the solution schemes will be presented. In every scheme or model there are some actors and each actor has different roles and functionalities. The purpose of this section is to list each possible actor and describe their role in the different schemes. The actors are the following:
      \begin{itemize}
        \item \textbf{Client.} The client is the user who wants to log in. The client's role will change in the different schemes. In the first \ref{Scheme_ISOL} and in the second \ref{Scheme_Second} scheme the client will not perform many operations. In these two models the client must send the necessary information in order to sign up and log in. Moreover, in the second scheme, the client carries out checks so as to be sure the other players do not cheat. Instead, in the third \ref{Scheme_III} scheme the client plays a key role because the Dealer logic is moved to the client side. 
        
        \item \textbf{Dealer.} The Dealer is the manager of the system. The dealer has a fundamental role in the system because they handle all the phases. Their main roles are the following:
        \begin{itemize}
          \item Dealer receives the secret $S$ from client, breaks $S$ and distributes the shares among the shareholders.

          \item Dealer receives the secret $\overline{S}$ from client. Dealer rebuilds the original secret $S$ from the shares and compares the two values. If $S$ is equal to $\overline{S}$, the client is authenticated.

          \item Dealer checks that Shareholders, Service and Client do not cheat.
        \end{itemize}
      
        \item \textbf{Shareholders.} The Shareholder acts like a container. They store the shares for each client and when it is necessary they send the piece of information to the Dealer. Moreover, the shareholder checks the share consistency, that is they verify the Dealer's correctness indirectly.

        \item \textbf{Service.} Logically, the Service is the machine where the service is implemented. The service is every type of online functionality offered to users, that is to say a social network, e-commerce web site, university web site, etc.

        \item \textbf{External Server.} The External Server is a server used to implement further features. These features are not essential to the functioning of the proposed solutions, but are like adds-ons, hence they could be included or not. Thus the proposed solutions are flexible to the various contexts in which they operate.
      \end{itemize}

    \subsection{Secret sharing schemes for authentication purposes}
      % ripresa del concetto visto nella sezione “A new way of thinking”. Introduzione della possibilità di usare SSS per spezzare il segreto e rendere più sicuro password-based authentication.
      The goal of this study is to design an authentication service that handles usernames and passwords in a new and extremely secure way. The basic idea consists in breaking the password down into $n$ pieces and distributing them on a set of servers. The key point is that only with at least $t$ pieces is it possible to reconstruct the secret. Thus if an attacker wants to obtain a user's password, they must obtain control of $t$-servers. 
      \\\\
      In order to implement the above idea a secret sharing scheme is used. In all three solutions Shamir and Pedersen's idea are adopted so that the goal fixed in \autoref{chap:Introduction} can be achieved. Shamir's work is used for key breaking, instead Pedersen's work is used to check shares correctness. Pedersen's solution and Shamir secret sharing are not sufficient because they are a general idea, not studied for authentication purposes. In order to implement the authentication through Shamir's and Pedersen's idea some precautions are necessary. In Shamir's original paper the Dealer is the holder of the secret and is the who breaks the secret in many pieces and distributes them among the Shareholders. In the password-based authentication it is no longer possible for the secret to be kept hidden because in order to verify the user's identity the verifier must have prior knowledge of the secret. This causes many problems for the privacy of the secret, that is the password in the password-based authentication. Hence the password cannot be sent in plaintext and some protection mechanisms must be adopted so that even if the Dealer is compromised and they will be impossible to retrieve any useful information.
      \\\\
      Finally, on a high level, the system could work in the following manner: username and password are not saved in a single server but when users sign up, their personal data, i.e. username and password, are broken down and distributed on a set of servers. At authentication time, the set of servers, with adequate minimum message exchanges, will recompose the necessary information in order to verify whether or not the incoming request is from a legitimate user.

    \subsection{First proposed solution with strong assumptions}
      \subsubsection{Attack Model}\label{sec:AttackModel_I_SOL}
        In this section a first study of the attack model against our system is examined. The most important class of attacks that a malicious person can perform in a general secret sharing scheme are introduced. The attack model is important because the knowledge about what the attacker can do is necessary in order to better understand how it is possible to react.
        \\\\
        Starting from this section an initial scheme is built in order to achieve the aim fixed in \autoref{chap:Introduction}. 
        The first solution is found in a relaxed form of attack model which concerns a trusted dealer. Trusted dealer means that the dealer is a trusted authority that will send the correct shares to the shareholders without stealing any information. Based on this strong assumption, a secret sharing scheme is designed, then the different levels of impairment are studied supposing that the previous assumptions are no longer true.
        \\\\
        In a second moment a relaxed attack model is no longer used and under this assumption the system will be corrected in order to mitigate the damage caused by those kind of attacks.
        \\\\
        Before illustrating the study it is necessary to explain the three levels of security:  
        \begin{itemize}
          \item \textbf{Safe:} obviously secure
          \item \textbf{Not completely safe:} the attacker can take specific information about some user, but other users' information is not touched.
          \item \textbf{Not safe:} not secure, the attacker can take all the dataset or they can compromise the protocol in order to create misunderstandings.
        \end{itemize}
        There are different classification, the attacker could be \textbf{passive} or \textbf{active} and in the distribution systems there are \textbf{honest} or \textbf{dishonest}(\textbf{malicious}) actors. The difference is subtle because an honest or dishonest player could respectively modify the protocol or not. The passive attacker cannot interact with any of the parties involved in the system but for example they could steal the actor's database. The active attacker could be seen as a dishonest player because they change the communication protocol. Hence there could be an honest player but a passive one, or a dishonest player who also acts as a passive one. However it is impossible to have an actor who is both honest and active.
        \\\\
        Since a trusted dealer is supposed, the hacker can attack only the participants, then a shareholder could be:
        \begin{itemize}
          \item \textbf{honest} - A shareholder is honest only if in the Reconstruction phase they send to the dealer D the share $s_i$ given to him at the Setup phase.

          \item \textbf{dishonest} - A dishonest shareholder could send altered shares to the dealer in order to compromise the Reconstruction phase.
            
          \item \textbf{passive} - A passive attacker takes control of the shareholder and they can only take information about the share from an infected player.
        \end{itemize}
        Another important fact should be noted, an attacker who wants to reconstruct the original secret $S$ by taking the control of at least $t$-players, in order to have $t$-shares, must know the abscissa vector. In the original Shamir's scheme this information is public, but nothing prevents it from being kept secret. In this way, though the attacker is able to take control of $t$-shareholders they cannot reconstruct the original secret $S$ without the abscissa vector.

        \paragraph{Passive attack}
          \textbf{Def. } \textit{The passive attacker takes control of \textbf{K} shareholders. The raider can only take some information from infected players.}
          \\\\
          There are two cases:
          \begin{itemize}
            \item \textbf{The attacker has the abscissa vector.} In this case there is a threshold t that:
              \begin{itemize}
                \item \textbf{K $<$ t:} in this case, thanks to the threshold scheme definition, the solution is \textbf{safe}.

                \item \textbf{K $\geq$ t:} in this case it is not possible to do anything because by definition with $t$-shares anyone can reconstruct the original secret $S$. Hence, if an attacker takes control of $t$-participants and  has the abscissa vector, they can rebuild the original secret $S$, and the solution is \textbf{not safe}.
              \end{itemize}

            \item \textbf{The attacker doesn't have the abscissa vector.} In this case \textbf{without} the vector of abscissas the attacker cannot rebuild the original secret $S$ even if they have K $\geq$ t, hence the solution is \textbf{safe}.
        
          \end{itemize}

        \paragraph{Active attack}
          \textbf{Def. } \textit{The active attacker takes control of \textbf{K} $\geq$ 1 shareholders. The aggressor can send \textbf{altered} information to the dealer in order to compromise the protocol.}
          \\\\
          The attacker could send altered information to the dealer during the Reconstruction phase. The altered information could be the share and/or the abscissa.
          \\\\
          In this case we need some extra information in order to understand if a share $s_i$ is not wrong, that is if it corresponds to a piece of the original secret $S$.
          This is a typical case in which \textbf{commitment scheme} is needed. The message in the box is hidden from the receiver, who cannot open the lock itself. Since the receiver has the box, the message inside can neither be changed nor merely revealed. Only when, in a second moment, the sender chooses to give the receiver the key, is it possible to open the locked box.
          \\\\
          Hence thanks to \textbf{Pedersen's Verifiable Secret Sharing Scheme} the dealer can check if some participants are malicious and if they are trying to cheat. Pedersen's checking is useful also to verify that the given abscissa is correct. Therefore the solution is \textbf{safe}.
          \\\\
          In order to maximize the security level of the system with this attack model, a scheme should be designed where the shareholder keeps only the share $s_i$. Instead the abscissa vector is kept secure somewhere and somehow but not in the shareholder side.

    \subsection{Scheme}\label{Scheme_ISOL}
      In this section a secure solution against the attack model seen in the previous chapter will be presented. The dealer is assumed to be honest and it runs the protocol without changing anything. Therefore, the dealer cannot steal any information and it is safe against passive attacks.  
      \\\\
      In the scheme presented by Shamir the dealer figure holds and knows the information $S$ and wants to keep it secret. In our case the dealer concept is moved and the client no longer performs the dealer actions. The dealer has an intermediate role, that is, upon the arrival of the secret S, they break the secret and transmit the shares $s_i$ to each shareholder $P_i$. Then, in a second moment, through the reconstruction phase, the dealer checks if the client who is trying access to the service is authorized and verifies that the shareholders are not cheating. The dealer verifies the system correctness.
      \\\\
      Once the protocol has been presented, the concept seen in the previous chapter will be resumed in a formal way. In order to improve the scheme security level it is possible to keep the abscissas $x_i$ secret and to allow the dealer to verify the share correctness. This is a big step forward for the Shamir Secret Sharing scheme because without the abscissa vector it is not possible to reconstruct the original secret $S$. The abscissa vector can be saved in different places, as we will see in the following paragraph.
      \\\\
      Another important concept is represented by the crystal box, which enriches the schemes with a new feature. This characteristic is used by those systems that need to be always notified when a secret $S$ is recomposed. In this way the system is able to keep track of when and who opens the original secret. In order to reach this feature the protocol has to be designed in such a way that the system keeps track of indispensable information for the secret reconstruction. Hence, without that information it is not possible to recompose the secret $S$, in this way the system can know when the secret is open and who opens it.
      \\\\
      An attack model check will be made in order to verify that the proposed scheme is safe against that kind of attacks. Next we will study what happens if the supposing initial hypotheses are no longer true, which is what happens if the dealer is no longer trusted. A risk analysis and the modification of the scheme to mitigate the attack are presented in order to understand what can be lost if the strong hypotheses are no longer true. The changes will not be substantial but they will be little cautions to attenuate the losses. 
      \\\\  
      Finally another real problem and a possible solution will be presented. The scenario is the following: the dealer is no longer honest and their new intention is to allow access to unauthorized users.

      \paragraph{Fixed Dealer Scheme}
        In this scheme the dealer is fixed and is not in the set of participants. The dealer's role is simple, they receive the information about the secret $S$, they have to break the secret and transmit the shares $s_i$ to each shareholder $P_i$. Then, in a second moment, through the reconstruction phase, the dealer checks if the client who is trying to gain access the services is authorized and ensures that the shareholders do not cheat. 
        \\\\
        Though the dealer is honest, they will not keep either the $S$ or the information about the shares $s_i$ once they send them. What they have to keep, after the sharing phase, is the information used to verify the shareholders' correctness in the reconstruction phase, in order to ensure that the participants do not cheat.
        \\\\
        Furthermore there is a setup phase in which the critical parameters are set up. In this phase it is supposed that nobody is supposed can cheat. This phase will run only once, that is when the system is up for the first time. At first, in this case, the critical parameters remain private on the dealer side because a trusted dealer is supposed. This means that the critical parameters are not broadcast between the system actors because it is not necessary to send them.
        \\\\
        The protocol is reported in the \ref{alg:caseA} and \ref{alg:FixedDelaer} procedures. There are two different algorithms that run separately in the dealer and in the set of shareholders.
        \\\\
        % forse è il caso di spiegare il network model??
        For this scheme a simple network model is supposed, where all the connections are assumed secure. The client is directly connected with the dealer, and the dealer is connected through a LAN network with the shareholders procedures can see in figure \ref{fig:DealerFixedNetworkModel} below.
        \begin{figure}[h]
          \centering
          \includegraphics[width=1.\textwidth]{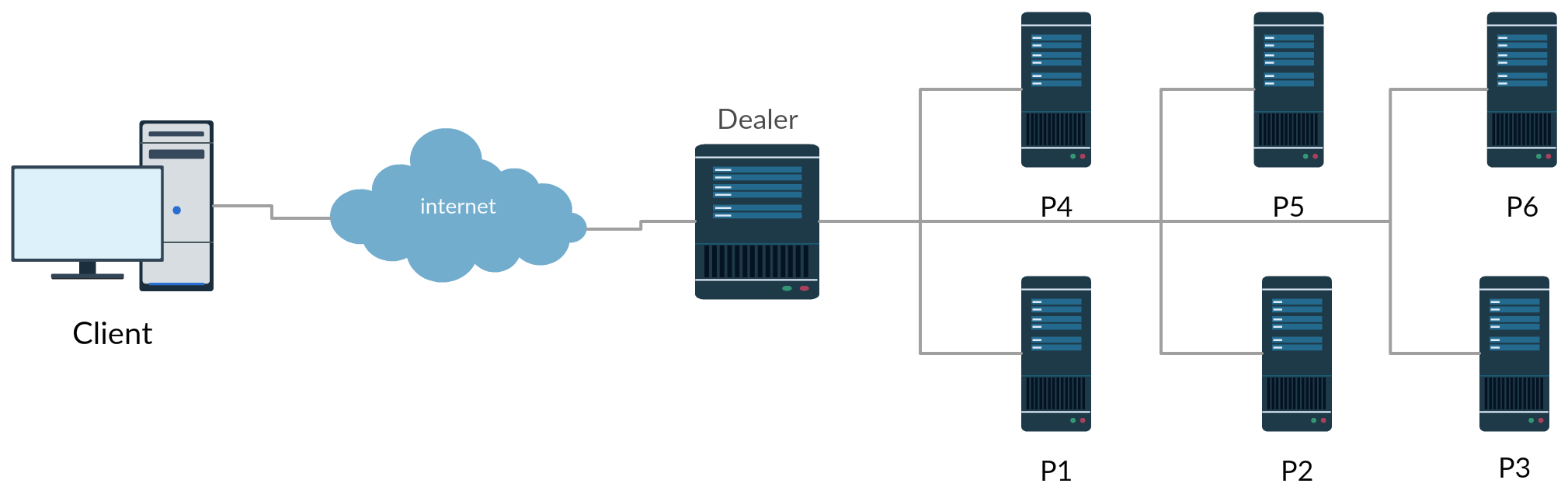}
          \caption{Dealer Fixed Network Model}
          \label{fig:DealerFixedNetworkModel}   
        \end{figure}
        \\\\
        In the following the two algorithms are shown:
        \begin{algorithm}
          \caption{Case B - Shareholder $P_i$}\label{alg:caseA}
          \begin{algorithmic}
            \Function{Sharing phase:}{share $s_i$}
              \begin{enumerate}
                \item $P_i$ saves in local the $s_i$ information. 
              \end{enumerate}
            \EndFunction

            \Function{Reconstruction phase:}{number $i$}
              \begin{enumerate}
                \item $P_i$ sends to the dealer the $s_i$ information.
              \end{enumerate}
            \EndFunction    
          \end{algorithmic}
        \end{algorithm}

        \begin{algorithm}
          \caption{Case B - Dealer}\label{alg:FixedDelaer}
          \begin{algorithmic}
            \Function{Setup phase:}{}
              \begin{enumerate}
              \item Dealer chooses the parameters $p,q,g \text{ and } h$.
              \end{enumerate}
            \EndFunction

            \Function{Sharing phase:}{original secret S from client}
            \begin{enumerate}
              \item  D secretly and randomly (uniformly and independently) chooses $t-1$ elements of $G_q$, such that:
              $a_1,a_2,\dots,a_{t-1} \in G_q$.

              \item set $a_0=S$

              \item D builds polynomial function s.t.: $F(x) = a_0+a_1x+...+a_{t-1}x^{t-1}\mod{p}$

              \item D secretly and randomly (uniformly and independently) chooses $t$ elements of $G_q$, so: $b_0,..., b_{t-1} \in G_q$. D builds polynomial function s.t.: $K(x) = b_0+b_1x+...+b_{t-1}x^{t-1}\mod{p}$

              \item D secretly and randomly chooses $n$ distinct elements s.t. $x_i=r$, where r=random number with $1\leq i \leq n$.  

              \item D computes the pair: $(s_i, t_i) = (F(x_i), K(x_i))$ for all $x_i$

              \item D computes $c_j = g^{a_j} h^{b_j}\mod{p} \ for \ j = 0,.., $t-1

              \item D outputs a list of n shares $(s_i)$ and distributes each share to corresponding participants $P_i$ privately.

              \item D only saves the abscissas vector $x_i$ and the $c_j$ information locally. 
            \end{enumerate}
              
            \EndFunction

            \Function{Reconstruction phase:}{secret S from client}
              \begin{enumerate}
                \item Dealer requests from at least $t$-shareholders their shares $s_i$ 

                \item Dealer gathers the replies with the $s_i$ information

                \item Dealer checks the shares correctness through the \ref{perdersenProof} formula, namely
                \begin{equation*}
                   g^{s_i} h^{t_i} \stackrel{?}{=} \prod_{j=0}^{t-1}{c_j^{x_i^j}} (\mod{p}) 
                \end{equation*}

                \item if the above equation is true, dealer rebuilds the secret $S^\prime$ thanks to \ref{eq:2} equation. 
                \begin{equation*}   
                S^\prime = \sum_{j=0}^{t-1}\big(s_j v_j\big)\mod{p}
                \end{equation*} 

                \item D checks if $S \stackrel{?}{=} S^\prime $, if $S$ is equal to $S^\prime$ then the user can access to the services.
              \end{enumerate} 
            \EndFunction    
          \end{algorithmic}
        \end{algorithm}

        \newpage
        \newpage
      \paragraph{New concept: abscissas vector information}\label{sec:newconcept}
        In order to improve the scheme security level there is a possibility to keep the $x_i$ information secret. This is a big step forward with respect to Shamir Secret Sharing scheme because without the abscissa vector it is not possible to reconstruct the original secret $S$ even with $t$-shares. The abscissas vector can be saved in different places:
        \begin{itemize}
          \item \textbf{Client side.} After the sharing phase, the abscissa vector is given back to the client. The dealer does not keep it. Later, in the reconstruction phase the client will send the secret $S$ and the abscissa vector information to the dealer.

          \item \textbf{Server side.} After the sharing phase, the abscissa vector is given to an external server that keeps it. Later, in the reconstruction phase the server will send the abscissa vector information to the dealer. In this case the network model could be as in Figure \ref{fig:DealerFixedNetworkModelExternalServe}.

          \begin{figure}[h]
            \centering
            \includegraphics[width=0.9\textwidth]{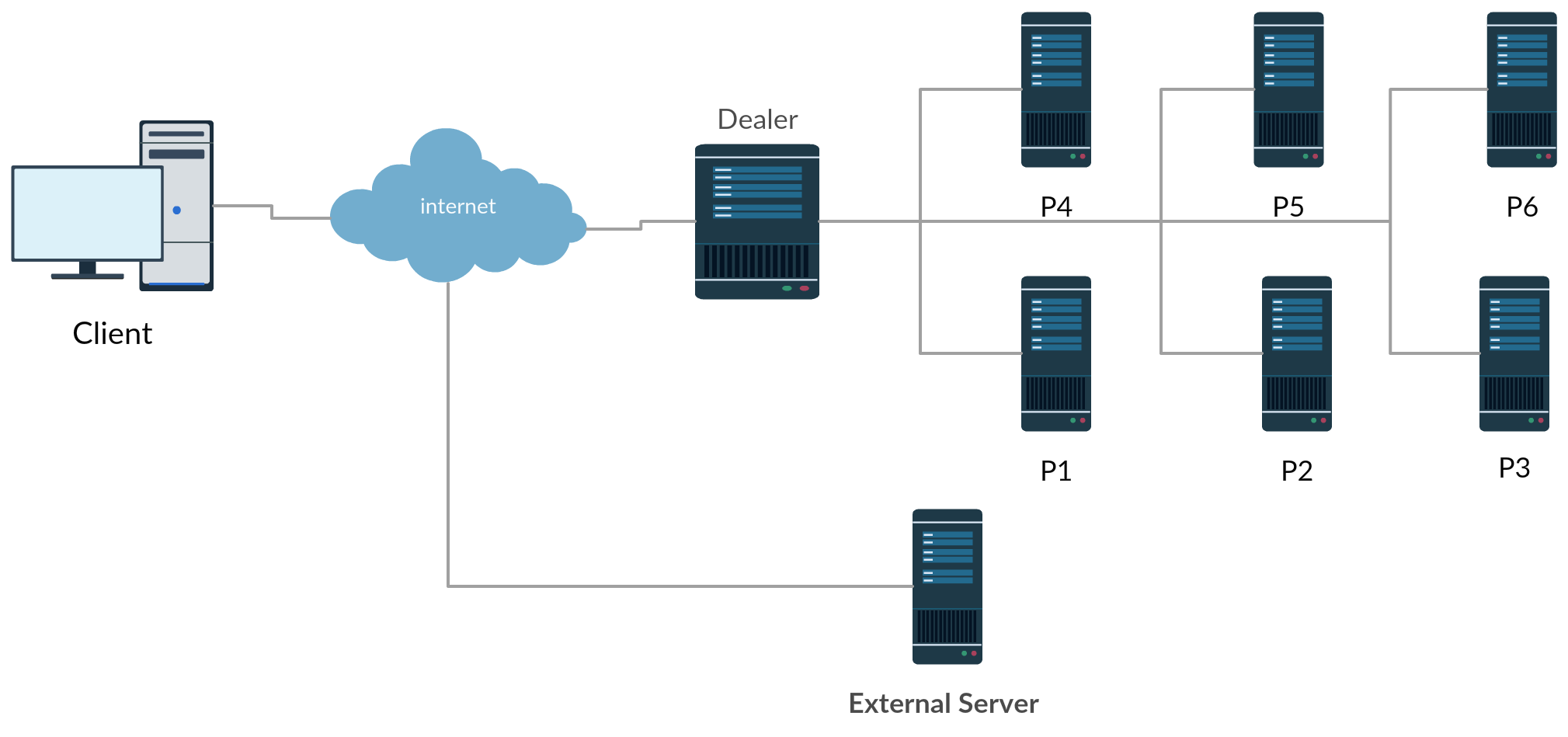}
            \caption{Dealer Fixed Network Model with External server}
            \label{fig:DealerFixedNetworkModelExternalServe}
          \end{figure}

          \item \textbf{Dealer side.} As in Algorithm \ref{alg:FixedDelaer} the abscissa vector information is kept by the dealer.
        \end{itemize}
        There are pros and cons for keeping the abscissas vector information secret. In our case the pros are the following:
        \begin{itemize}
          \item An attacker who succeeds in gathering al least $t$-shares up to $n$, is unable to rebuild the original secret $S$ because without the abscissa vector they cannot rebuild anything.
          \item The security level gets up.
        \end{itemize}
        while the cons are:
        \begin{itemize}
          \item The scheme needs to be modified in order to handle this extra information.

          \item From Pedersen's work, if the shareholder wants to check whether the given share is correct, they have to know the proper coordinate $x_i$. When the dealer is trusted, the shareholder has nothing to check, so the problem does not arise. However if the dealer is no longer trusted, we have to handle this case if we still want to hide the abscissa vector.
          \item Since the abscissas vector is necessary to rebuild the secret, there will be a point where this information is revealed to somebody.
        \end{itemize}
        The introduction of this new concept leads the system to a higher effort: the system will have to hide this new information and a new complexity will be introduced. All this work is done in order to improve the scheme security level hence the effort is rewarded with a better security. Since the benefits are bigger than the effort the system is designed in order to implement this new concept.

      \paragraph{Crystal box concept}
        In this section the crystal box concept will be presented. The crystal box concept enriches the schemes with another feature. This characteristic is used by the systems that need to be always notified when a secret $S$ is recomposed. In this way the system is able to keep track of who opens the original secret and when it is opened. In order to reach this feature the protocol has to be designed in such a way that the system keeps track of indispensable information for the secret reconstruction. Hence, without that information it is impossible to recompose the secret $S$, so in this way crystal box concept is implemented. This new feature may or may not be used, because it is not essential to the working of the model. Hence, this is an add-on. Thus the scheme is flexible so as to allow it to be used in different ways.
        \\\\
        Two different implementations are studied, the first is based on the abscissa vector, the second one is based on random numbers.

        \begin{itemize}

          \item \textbf{Case I. } In order to implement this concept it is necessary something that is indispensable for the rebuilding, if the information about the abscissa vector is used, the purpose is achieved.

          \item \textbf{Case II. } In this case, the breaking of the secret is done on $S^\prime$ instead of $S$. $S^\prime$ is done in the following way: $S^\prime = (S \oplus r)$,  where $r$ is a random number. Then the information about $r$ is given to an external trusted server. In this way in the reconstruction phase the number $r$ is essential in order to verify that the user is authorized to access the system. In the sharing phase the client will give the secret $S$ to the dealer. Before breaking $S$, the dealer will be dirtied with a random number $r$ and they will give the number $r$ to an external trusted server. Then, in a second moment during the reconstruction phase, the external server will send the number $r$ to the dealer in order to verify that the user is authorized. Since the number $r$ is necessary to rebuild the secret the system can track when the secret is rebuilt thanks to the external server because without the $r$-information the dealer cannot check the user's validity.
        \end{itemize}

      \paragraph{Problem about access of an unauthorized client}
        \label{par:Problem about access of an unauthorized client}

        In this last section another kind of problem is introduced. The problem can arise when a dealer is under attack. In this case the attacker's aim is no longer to steal information but to allow access to the services to an unauthorized client. 
        \\\\
        A possible solution is the following: once the dealer allows the client to access the services, after they proved their identity through the secret $S$, the system will start communicating with the client with encrypted messages. This means that the packets will be encrypted, and the system will use the secret $S$ as key. In order to communicate with the services the client must know the secret $S$ because all messages sent to them will be encrypted with a key $k$ and the key will be the secret $S$ itself.
          \[ \big[message\big]_k \text{, where k=S}\]
        Thus although the malicious dealer wants to allow access to an unauthorized client, the latter will not know the key $k$ and it will be unable to decrypt messages or access the services because they do not know the secret $S$. Without the secret $S$ the unauthorized client will not be able to encrypt or decrypt messages. 
        \\\\
        Thus, although the malicious dealer may cheat to allow access to an unauthorized client, the latter will not be able to do anything because they do not have the knowledge to communicate. 
            
      \subsubsection{Attack Model check}
        Now the security of the presented scheme against an attacker that behaves as described in the \autoref{sec:AttackModel_I_SOL} will be discussed. The attacker tries to enter into the system steal some useful information or tries to alter the system behaviour in order to create misunderstanding between the systems' actors. We have supposed a trusted dealer, that is the dealer is a trusted authority, hence by definition it is impenetrable. An attacker could try to attack the dealer but without any result. In this case it is the key point of the system. The adversary could only take possession of the shareholders.
        \\\\
        In the real world when the system is put online it is secure, before being attacked it will pass a time $l$. This means that from the point $0$, when the system is accessible to the point $l$ when the system is attacked, there is a period of time $l$ when the system is secure and in this period the system will work in a correct way without interference. 
        \\\\
        Obviously this is an assumption, but in order to attack a real system a person cannot play as hacker. Knowledge, experience, specific skills and above all a motivation and a reward are needed. When an attacker's goal is to steal information because it is useful and commercially saleable, the adversary will wait for sensitive information to be collected in the system. The same dynamic occurs when a thief wants to steal in a physical shop: he will wait until the end of the day or until the end of the week in order to carry out the theft, that is when all the money is there and not on Monday morning when the cash desk is empty.
        \\\\
        In this case the core of the system is the dealer because they manage the original secret $S$ during the sharing and reconstruction phases and the abscissa vector. A premature attack is useful to gather information when the dealer is attacked. However even in this case it is useless because the adversary, after being penetrated in the scheme, should wait for somebody to uses the system. In any case for this first study we assume a trusted authority so that the dealer cannot be penetrated.
        \\\\
        In order to check the attack model two cases must be be distinguished: a passive and an active attack. In this first case the adversary can only attack the shareholders. 

        \begin{itemize}
          \item \textbf{Passive attack. } The passive adversary takes control of $K \geq t$ shareholders in order to steal the shares and reconstruct the secret $S$. In this case too with $n$ shares the secret cannot be reconstructed because the attacker does not know the abscissa vector. Since the dealer is trusted the information about coordinates is not accessible. 

          \item \textbf{Active attack. } The active adversary takes control of $K \geq 1$ shareholders in order to create misunderstanding in the system. The misunderstanding is achieved by sending wrong shares to the dealer. Thanks to Pedersen's work the dealer can verify the correctness of the shares and discard the false one and notify the problem. In this case the problem arises when the attacker takes control of $K \geq n-(t-1)$ shareholders. In this case the attacker can perform a Ddos attack because the right shares will be less than $t$ and the dealer cannot reconstruct the original secrets. Obviously thanks to Pedersens'VSS the dealer can verify the shares' correctness and instantly identify the problem and notify the administrators.
        \end{itemize}

    \subsection{Risk analysis: the strong assumptions are no longer valid}
      What happens if the dealer is no longer trusted? In this section what happens if the strong assumption is no longer true will be seen. First of all what a malicious dealer can do will be introduced, then the changes to mitigate the damage will be shown. There won't be any substantial changes but simple attention that should be paid to resolve some issues to preserve the greatest amount of data. A basic risk analysis must answer the following question:
      \begin{itemize}
        \item What can happen to the system?
        \item How likely is it that it will happen?
        \item If it does happen, what are the consequences?
      \end{itemize}
      A malicious dealer can perform an active or passive attack:
      \begin{itemize}
        \item \textbf{Passive. } The passive attacker takes control of the dealer and can only take some information. The dealer deals (in the two phases) only with user specific information. In the Setup and in the Reconstruction phase the attacked dealer can steal $S$. 

        \item \textbf{Active. } The dishonest dealer can act in different ways depending on the phase they are running. 
        \begin{itemize}
          \item \textbf{Sharing phase.} In the Sharing phase the misleading dealer may be distributing shares $s_1, ... , s_n$ to the players so that when players $i_1, ... , i_k$ put their shares together, they get the secret $S$, but when players $j_1, ... , j_k$ put their shares together, they get the secret $S' \neq S$.

          \item \textbf{Reconstruction phase.} In the Reconstruction phase a malicious dealer can force shareholders to send it other shares $s_i$ of another original secret $S_1$ in order to rebuild another original secret. Looping this procedure the attacker can rebuild the whole dataset. This is the worst case because with just a few messages the dealer could rebuild the whole dataset of the secrets.
        \end{itemize}
      \end{itemize}
      In the presented algorithm \ref{Scheme_ISOL} many problems can arise if the dealer is no longer trusted. In this case a passive attack on the dealer would lead to the loss of the secret $S$ for the new users that submit their secret for the first time and also of the secret $S$ in the reconstruction phase of older users.  
      \\\\
      If the dealer performs an active attack they could cheat by sending wrong shares because the shareholders do not have any tools to check the shares correctness. However the worst case is when the malicious dealer forces shareholders to send them other shares $ s_i $ of another original secrets $ S_1, S_2, \dots, S_n $ in order to rebuild the entire dataset.
      \\\\
      The scheme could be changed in order to mitigate the damage, but not to completely resolve it. At this stage nothing can be done against the passive attacks because the secret is given to the dealer who handles the sharing and reconstruction phases.
      \\\\
      The active attack is the most dangerous because the malicious dealer can request the shares of all users and discover the whole database. To mitigate the active attack the following changes can be made:
      \begin{itemize}
        \item The malicious dealer could send wrong shares. In this case the shareholders should be able to verify the share correctness. The only way is through Pedersen's work.

        \item Against the fake dealer request there is the need of some information that verifies that the user is truly asking for the reconstruction phase. To solve this issue the scheme has to be changed as follows:\\\\
        
        \textbf{Sharing phase. }
          \begin{itemize}
            \item At the end of the sharing phase, the dealer gives back the coordinates $(x_1, x_2, \dots, x_n)$ to the client and does not save them locally.

            \item The dealer sends the shareholder the triple $(s_i,t_i,x_i)$ in order to give the participants the opportunity to check the share correctness.
          \end{itemize}
        \textbf{Reconstruction phase. }
          \begin{itemize}
            \item During the Reconstruction phase the client will give the abscissa vector to the dealer. The coordinates prove that the client is truly asking for the reconstruction phase. The shareholder checks if the coordinate $x_i$ is correct, in an affirmative case it sends back the share $s_i$ to the dealer, otherwise not. 
          \end{itemize}
      \end{itemize}
      In this way the attack is mitigated because the malicious dealer will be able to gather the $S$ information only in two ways:
      \begin{itemize}
        \item When a new client wants to use the service for the first time, it will require the sharing phase.

        \item When an old client uses the reconstruction phase.
      \end{itemize}
      When the system is put online it is assumed to be secure. Before being attacked an amount of time $l$ passes. This means that from the point $0$, when the system is accessible, to the point $l$ when the system was attacked, there is a period of time $l$ when the system is secure and in this period the system will work in a correct way without interference. In that period the dealer will not save the coordinate $x_i$, so if it is comes under attack the hacker cannot retrieve the whole dataset.
      \\\\
      In this way the attack is mitigated because compared to the previous situation, in this case the attacker will wait for all the clients to use the reconstruction in order to rebuild the whole dataset. The longer the attacker has to wait, the easier it is to patch the leak or discover the attack. Hence, the loss of data is slowed down. The network model remains the same as in the previous case \ref{fig:DealerFixedNetworkModel}, but the protocol is the underlying in the \ref{alg:FixedDelaer2} Figure.
      \\\\
      Also in this case a cheat free setup phase is supposed where the parameters $p,\ q,\ g\text{ and }h$ are chosen. The parameters are public to every actor of the system.
      \\\\\\\\
      \begin{algorithm}[H]
        \caption{Case B - Shareholder $P_i$}\label{alg:fixedDealer-share2}
        \begin{algorithmic}
          \Function{Sharing phase:}{share $s_i$, value $t_i$, coordinate $x_i$, information $c_j$}
            \begin{enumerate}
              \item The shareholder $P_i$ checks the shares correctness through the \ref{perdersenProof} formula, that is
              \begin{equation*}
                 g^{s_i} h^{t_i} \stackrel{?}{=} \prod_{j=0}^{t-1}{c_j^{x_i^j}} (\mod{p}) 
              \end{equation*}

              \item If the above equation is true, the shareholder saves locally the triple as follow: 
                \[(s_i,\ t_i,\ h(x_i))\]
              Where h is a cryptographic hash function.
            \end{enumerate}
          \EndFunction

          \Function{Reconstruction phase:}{coordinate $h(x_i)^\prime\ $}
            \begin{enumerate}
              \item $P_i$ sends to the dealer the $s_i$ information only if:
              \[ h(x_i)^\prime =  h(x_i)\]
            \end{enumerate}
          \EndFunction    
        \end{algorithmic}
      \end{algorithm}

      \begin{algorithm}
        \caption{Case B - Dealer}\label{alg:FixedDelaer2}
        \begin{algorithmic}
          \Function{Setup phase:}{}
            \begin{enumerate}
            \item Dealer and shareholders choose parameters $p,q,g \text{ and } h$.
            \item The parameters $p,q,g \text{ and } h$ are public, everyone knows them.
            \end{enumerate}
          \EndFunction

          \Function{Sharing phase:}{original secret S from client}
          \begin{enumerate}
            \item  D secretly and randomly (uniformly and independently) chooses $t-1$ elements of $G_q$, such that:
            $a_1,a_2,\dots,a_{t-1} \in G_q$

            \item set $a_0=S$

            \item D builds polynomial function s.t.: $F(x) = a_0+a_1x+...+a_{t-1}x^{t-1}\mod{p}$

            \item D secretly and randomly (uniformly and independently) chooses $t$ elements of $G_q$, so that: $b_0,..., b_{t-1} \in G_q$. D builds polynomial function s.t.: $K(x) = b_0+b_1x+...+b_{t-1}x^{t-1}\mod{p}$

            \item D secretly and randomly chooses $n$ distinct elements s.t. $x_i=r$, where r=random number with $1\leq i \leq n$.  

            \item D computes the pair: $(s_i, t_i) = (F(x_i), K(x_i))$ for all $x_i$

            \item D computes $c_j = g^{a_j} h^{b_j}\mod{p} \ for \ j = 0,.., $t-1

            \item D outputs a list of tries $(s_i,\ t_i,\ x_i)$ and distributes each share to corresponding participants $P_i$ \textbf{privately}.

            \item D broadcasts the $c_j$ information.

            \item D saves only the $c_j$ information locally.
          \end{enumerate}
            
          \EndFunction

          \Function{Reconstruction phase:}{secret S from client, $(x_1,x_2,\dots,x_n)$ abscissas vector}
            \begin{enumerate}
              \item Dealer requests to at least $t$-shareholders their shares $s_i$. In the request the Dealer sends the $h(x_i)$ information to prove that the client is actual asking for the reconstruction phase.

              \item Dealer collects the replies with the $s_i$ information

              \item Dealer checks the shares correctness through the \ref{perdersenProof} formula, that is
              \begin{equation*}
                 g^{s_i} h^{t_i} \stackrel{?}{=} \prod_{j=0}^{t-1}{c_j^{x_i^j}} (\mod{p}) 
              \end{equation*}

              \item If the above equation is true, the Dealer rebuilds the secret $S^\prime$ thanks to \ref{eq:2} equation. 
              \begin{equation*}   
              S^\prime = \sum_{j=0}^{t-1}\big(s_j v_j\big)\mod{p}   
              \end{equation*} 

              \item D checks if $S \stackrel{?}{=} S^\prime $, if $S$ is equal to $S^\prime$ then the user can access the services.
            \end{enumerate} 
          \EndFunction    
        \end{algorithmic}
      \end{algorithm}

    \newpage
    \subsection{Second proposed solution}
      \subsubsection{Attack Model} \label{Attack_mode_II}
        % --------------- INTRODUZIONE ---------------
        In this chapter a second study of the attack model against the system will be discussed. A summary of all the possible classes of attacks that a malicious agent can perform in a general secret sharing scheme will be presented. In the next section we will explain how the system can handle these attacks. All the concepts have already been seen in the previously chapters but in this one we want to put them together in an ordered and formalized way.
        \\\\
        In this second attack model a new concept will be discussed: there is the possibility to have a untrusted dealer. In this study we consider the possibility of a malicious dealer and a malicious shareholder. In the next section a second proposed solution will be shown in order to protect oneself against this kind of attack.
        \\\\
        Before illustrating the study we must explain the three levels of security: 
        \begin{itemize}
          \item \textbf{Safe:} obviously secure
          \item \textbf{Not completely safe:} the attacker can take specific information about some user, but other users' information is not touched.
          \item \textbf{Not safe:} not secure, the attacker can take all the dataset or they can compromise the protocol in order to create misunderstandings.
        \end{itemize}
        Furthermore the different possible categorizations are renewed: the attacker could be \textbf{passive} or \textbf{active} and in the distribution systems there are \textbf{honest} or \textbf{dishonest}(\textbf{malicious}) actors. The difference is subtle because an honest or dishonest player could respectively modify the protocol or not. The passive attacker cannot interact with any of the parties involved in the system but for example they could steal the actor's database. The active attacker could be seen as a dishonest player because they change the communication protocol. Hence there could be an honest player who is passive, or a dishonest player who is passive. However it is impossible to have an actor who is both honest and active.
        \\\\
        Since a trusted dealer is no longer supposed, a hypothetical attacker can attack the participants and the dealer. Depending on which actor is under attack the aggressor can behave in a different way.\\
        A shareholder could be:
        \begin{itemize}
          \item \textbf{honest} - A shareholder is honest only if in the Reconstruction phase they send back to the dealer D the share $s_i$.

          \item \textbf{dishonest} - The dishonest shareholder could send altered shares to the dealer in order to compromise the Reconstruction phase.
            
          \item \textbf{passive} - The passive attacker takes control of the shareholder and  can only take information about the share from an infected player.
        \end{itemize}
        As we saw in previous chapters in the previous chapters, an attacker who wants to reconstruct the original secret $S$ must take the control of at least $t$-players in order to have t-shares and must also know the abscissa vector. In the original Shamir's scheme this information is public, but nothing prevents it from being kept secret. Thus even if the attacker is able to take control of $t$-shareholders they cannot reconstruct the original secret $S$ without the abscissa vector.
        \\\\
        Instead a malicious dealer could be:
        \begin{itemize}
          \item \textbf{honest} - A dealer is honest only if they send the right shares to the shareholders and do not keep any information about the original secret, not request shares when there is no need or try to allow the access of an unauthorized client.

          \item \textbf{dishonest} - The dishonest dealer can act in different ways depending on the phase they are dealing with. 
          \begin{itemize}
            \item \textbf{Sharing phase.} In the Sharing phase the misleading dealer may be distributing shares $s_1, ... , s_n$ to the players so that when players $i_1, ... , i_k$ put their shares together, they get the secret $S$, but when players $j_1, ... , j_k$ put their shares together, they get the secret $S' \neq S$.

            \item \textbf{Reconstruction phase.} In the Reconstruction phase a malicious dealer can force shareholders to send them other shares $s_i$ of another original secret $S_1$ in order to rebuild another original secret. Looping this procedure the attacker can rebuild the whole dataset. This is the worst case because with a few messages the dealer could rebuild the whole dataset of the secrets. Furthermore the malicious dealer could try to allow the access of an unauthorized client.
          \end{itemize}

          \item \textbf{passive} - The passive attacker takes control of the dealer and can only steal some information. The dealer deals (in the two phases) only with user specific information. In the Sharing and in the Reconstruction phases the attacked dealer can steal $S$. 
        \end{itemize}
        In the following sections the two different cases will be studied separately but since the shareholder case was examined in \autoref{sec:AttackModel_I_SOL} section it will not be repeated.

        % --------------- DEALER COMPROMISED ---------------
        %\paragraph{Dealer}
        \paragraph{Passive attack}
          \textbf{Def. } \textit{The passive attacker takes control of the dealer. The raider can only take some information. The dealer deals (in the two phases) only with user specific information.}
          \\\\
          In order to understand we must consider the two phases separately:
            \begin{itemize}
            \item   \textbf{Sharing phase:} in this phase the dealer has to break the original secret $S$ into different pieces $s_i$ and add extra check information. The user sends the original secret $S$ to the dealer, so if the dealer is compromised the attacker can steal $S$.  Hence the sharing phase is \textbf{not completely safe.}
            
            \item \textbf{Reconstruction phase:} in this phase the dealer has to take $s_i$ from shareholders in order to reconstruct the original secret $S$ and verify if the user can access the service. Therefore, the malicious dealer is able to retrieve the original secret $S$. Hence, the reconstruction phase is \textbf{not completely safe.}
          \end{itemize}

        \paragraph{Active attack}
          \textbf{Def. } \textit{The active attacker takes control of the dealer. The dealer can modify the protocol and the content of the messages as they wish. The assaulter can act in different way depending on the phase they are dealing with.}
          \\\\
          In order to understand we have to consider the two phases separately:
          \begin{itemize}
            \item In the \textbf{Sharing phase} the misleading dealer may be distributing shares $s_1, ... , s_n$ to the players so that when players $i_1, ... , i_k$ put their shares together, they get the secret $s$, but when players $j_1, ... , j_k$ put their shares together, they get the secret $s' \neq s$. A dealer is honest only if the secret reconstructed by any combination of k players is the same. In this case, we say that the players' shares are \textbf{consistent.} Another problem arises when we want to keep the abscissas vector secret, so we must consider the possibility that a dealer sends the wrong $x_i$ in order to alter the reconstruction phase, in this way the rebuilt secret will not be the original S sent by the user. If the system uses the Pedersen's work it can be considered \textbf{safe} otherwise not. 
            
            \item In the \textbf{Reconstruction phase} a malicious dealer can force shareholders to send them other shares $s_i$ of another original secret $S'$ in order to rebuild another original secret. Looping this procedure the attacker can rebuild the whole dataset.  If the system uses a way to avoid the looping requests it can be considered \textbf{safe} otherwise not.
          \end{itemize}

        \paragraph{Access of an unauthorized client}
          \textbf{Def. } \textit{The attacker's aim is no longer to steal information but the new purpose of the malicious dealer is to allow an unauthorized client access to the service.}
          \\\\
          This problem can arise when the dealer cheats and tries to bypass the tests in order to allow the access of an unauthorized client. An unauthorized client is one who is not registered in the system but who tries to access the services in any case. This could happen only if the dealer does not check the unauthorized client and cheats about his identity. In our digital system the client's identity is given by the original secret $S$, so in order to be protected from this kind of attack we could use the original secret $S$ information also in  future connections in order to ensure that the client is authorized. An example of solution is treated in \ref{par:Problem about access of an unauthorized client} section.

        \paragraph{Conclusion}
          In order to maximize the security level of the system with this attack model, we should design a scheme where the shareholder keeps only the share $s_i$ and the abscissas vector is kept secure somewhere and somehow and only in the Reconstruction phase is it revealed to the shareholders in order to allow them to check the shares and the abscissas correctness thanks to the Pedersen's works. In this way the check will be done also from the shareholder side in order to prevent an active attack by a dealer. 
          \\\\
          In the same way the dealer too can check the shares and the abscissa's correctness. Every time that the abscissa vector is revealed the secret $S$ is exposed to the dealer, because they manage all the information that is needed to rebuild the original secret $S$. Hence, the secret must change every time it is rebuilt. In this way the secret has a one-time-validity, that is once it is opened it is no longer useful. In the next section we will see how it is possible to achieve this.
      
      \subsubsection{Scheme}\label{Scheme_Second}
        In this section one of the two final solutions will be presented. Compared to the second one this one has a greater complexity in terms of messages this is because there is a dealer, the service, and shareholders and many checks are needed in order to ensure a good level of security.
        \\\\
        Before we discuss the scheme, some important considerations regarding the proposed solution are necessary: 

        \begin{itemize}
          \item The scheme is based on the idea that a secret is a secret as long as nobody knows it, in the moment the secret is revealed to someone it is no longer a secret and will never be a secret again. For this reason the system does not manage $S$ but $S^\prime$ where $S^\prime = (g^S h^r)$. In this way, thanks to $r$, that is a random number, if an attacker is able to obtain $S^\prime$ they cannot extract $S$. Thanks to this mechanism in $S$ we have a user's password, and starting from that and $r$ it is easy to reconstruct $S^\prime$. In this way the information about $r$ can be saved permanently in a digital form and the password is typed every time.

          \item The proposed solution is a kind of proactive scheme and uses a one-time-password logic. A one-time password is a password that is valid for only one login session or transaction, on a computer system or other digital device. This result is obtained because given $S^\prime = (g^S h^r)$, changing $r$ changes $S^\prime$ and so the shares. The idea is that once $S^\prime$ is rebuilt, it is no longer valid, that is the shares of another secret $S^{\prime\prime}$ must be redistributed. Being $r^\prime \ne r$ it is possible to calculate $S^{\prime\prime} = (g^S h^{r^\prime})$ so that $S^{\prime\prime} \ne S^\prime$. The new secret $S^{\prime\prime}$ will be used for the next session. In this way a sort of token access is created, where each time the token is given a different value of $r$.

          \item The verification of the correctness of shares by the shareholders is eventually verified during the Reconstruction phase. It is performed in the reconstruction phase, in this way the revelation of the coordinates to the shareholder is done only at the end. In this way also if an attacker is able to take $k$-shareholders with $k \ge t$ they can take no information about $S^\prime$ because they do not know the abscissas vector. In any case, if the attacker has $k \ge t$ shares and the abscissas vector they can only know $S^\prime$ for that session, that is the one in progress and therefore they cannot do anything with the retrieved information. 
        \end{itemize} 
        In order to be safe from the access of unauthorized clients we cannot give $S^\prime$ to the service because it does not make sense. Some information will be given to the service so that it can verify the user's validity but without keeping any sensitive information.
        \\\\
        Now the protocol will be shown. There are the two classical phases, sharing and reconstruction. For each phase there is a simple exchange messages scheme and a brief description. This part does not focus on problems and limits (this will be done in \ref{subsection:Pros, contras and limits}) but it focuses on the presentation of the scheme. The following is a network model where the protocol works.    
          \begin{figure}[!htb]
            \centering
            \includegraphics[width=0.8\textwidth]{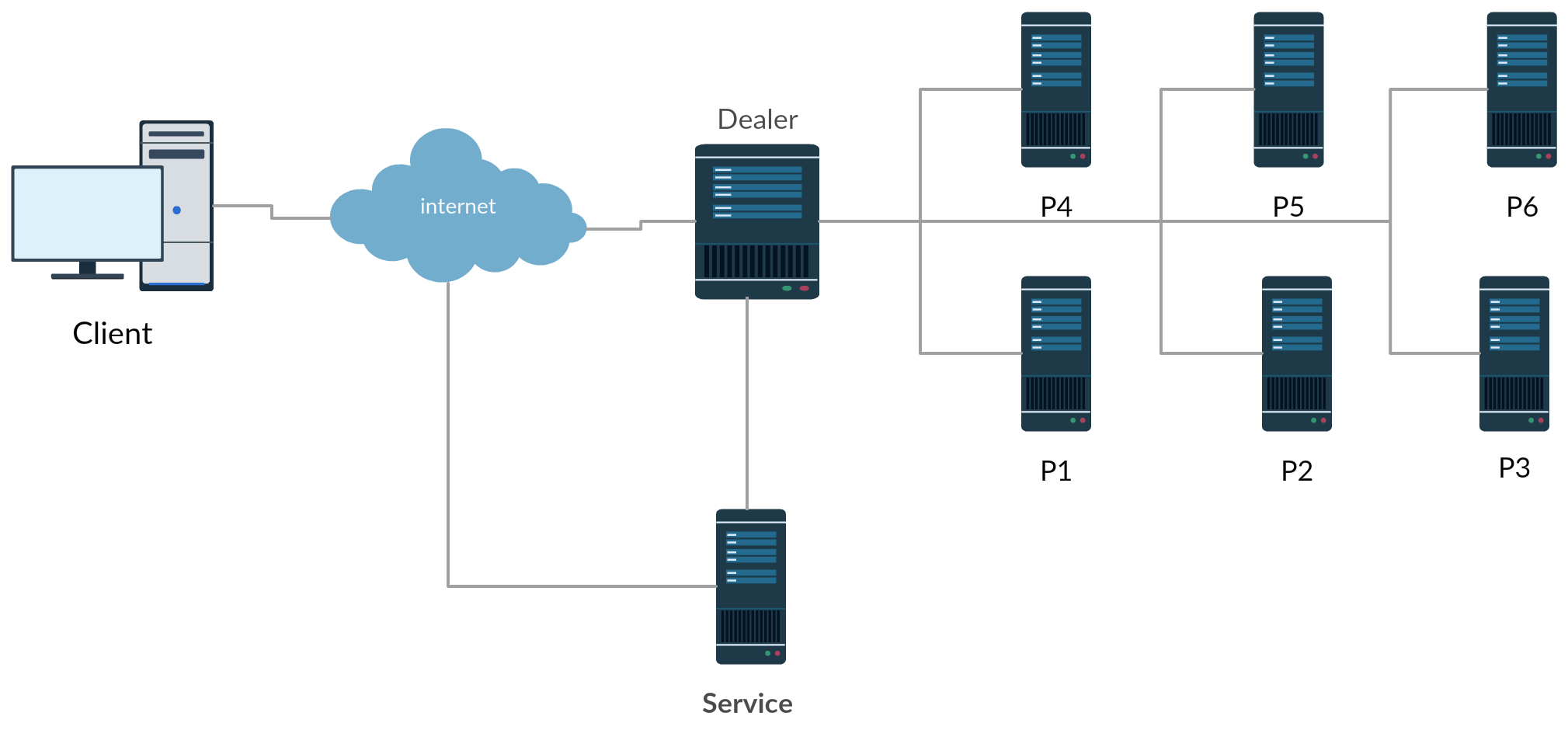}
            \caption{Network Model with Service}
            \label{fig:DealerFixedNetworkModelExternalServe2}
          \end{figure}
          \\\\
        We suppose that each channel is private with advanced techniques and that there is a broadcast channel between shareholders and dealer.

\begin{figure}[!htb]
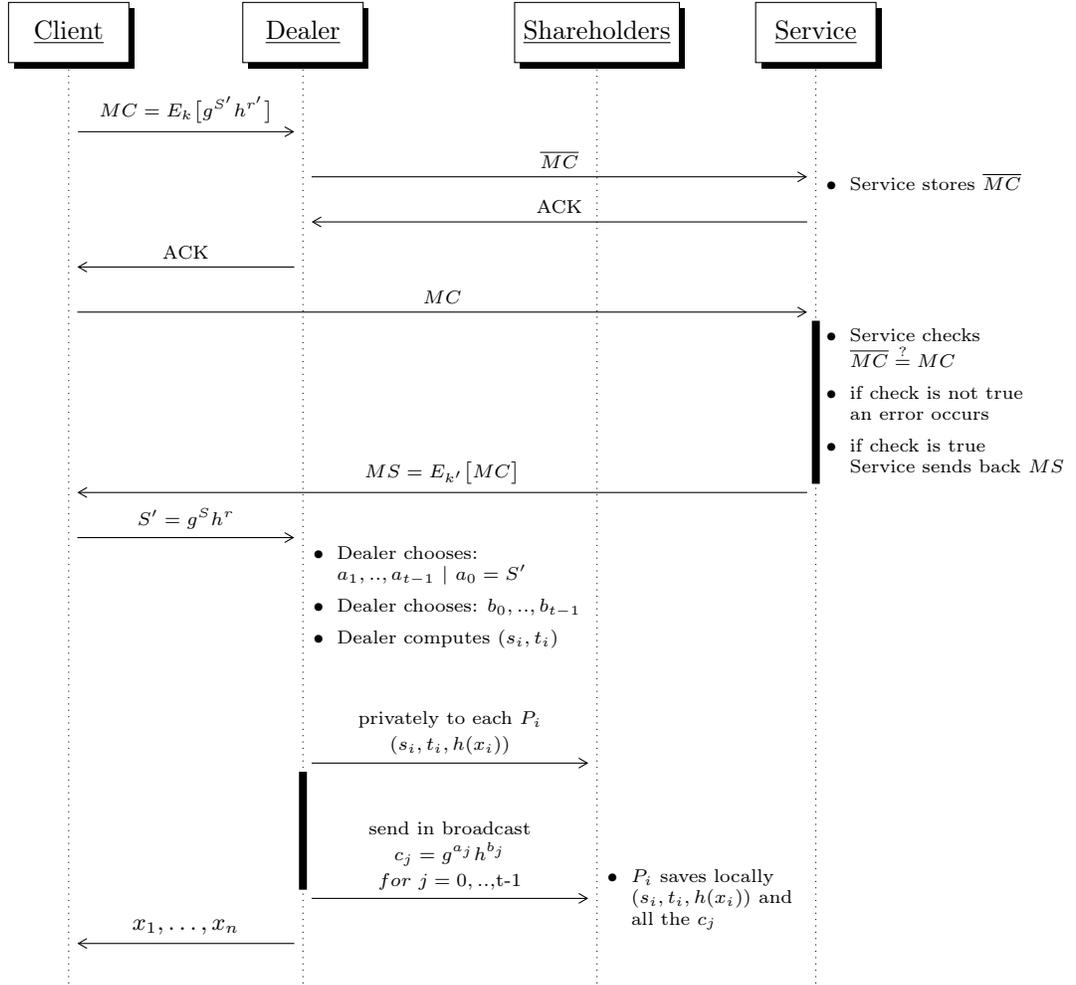

          \centering
          \begin{sequencediagram}
                \newinst{C}{Client}
                \newinst[1.5]{D}{Dealer}
                \newinst[2]{P}{Shareholders}
                \newinst[1]{ES}{\shortstack{Service}}
               
              \mess[0,font=\small]{C}{\scriptsize\shortstack{$MC=E_k\big[g^{S^\prime} h^{r^\prime} \big]$}}{D}

              \mess[0,font=\small]{D}{\scriptsize\shortstack{$\overline{MC}$}}{ES}
                \node (t68) at (mess to) {}; 
                \node [right=-7mm,align=left](t69) at (t68){\\
                \Large{
                \begin{varwidth}{\linewidth}\scriptsize \begin{itemize}
                    \item Service stores $\overline{MC}$            
                \end{itemize}\end{varwidth}
              }};
              
              \mess[0,font=\small]{ES}{\scriptsize\shortstack{ACK}}{D}
              
              \mess[0,font=\small]{D}{\scriptsize\shortstack{ACK}}{C}           

              \mess[0,font=\small]{C}{\scriptsize\shortstack{$MC$}}{ES}
                \node (t67) at (mess to) {}; 
                
                \node [right=-7mm,align=left](t70) at (t67){\\\\\\\\\\\\\\
                \Large{
                \begin{varwidth}{\linewidth}\scriptsize \begin{itemize}
                    \item Service checks\\$\overline{MC}  \stackrel{?}{=} MC$ 
                    \item if check is not true\\an error occurs
                    \item if check is true\\Service sends back $MS$        
                \end{itemize}\end{varwidth}
              }};

            \stepcounter{seqlevel}
            \stepcounter{seqlevel}
            \stepcounter{seqlevel}

              \mess[0,font=\small]{ES}{\scriptsize\shortstack{$MS=E_{k^\prime}\big[MC\big]$}}{C}
                \node (t71) at (mess from) {};

              \draw[line width=3pt] (t67) -- (t71);
             
              \mess[0,font=\small]{C}{\scriptsize\shortstack{$S^\prime = g^{S} h^r $}}{D}
                \node (t0) at (mess to) {};

              \stepcounter{seqlevel}
              \stepcounter{seqlevel}
              \stepcounter{seqlevel}
              \stepcounter{seqlevel}

              \mess[0,font=\small]{D}{\scriptsize\shortstack{privately to each $P_i$\\$(s_i,t_i,h(x_i))$}}{P}
                \node (t1) at (mess from) {}; 

                \node [right=-7mm,align=left](t5) at (t0){\\\\\\\\\\
                \Large{
                \begin{varwidth}{\linewidth}\scriptsize \begin{itemize}
                    \item Dealer chooses: \\$a_1, .., a_{t-1}\ |\ a_0=S^\prime$
                    \item Dealer chooses: $b_0, .. ,b_{t-1}$
                    \item Dealer computes $(s_i,t_i)$             
                \end{itemize}\end{varwidth}
              }};

              \stepcounter{seqlevel}
              \stepcounter{seqlevel}

              \mess[0,font=\small]{D}{\scriptsize\shortstack{send in broadcast\\$c_j = g^{a_j} h^{b_j}$ \\ $for \ j = 0,.., $t-1}}{P}
                \node (t2) at (mess from) {}; 
                \node [right=-7mm,align=left](t6) at (mess to){\\
                \Large{
                \begin{varwidth}{\linewidth}\scriptsize \begin{itemize}
                    \item $P_i$ saves locally\\$(s_i,t_i,h(x_i))$ and \\all the $c_j$
                \end{itemize}\end{varwidth}
              }};
                
              \draw[line width=3pt] (t1) -- (t2);

              \mess[0,font=\small]{D}{$x_1,\dots,x_n$}{C}
            
          \end{sequencediagram}
            \caption{Exchange messages of Sharing phase for second proposed scheme}
            \label{protocol:ExchangeMessagesSharing2_0}
        \end{figure}

        \paragraph{Sharing phase}
          In figure \ref{protocol:ExchangeMessagesSharing2_0} exchange messages scheme of the sharing phase are presented. This phase is very simple, the original secret $S$ is never given to the dealer, but it is given $S^\prime$, where $S^\prime=(g^S h^r)$. Therefore thanks to the $S^\prime$, $S$ can never discovered. Note that in the reconstruction phase $r$ will be changed every time that the client logs in. In this way a kind of one-time-password is implemented. Once the secret $S^\prime$ is opened, after the appropriate checks, it will never be valid for the next authentications. Thanks to this mechanism also if the attacker is able to take control of the $k \ge t$ shareholders they can discover the secret only in the reconstruction phase and the secret is valid only for that session. In the other case instead if the attacker can reach the dealer's control they discover only $S^\prime$ but not $S$. In this last case even if the attacker they discover $S^\prime$ they cannot log into the service. In order to log in it they must know other additional information that are saved on the client-side. Hence in order to enter and take useful information the attacker must steal a lot of information saved on two different places: client and dealer or shareholders side.
          \\\\
          The first three messages serve to guarantee protection against a cheat dealer who tries to access unauthorized clients. Encrypting $g^S h^{r^\prime}$ with a key $k$ that only a user knows will guarantee to the service a an user who is trying to access is authorized. Moreover this approach guarantees the dealer that a service can not cheat.
          \\\\
          In this phase no verification of the shares is made, the checks are postponed in the reconstruction phase in order to keep as less information as possible on the shareholders' side. Then in the reconstruction phase the share given to the shareholders is checked with the $c_j$ and the $x_i$ information. If the test is successful the shareholder can send back $s_i$ to the dealer. In this phase the dealer should not save any information about $x_i,\ a_i$ or $b_i$, they only save the $c_j$ information.
          
        \paragraph{Reconstruction phase}
          We will now consider what happens in the reconstruction phase. When the user wants to log in he sends the secret $S^\prime$ with the abscissas vector. Together with this information he sends also the $S^{\prime\prime}$ where $S^{\prime\prime}= g^{S} h^{r^{\prime\prime}}$ with $r \ne r^{\prime\prime}$. The information $S^{\prime\prime}$ will be the secret to break if the $S^\prime$ information is correct and will be the secret that the user will use next time in order to log in. In this way each time that the user calls the reconstruction phase he will send also the information about $S^{\prime\prime}$ that at the end of this phase, if everything goes well, will be broken and it will be used next time for authentication.
          \\\\
          Once the information has been sent, the dealer will provide to reconstruct $\overline{S}^\prime$ over the $t$ shares $s_i$ and they will check if it is equal to $S^\prime$. The shareholder $P_i$ will check the $x_i$ and $s_i$ validity thanks to Pedersen's work. In this way the dealer will not be able to cheat, because when they send $x_i$ to the shareholders, if the coordination is not correct with the Pedersen's equation the participant will notice it. Otherwise, if the coordination is correct the player will send back the $s_i$ and $t_i$ information.
          \\\\
          If the $\overline{S}^\prime$ is equal to $S^\prime$ the dealer sends the $E_{MS}\big[k,\ (g^{S^\prime}h^{r^\prime})\big]$ information to the service. This information was sent by the user in the first message and it is encrypted with the $MS$ key. $MS$ was exchanged in the sharing phase and only the client and the service know it. In this way if the secret  $\overline{S}^\prime$ is equal to the $S^\prime$ the dealer sends $\big[k,\ (g^{S^\prime} h^{r^\prime})\big]$ in a way that only the service or client can open it. The message will be encrypted with $MS$ so only those who know $MS$ can extract the contents of the message. This is done to make sure that even if the dealer wants to cheat with a false content, not knowing MS, they cannot do that. In this way the dealer can neither alter nor open the message.
          \\\\
          The $k$ and $(g^{S^\prime} h^{r^\prime})$ data serve to verify the client's identity. Thanks to $MC$ given in the sharing phase the service can check the client's identity. In fact the service decrypts $MC$ with the $k$-information and checks if it is equal to $(g^{S^\prime} h^{r^\prime})$. If everything goes well, the service will reply with $k^\prime$ in order to demonstrate to the user who he is talking to. With $k^\prime$ the client can decrypt $MS$ and verify if it is equal to $MC$.
          \\\\
          After that the client sends $MC^\prime$ to the service, and the latter sends the $MS^\prime$ for the next login. At this point the service sends $k,\ k^\prime$ and $(g^{S^\prime}h^{r^\prime})$ to the Dealer. With this information the dealer can check the user and the service correctness. If everything goes well the dealer must replace the old information $S^\prime$ with the new one $S^{\prime\prime}$.
          %Since the dealer has to replace the old information $S^\prime$ with the new one $S^{\prime\prime}$, in order to do that 
          %it waits that the user is correct and it waits the information that certifies the client's and service's correctness. 
          %The data are $k,\ k^\prime$ and $(g^{S^\prime}h^{r^\prime})$, with these information the dealer can check the user and the service correctness and if so it will start an internal sharing phase, that is skipping the first four messages of the original one, to replace the shares that have become old.
          \\\\
          The reconstruction phase is rather delicate and it is a little more complicated. The $MC$ and $MS$ mechanism is made in order to neutralize access to unauthorized users when the dealer is compromised. Another important fact is that $S$ is never released, so even if an attacker could discover $S^\prime$ or $S^{\prime\prime}$ they would be unable to get $S$.
          \\\\
          A doubt that could be raised concerns the management of $k^\prime$ by the service. If the $k^\prime$ was to be saved locally the service should load the burden of managing that data. The idea that we will see at \ref{chap:How to manage sensible information in a secure way} is that the information is used by the service on-the-fly and the information is saved in the system. In this way the service uses $k^\prime$ only when it is needed. We can use the Shamir's theory at his primary study. The service breaks $k^\prime$ and gives the shares to the shareholders without releasing the abscissas information. In this way when it needs to use $k^\prime$ it can ask the shares and reconstructs the secret on-the-fly without saving it locally. 
          \begin{figure}[H]
          \tikzset{
            % add this style to all tikzpicture environments
            every picture/.append style={
              % enable scaling of nodes
              transform shape,
              % set scale factor
              scale=0.8
            }
          }
          \begin{sequencediagram}
                \newinst{C}{Client}
                \newinst[1]{D}{Dealer}
                \newinst[2]{P}{Shareholders}
                \newinst[2]{ES}{\shortstack{Service}}

                \stepcounter{seqlevel}
                \stepcounter{seqlevel}
                \stepcounter{seqlevel}

              \mess[0,font=\small]{C}{\scriptsize\shortstack{$S^\prime = g^{S} h^r $\\$x_1,\dots,x_n$\\$S^{\prime\prime} = g^{S} h^{r^{\prime\prime}}$\\$MC^\prime =E_{k^{\prime\prime}}\big[g^{S^{\prime\prime}} h^{r^{\prime\prime\prime}} \big]$
              \\ $E_{MS}\big[k,\ (g^{S^\prime} h^{r^\prime})\big]$}}{D}

            \mess[0,font=\small]{D}{\scriptsize\shortstack{privately to each $P_i$\\$\overline{x_i}$}}{P}
                \node (t0) at (mess to) {}; 

              \stepcounter{seqlevel}
              \stepcounter{seqlevel}
              \stepcounter{seqlevel}
              \stepcounter{seqlevel}

              \mess[0,font=\small]{P}{\scriptsize\shortstack{$(s_i,t_i)$}}{D}
                \node (t1) at (mess from) {}; 
                \node (t2) at (mess to) {}; 
                \node [right=-7mm,,align=left](t5) at (t0){\\\\\\\\\\\\\Large{
                \begin{varwidth}{\linewidth}\scriptsize \begin{itemize}
                  \item $P_i$ checks\\$h(\overline{x_i}) \stackrel{?}{=} h(x_i)$
                  \item \scriptsize if check is not true \\an error occurs$^1$
                     
                    \item \scriptsize otherwise $P_i$ checks\\ $ g^{s_i} h^{t_i} \stackrel{?}{=} \prod_{j=0}^{t-1}{c_j^{x_i^j}}$
                    \item \scriptsize if check is true the $P_i$\\sends back the $(s_i,t_i)$
                    \item \scriptsize otherwise an error occurs$^2$
                \end{itemize}\end{varwidth}
              }};

              \draw[line width=3pt] (t0) -- (t1);

              \stepcounter{seqlevel}
              \stepcounter{seqlevel}
              \stepcounter{seqlevel}
              \stepcounter{seqlevel}
              \stepcounter{seqlevel}
              \stepcounter{seqlevel}
              \stepcounter{seqlevel}
              \stepcounter{seqlevel}
              
              \mess[0,font=\small]{D}{\scriptsize\shortstack{$\overline{E_{MS}}\big[k,\ (g^{S^\prime} h^{r^\prime})\big]$}}{ES}
                \node (t3) at (mess from) {}; 
                
                \node [right=-7mm,align=left](t6) at (t2){\\\\\\\\\\\\\\\\\\\\\\\\
                \Large{
                \begin{varwidth}{\linewidth}\scriptsize \begin{itemize}

                    \item \scriptsize$D$ checks\\ $ g^{s_i} h^{t_i} \stackrel{?}{=} \prod_{j=0}^{t-1}{c_j^{x_i^j}}$
                    \item if check is not true\\an error occurs$^{3}$
                    \item if check is true D rebuilds $\overline{S^\prime}$
                    \item D checks if $\overline{S^\prime} \stackrel{?}{=} S^\prime$
                    \item if check is not true\\an error occurs$^{4}$
                    \item if check is true D sends $E_{MS}$\\to the External Server
                \end{itemize}\end{varwidth}
              }};
               
               \draw[line width=3pt] (t2) -- (t3);

                \node (t7) at (mess to) {};

              \mess[0,font=\small]{ES}{\scriptsize\shortstack{ACK}}{D}

                \node [right=-7mm,align=left](t78) at (t7){\\\\\\\Large{
                \begin{varwidth}{\linewidth}\scriptsize \begin{itemize}
                    \item Service stores\\$\overline{E_{MS}}\big[k,\ (g^{S^\prime} h^{r^\prime})\big]$
                     \end{itemize}\end{varwidth}
              }};
              
            \mess[0,font=\small]{D}{$x_1^\prime,\dots,x_n^\prime$}{C}

            \mess[0,font=\small]{C}{\scriptsize\shortstack{$E_{MS}\big[k,\ (g^{S^\prime} h^{r^\prime})\big]$}}{ES}
              \node (t79) at (mess to) {}; 
            
            \stepcounter{seqlevel}
            \stepcounter{seqlevel}
            \stepcounter{seqlevel}
            \stepcounter{seqlevel}
            \stepcounter{seqlevel}

              \mess[0,font=\small]{ES}{\scriptsize\shortstack{$E_{MS}\big[k^\prime\big]$}}{C}
                \node (t8) at (mess from) {}; 
                \node (t10) at (mess to) {}; 
                
                \node [right=-7mm,,align=left](t9) at (t79){\\\\\\\\\\\\\\\\\\\\\Large{
                \begin{varwidth}{\linewidth}\scriptsize \begin{itemize}
                  \item Service checks\\$\overline{E_{MS}} \stackrel{?}{=} E_{MS}$
                    
                    \item if check is true continue\\with others

                    \item Service checks\\$ (g^{S^\prime} h^{r^\prime}) \stackrel{?}{=} D_k\big[MC\big]$
                    \item if check is not true\\an error occurs$^{5}$
                    \item if check is true ES\\sends back $k^\prime$
                \end{itemize}\end{varwidth}
              }};

              \draw[line width=3pt] (t79) -- (t8);

              \stepcounter{seqlevel}
              \stepcounter{seqlevel}
              \stepcounter{seqlevel}
              \stepcounter{seqlevel}

              \mess[0,font=\small]{C}{\scriptsize$MC^\prime$}{ES}
                \node (t11) at (mess from) {}; 
                \node [right=-7mm,,align=left](t12) at (t10){\\\\\\\\\\\\\\\Large{
                \begin{varwidth}{\linewidth}\scriptsize \begin{itemize}

                    \item C checks\\$MC \stackrel{?}{=} D_{k^\prime}\big[ MS \big]$
                    \item if check is not true\\an error occurs$^{6}$
                    \item if check is true C\\sends back $MC^\prime$
                \end{itemize}\end{varwidth}
              }};

               \draw[line width=3pt] (t10) -- (t11);

              \mess[0,font=\small]{ES}{\scriptsize$MS^\prime = E_{k^{\prime\prime\prime}} \big[ MC^\prime \big]$}{C}
                \node (t19) at (mess to) {}; 
                \node [right=0mm,align=left,font=\small](t18)at (t19){\scriptsize\shortstack{\\\\\\\\\\\\established\\connection}};

              \mess[0,font=\small]{ES}{\scriptsize$k,\ k^\prime,\ MS, (g^{S^\prime} h^{r^\prime}),\ \overline{MC^\prime}$}{D}
                \node (t13) at (mess to) {};

                \node [right=-7mm,,align=left](t14) at (t13){\\\\\\\\\\\\\\\\\\\Large{
                \begin{varwidth}{\linewidth}\scriptsize \begin{itemize}

                    \item D checks $ (g^{S^\prime} h^{r^\prime}) \stackrel{?}{=} D_k\big[MC\big]$ 
                    \item D checks $MC \stackrel{?}{=} D_{k^\prime}\big[ MS \big]$
                    \item D checks $MC^\prime \stackrel{?}{=} \overline{MC^\prime}$
                    \item if check is not true an error occurs$^{7}$
                    \item if check is true D starts with a\\ sharing phase on $S^{\prime\prime}$, $MC^\prime$and $x_1^\prime,..,x_n^\prime$
                \end{itemize}\end{varwidth}
              }};
              
          \end{sequencediagram}
          \label{protocol:ExchangeMessagesRec2_0}
            \caption{Exchange messages of Reconstruction phase for second proposed scheme}
          \end{figure}
          \newpage
          Once $k^\prime$ is used it would no longer be valid, so the service could take another secret $k^{\prime\prime\prime}$ and repeat the same steps. In this way the information is not saved locally but only in RAM. The only information that the service has to keep is the abscissas vector. 
          \\\\
          A possible structure could be that it does not exist only one Service but a lot of them. Hence the system will have many clusters of users to manage. The number of clusters is equal to the number of the Services. Thus if a Service is compromised the only users affected are those related to that Server. 
          \\\\
          The concepts presented above will be revised better in the next sections and we will see the pros, cons and limits of this scheme. In the next section,  another final solution different from this one will be described.

      \subsubsection{Attack Model check}
        In the previous section the scheme was presented. In this section the security of the scheme will be verified. 
        \\\\
        Compared to the previous attack model in this case every actor in the system can be attacked and therefore can lie. In the last attack model proposed the dealer, the shareholder and the service may be subject to attacks and may lie. First of all the correctness of the scheme regarding the dealer and shareholder is verified, then the role of the service will be examined.

        \paragraph{Dealer analysis}
          The analysis will start by studying the dealer's behaviour. As we have seen, once the attacker takes the control of the dealer they can operate in two ways: actively or passively. Regarding \textit{passive attack} the two different phases will be treated separately:
          \begin{itemize}
            \item \textbf{Sharing phase.} In this phase, once the dealer has received the secret, they have the task of breaking it. If they operated in a passive way they could steal the secret $S^\prime$. Since $ S^\prime = (g^S h^r)\mod{p} $ it is infeasible to open it. The hypothetical hacker is unable to obtain $S$ because the information $S$ is information-theoretic secure \footnote{Information-theoretic security is a cryptosystem whose security derives purely from information theory. In other words, it cannot be broken even if the adversary had unlimited computing power. The adversary simply does not have enough information to break the encryption and so the cryptosystems are considered cryptanalytically-unbreakable. \cite{Information-theoretic_security_DEF}}. 

            \item \textbf{Reconstruction phase.} The same is also for the Reconstruction phase. Once the attacker rebuilds $S^\prime$ it would not be able to obtain $S$.
          \end{itemize}
          Hence if an attacker operates passively they could not retrieve any useful information from  $ S^\prime $. This is also true for the next phase, that is in the reconstruction phase. This means that with only $S^\prime$ the attacker cannot perform malicious actions. Another important thing to observe is that $S^\prime$ changes every time the user logs in. In this way the stolen information immediately becomes useless.
          \\\\
          On the other hand, as regards an \textit{active attack}, a dealer can behave in different ways depending on the phase in which they are operating:
          \begin{itemize}
            \item \textbf{Sharing phase.} In the sharing phase the dealer could send different and non-consistent shares so that when players $i_1,..,i_k$ put their shares together, they get the secret s, but when players $j_1,..,j_k$ put their shares together, they get the secret $s^\prime \ne s$.
            \\\\
            In the sharing phase, since the $x_i$ information is not given to the shareholders, it is not possible to verify the share correctness. This is because if the $x_i$ information had been given to the participants it could be stolen: if the $t$-shareholders were under control of the hacker, they could have the possibility of rebuilding the secret of all users. Hence the check is made by the participants in the reconstruction phase when the dealer sends the coordinate $x_i$. With this check the participant can verify if the share is consistent. If the check is not successful it means that the dealer in the sharing phase did something wrong and they could be under attack. Hence the shareholders must notify the issue to the administrator so that they can intervene and fix the problem. 
            
            \item \textbf{Reconstruction phase.} In the reconstruction phase the dealer could force the sending of the shares of other clients. Therefore, in the sharing phase the $h(x_i)$ information is given to the shareholder in order to be used to solve this problem. In this way the shareholder can check if the incoming request is valid or not. Hence the dealer can not looping request the shares of other clients and the dataset can not be rebuilt.
          \end{itemize}
          Hence if an attacker operates actively, if they lied in the sharing phase this is detected by the shareholders in the reconstruction phase. In this way the appropriate countermeasures can be taken.

        \paragraph{Shareholders analysis}
          The shareholder can operate actively or passively. We shall begin with the \textit{passive attacks}.
          \\\\
          In the passive attack, $K$ participants are under attack. The distinction is whether the attacker has the coordinates or not. The protocol is set in a way that the coordinates are never saved locally. The only information that is saved is $h(x_i)$, obviously if it seems to not very secure it could also be saved in other more ways, but for now  $h(x_i)$ is sufficient. The key point is that $x_i$ is sent only for  verification, and if the test is positive, the corresponding share is returned. The coordinate is not saved locally, so the hacker should take control of the machine and it should read the data sent to it or the data in RAM. Since $S^\prime$ changes at every login, the hacker must be fast to steal $S^\prime$ and open it.
          \\\\
          In any case, even if they could take $t$-share and $t$-coordinate information and rebuild $S^\prime$ the attacker would not do anything whit this information. The hypothetical hacker is unable to obtain S because the information $S^\prime$ is information-theoretic secure. Even if the hacker wanted to try to trick the Service this is not possible with only the $S^\prime$ information. Ultimately if the attacker was able to attack $ K \ge t $ shareholders they would not have the coordinates to rebuild the original secret. Even if they were able to obtain $S^\prime$, they cannot do anything with $S^\prime$ and they cannot retrieve any useful information.
          \\\\
          As regards \textit{active attacks} if a shareholder sends a wrong share the dealer can verify its correctness. Hence the dealer could disqualify that participant until their correctness is restored.

        \paragraph{Service analysis}
          We shall now look at service analysis. In the protocol presented in section \ref{Scheme_Second} a new actor called Service was introduced in the scheme. 
          \\\\
          The Service does not save any sensitive user's information but only $k$, although as we will see in section  \ref{chap:How to manage sensible information in a secure way} this information may not be saved permanently by the service but only requested when needed.
          \\\\
          In any case, if the dealer tries to lie and authenticate an unauthorized user, since the information $MS = E_ {k^\prime} \big[MC \big]$ is sent directly to the user in the sharing phase, the dealer can not cheat because they do not know that information. Besides, when the $\big[k,\ (g^{S^\prime} h^{r^\prime})\big]$ information is sent to the dealer by the user it is encrypted with $MS$ that it is the only information that the Service and the Client know. To achieve its purpose, the malicious dealer has to know $MS$ information and in order to do that they must hack that specific user. If the verification from the service side of this information is not successful it means that the dealer or the client are lying, so the service will not accept that connection.
          \\\\
          The same is true for the client, if the check on $k^\prime$ does not succeed, it means that the dealer or the service are lying and therefore the client blocks the connection because it is not secure.
          \\\\
          The last check also takes place on the dealer side: if the service is lying  the dealer will not update the information regarding the shares. 
          \\\\
          With these three last tests each actor, such as client, service and dealer check the correctness of the others two parts:
          \begin{itemize}
            \item The service tests the information about $E_{MS}\big[k,\ (g^{S^\prime} h^{r^\prime})\big]$ in order to verify the dealer and client correctness.

            \item The client tests the information about $E_{MS}\big[k^\prime\big]$ in order to verify the dealer and service correctness.

            \item The dealer tests the information about $k,\ k^\prime,\ MS,\ (g^{S^\prime} h^{r^\prime})$ in order to verify the service and client correctness.

          \end{itemize}

      \subsubsection{Errors classification}
        In this section an error classification will be presented. In scheme \ref{Scheme_Second} there are many possibilities to make mistakes. This happens because the parties involved are many thus the risk of cheating by the different actors increases. Since there is the possibility that someone can cheat some tests are needed. If there are tests there can be errors when the test fails. The tests are done only in the reconstruction phase, so the sharing phase is free of errors, that is since there are no checks, the errors cannot be discovered, since they aren't being looked for.
        \\\\
        The errors can be of two types:
          \begin{itemize}
            \item \textbf{Fatal}: this means that if this error arises the system must not move on.
            \item \textbf{Not-fatal}: this means that if this error arises the system can move on with some precautions.
          \end{itemize}
        In \ref{Scheme_Second} the tests are made only in the reconstruction phase so the errors are detected only in this last phase. Every test is made to check that the other parts are not lying. So whenever the server finds an error it should understand who was wrong or whoever is lying and communicate the error to the system administrator in order to intervene in case of attack. In this case we have seven checks and then seven possible errors.
          \begin{enumerate}
            \item The first error occurs when a wrong $x_i$ is sent to a shareholder. This happens for three reasons:
            \begin{itemize}
              \item a user sent the wrong coordinates.
              \item an unauthorized user is trying to cheat.
              \item the dealer is trying to cheat in order to gather as many shares as possible of another secret.
            \end{itemize}
            Whatever the case thanks to the hash of the $x_i$ information this test serves to prevent this kind of attack or misunderstanding. This type of error is \textbf{not fatal} but it is an alarm bell.
            \\\\
            After receiving a number of wrong attempts, the shareholder becomes suspicious and stops to reply and  log the error for further verification. The worst case is when a dealer is attempting to reconstruct others secrets through other shares. To overcome this problem and mitigate the attack we could think of some mechanism so as to inform the other shareholders of the problem and contact the system administrator to proceed to check on the correctness of the dealer.

            \item The second error occurs when a $x_i$ sent to a shareholder is right but through Pedersen's works a participant is able to discover the wrong share. This can happen only if the dealer in the sharing phase sent the wrong share in order to create misunderstandings between shareholders. In this case the participants do not send the share back because it not consistent with the original secret $S^\prime$ and the others shares. The error is \textbf{not fatal} for the system, but it could become fatal if the number of non-consistent shares exceeds $(n-t)$. %This is because we need at least $t$-shares consistent to rebuild the secret. If there are less, it is not possible to reconstruct the secret.
            \\\\
            In this case too in view of solving this problem and mitigate the attack the  shareholders could contact the system administrator to proceed to check on the correctness of the dealer.

            \item The third error is related to the second one. In this case too the error is \textbf{not fatal} unless the wrong shares are less than $(n-t)$. Unlike before, the verification is done to check shareholders correctness. A participant is correct if they send the share received in the sharing phase and if the share is correct with the coordinate $x_i$ previously sent to them.
            \\\\
            In this case too when a shareholder is lying the dealer should disqualify them and otherwise log the error and alert the system admin for further verification in order to check and fix them if the shareholder is compromised.
            
            \item The fourth error concerns the verification of the $S^\prime$ correctness. The check is made because even when the attacker guesses the coordinates they must know also $r$ and $S$ in order to build $\overline{S^\prime}$. So in the case where $\overline{S^\prime} \ne S^\prime$ the error is \textbf{fatal} and the system stops the protocol and logs the error. In this case the liar may have been:
            \begin{itemize}
              \item an unauthorized user trying to cheat.
              \item the client that has sent the wrong $\overline{S^\prime}$.
              \item the dealer that in the sharing phase has set the wrong shares of another secret $S^{\prime\prime\prime}$, that is it has set $a_0=S^{\prime\prime\prime}$ with $\overline{S^\prime} \ne S^{\prime\prime\prime}$.
            \end{itemize}
            To solve this problem and mitigate the attack we could think of some mechanism to inform the other shareholders of the problem and contact the system administrator to proceed to check on the correctness of the dealer. The administrator verification is needed in order to prevent other attacks. But in any case the dealer stops the protocol because the error is \textbf{fatal}.

            \item The fifth verification is done because the service wants to be sure that whoever is accessing is an authorized client. Since the content of the message is encrypted with the $MS$ key only the service can read the contents of the message because $MS$ is an information that in theory only the client and the service know. Once the message is opened the $k$ and $(g^{S^\prime} h^{r^\prime})$ information are extracted. Thanks to these data the service can check the client correctness. If the verification does not go well it means that someone is cheating:
            \begin{itemize}
              \item an unauthorized user is trying to cheat without the dealer's help.
              \item a client has sent the wrong information on $E_{MS}\big[k, (g^{S^\prime} h^{r^\prime}) \big]$.
              \item the dealer is trying to cheat and bypass the checks in order to pass an unauthorized user.
            \end{itemize}
            In this case too in order to solve this problem and mitigate the attack we could think of some mechanism to inform the other shareholders of the problem and contact the system administrator to proceed to check on the correctness of the dealer. The service stops the protocol because the error is \textbf{fatal}. The administrator verification is needed in order to prevent other attacks and issues.

            \item The sixth verification is done on the information sent to the client from the service. This test is done because:
            \begin{itemize}
              \item a compromised dealer could put a fake service online and impersonate  the right service. With this test the client wants to be sure that the service with whom they are talking is the same as the one of the sharing phase. 
              \item the service is not the right server or it has sent a wrong data. 
            \end{itemize}
            In this case too in order to solve this problem and mitigate the attack we could think of some mechanism to inform the system administrator about the problem and proceed to checks on the correctness of the dealer or the service. The client stops the protocol because the error is \textbf{fatal}. The administrator verification is needed in order to prevent other attacks and issues.
            
            \item The last check is made because once the connection between client and service has occurred the information concerning the old shares must change. Service sends the information in order to show the dealer that everything has been successful between the client and the service, it like a proof of correctness. The dealer can verify that the service is not lying through the data sent. In a negative case, that is when the service is trying to cheat, the dealer stops the protocol and does not change the old share. The error is \textbf{fatal}.  In this case a report with the log of the error is made and the system administrator is contacted in order to verify the service's correctness. 
          \end{enumerate}
        This is the errors classification. Whenever an error occurs in scheme \ref{Scheme_Second} it means that someone could try to cheat. Hence every time an error arises, the system has to perform serious checks to deeply understand who is trying to break the system.

      \subsubsection{How to manage sensitive information in a secure way}\label{chap:How to manage sensible information in a secure way}  
        In this section an important problem is discussed. In scheme \ref{Scheme_Second} the information of $k^\prime$ is saved in the service; $k^\prime$ is sensitive information and the service should keep it stored locally. Hence service should take the responsibility of that data. 
        \\\\
        A possible solution to this problem could be to "reuse" parts of the system, so the resources used can be reused for other purposes such as using the primary Shamir's idea. That is, as in the case of scheme \ref{Scheme_III}, the service breaks the secret $k^\prime$ and leaves the management of the shares to the system. The coordinates are saved locally in the service. In this way, when the service needs to use the $k^\prime$ information it can ask the system for the shares and rebuild the secret about $k^\prime$. In this way the key $k^\prime$ is not stored locally but it is just used on-the-fly. Everything will remain in volatile memory. In this way the service does not have to memorize any sensitive info and even if it is under attack the attacker can not obtain any useful information but just the coordinates.

        \paragraph{Sharing phase}
          This scheme is very similar to the \ref{Scheme_III}  solution with the only difference that the coordinates $x_i^\prime$ and the various $c_j^\prime$ are saved on the service side. At this stage the service takes $k^\prime$, breaks it and sends the shares and commitments $c_j^\prime $ to the dealer. When the service needs $k^\prime$ it contacts the dealer.

          \begin{figure}[!htb]
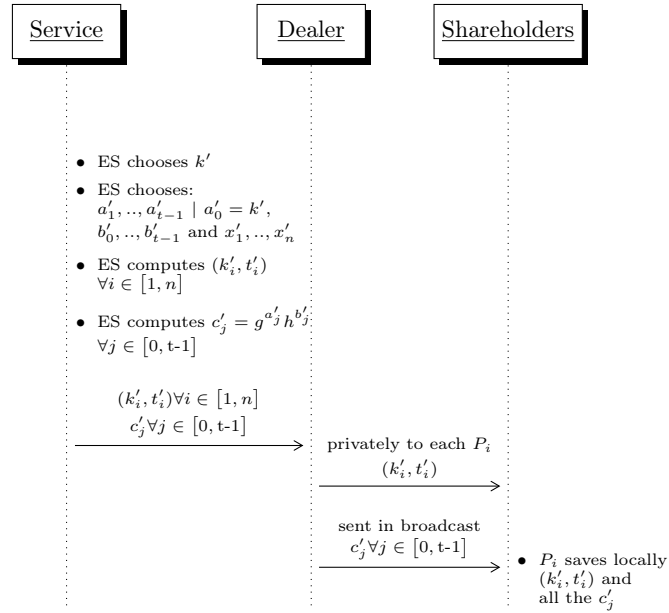

            \tikzset{
              % add this style to all tikzpicture environments
              every picture/.append style={
                % enable scaling of nodes
                transform shape,
                % set scale factor
                scale=0.9
              }
            }
            \centering
            \begin{sequencediagram}
                  \newinst{ES}{\shortstack{Service}}
                  \newinst[2]{D}{Dealer}
                  \newinst[1]{P}{Shareholders}

                  \stepcounter{seqlevel}
                \stepcounter{seqlevel}
                \stepcounter{seqlevel}
                \stepcounter{seqlevel}
                \stepcounter{seqlevel}
                \stepcounter{seqlevel}
                \stepcounter{seqlevel}
                \stepcounter{seqlevel}

                  \mess[0,font=\small]{ES}{\scriptsize\shortstack{$(k_i^{\prime},t_i^{\prime})\forall i \in \big[1,n\big]$\\$c_j^{\prime} \forall j\in \big[0,\text{t-1}\big]$}}{D}
                    \node (t0) at (mess from) {};
                    \node [right=-7mm,align=left,anchor=south west](t1) at (t0){\Large{
                  
                  \begin{varwidth}{\linewidth}\scriptsize \begin{itemize}
                      \item ES chooses $k^\prime$ 
                      \item ES chooses: \\$a_1^\prime, .., a_{t-1}^\prime\ |\ a_0^\prime=k^\prime$, \\$b_0^\prime, .. ,b_{t-1}^\prime$ and $x_1^\prime,..,x_n^\prime$
                   
                      \item ES computes $(k_i^\prime,t_i^\prime)$\\ $\forall i \in \big[1,n\big] $       
                      \item ES computes $c_j^\prime = g^{a_j^\prime} h^{b_j^\prime}$ \\ $\forall j \in \big[0,\text{t-1}\big]$
                      
                  \end{itemize}\end{varwidth}
              }\\\\\\\\};

              \mess[0,font=\small]{D}{\scriptsize\shortstack{privately to each $P_i$\\$(k_i^{\prime},t_i^{\prime})$}}{P}
                  
                  \stepcounter{seqlevel}

                  \mess[0,font=\small]{D}{\scriptsize\shortstack{sent in broadcast\\$c_j^{\prime} \forall j\in \big[0,\text{t-1}\big]$}}{P}
                  \node [right=-7mm,align=left](t6) at (mess to){\\\\
                  \Large{
                  \begin{varwidth}{\linewidth}\scriptsize \begin{itemize}
                      \item $P_i$ saves locally\\$(k_i^{\prime},t_i^{\prime})$ and \\all the $c_j^{\prime}$
                  \end{itemize}\end{varwidth}
                }};

              %\draw[line width=3pt] (t1) -- (t0);

            \end{sequencediagram}
            \label{protocol:ExchangeMessagesSharing_SensibileInfo}
              \caption{Sharing phase for how to manage sensitive information in a secure way}
          \end{figure}

        \paragraph{Reconstruction phase}
          First of all, the service sends the info of the $c_j^\prime$ in order to prove who is the requester of the shares. The shares are sent back to the dealer if the $c_j^\prime$ are correct and so they are sent to the service. At this point the service checks the shares correctness thanks to the Pedersen's work and in a positive way the service rebuilds the secret $k^\prime$. When the service wants to replace the old secret $k^\prime$ with a different $k^{\prime\prime}$ he could send the new shares in order to replace the older ones.
          \begin{figure}[H]
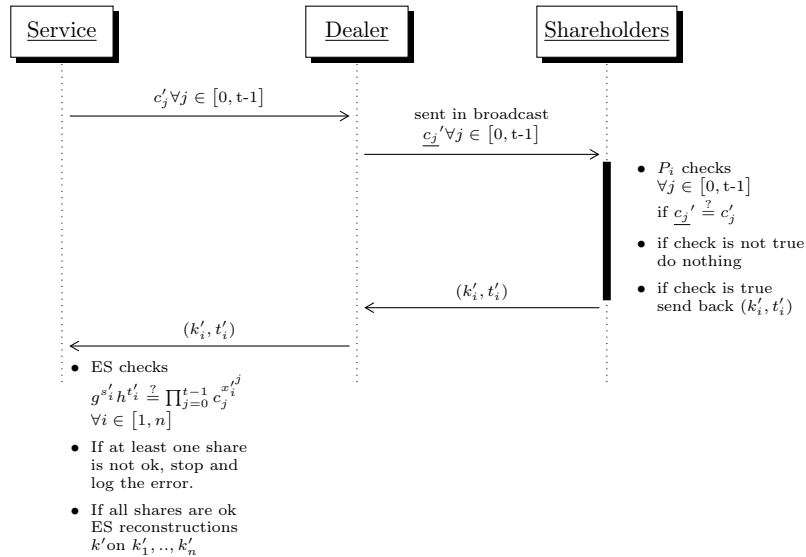

            \tikzset{
              % add this style to all tikzpicture environments
              every picture/.append style={
                % enable scaling of nodes
                transform shape,
                % set scale factor
                scale=0.85
              }
            }
            \centering
            \begin{sequencediagram}
                  \newinst{ES}{\shortstack{Service}}
                  \newinst[3]{D}{Dealer}
                  \newinst[2]{P}{Shareholders}

                  \mess[0,font=\small]{ES}{\scriptsize\shortstack{$c_j^{\prime} \forall j\in \big[0,\text{t-1}\big]$}}{D}

                  \mess[0,font=\small]{D}{\scriptsize\shortstack{sent in broadcast\\$\underline{c_j}^{\prime} \forall j\in \big[0,\text{t-1}\big]$}}{P}
                    \node (t0) at (mess to) {};

              \stepcounter{seqlevel}
              \stepcounter{seqlevel}
              \stepcounter{seqlevel}

              \mess[0,font=\small]{P}{\scriptsize $(k_i^{\prime},t_i^{\prime})$}{D}
                    \node (t1) at (mess from) {};
                    %,draw=blue!80,
                    \node [right=15mm,align=left,below](t2) at (t0){\Large{
                  
                  \begin{varwidth}{\linewidth}\scriptsize \begin{itemize}
                      \item $P_i$ checks \\$\forall j \in \big[0,\text{t-1}\big]$\\ if $\underline{c_j}^{\prime} \stackrel{?}{=} c_j^\prime$
                      \item if check is not true \\ do nothing
                      \item if check is true\\ send back  $(k_i^{\prime},t_i^{\prime})$
                  \end{itemize}\end{varwidth}
              }};
              
              \draw[line width=3pt] (t1) -- (t0);

              \mess[0,font=\small]{D}{\scriptsize $(k_i^{\prime},t_i^{\prime})$}{ES}
                \node (t4) at (mess to) {};
                \node [right=-7mm,align=left](t5) at (t4){\\\\\\\\\\\\\\\\\\\\\Large{         
                  \begin{varwidth}{\linewidth}\scriptsize \begin{itemize}
                      \item ES checks \\ $ g^{s_i^\prime} h^{t_i^\prime} \stackrel{?}{=} \prod_{j=0}^{t-1}{c_j^{x_i^{\prime^j}}}$\\$ \forall i \in \big[1,n\big]$

                    \item If at least one share\\ is not ok, stop and\\ log the error.

                    \item If all shares are ok 
                    \\ES reconstructions\\
                     $k^\prime $on $k_1^\prime,..,k_n^\prime$
                  \end{itemize}\end{varwidth}
                  }};
            \end{sequencediagram}
            \label{protocol:ExchangeMessagesReconstruction_SensibileInfo}
              \caption{Reconstruction phase for how to manage sensitive information in a secure way}
          \end{figure}

    \subsection{Third proposed solution}
      \subsubsection{Scheme}\label{Scheme_III}
        In this section another secure solution against the last attack model will be presented. As anticipated before this solution is less complex in terms of messages but it requires a greater effort from the client. This is because some cryptographic operations are done by the client. Thus we will have a benefit in terms of message complexity but we have to be aware that the client has to perform special operations.
        \\\\
        The total number of messages exchanged between the actors in the sharing and reconstruction phase is just five. The basic idea is that the server becomes a container of information. The point is that the information is not sensitive so if it is lost the thief cannot extract any useful data. The information kept in the server are the shares but without the coordinates $x_i$, in this way even if the attacker can penetrate the machine and steal the share they cannot extract any useful data. Thus in order to achieve their purpose the attacker must steal the shares in the server and also steal the coordinates in the client.
        \\\\
        The basic idea is that who breaks the secret is the owner of the secret, hence the client. Also in this case the shared secret is not $S$ but $S^\prime$ where $S^\prime = (g^S h^{r^\prime})$, in this way changing $r^\prime$ changes $S^\prime$.
        Once the information $S^\prime$ is broken, it will be reconstructed in a second moment in order to give an evidence of the client's identity to the server. Thus the server will only have to manage the shares but without the coordinates the information are useless in order to rebuilt $S^\prime$. 
        In the next paragraphs the two phases will be presented. This scheme is very simple and does not require a network model as before. The network model used is backward compatible with the client-server one. This is because the dealer figure and the dealer actions will be performed by the client. The logic is focused on saving the shares in a server without the abscissas coordinator. Obviously the various shares could be distributed in other various servers to have a greater security, but the point is that it is not possible to reconstruct the secret even with $n$-share. Also the coordinates are saved server side but in an encrypted way, where the encryption key is the hashed secret $S$ that is the $password$. In this way the client will ask for the information of the encrypted coordinates, and will open the content because only they own the encryption key, that is the password. To authenticate they will send the plaintext coordinates so the server will be able to verify their correctness. 
        \\\\
        A new secret $S^{\prime\prime}$ will be sent where $S^{\prime\prime} = (g^S h^{r^{\prime\prime}})$ with $r^\prime \ne r^{\prime\prime}$. Once the secret $S^\prime$ is rebuilt it is no longer valid and $S^{\prime\prime}$ is needed. In the end, each time the client logs in, the shares change.
        \\\\
        Hence, in this case too the scheme continues to change shares every time that the client logs in, coordinates are never given to the server in plaintext until the reconstruction phase. The check on the shares correctness is done at the end of the reconstruction phase.

        \paragraph{Sharing phase}
          The sharing phase consists of only three messages. In the first message the client sends the information about shares to the server, the encrypted coordinates and the $c_j$ contributes. In this first phase the client must choose all the parameters to built the two polynomials of $t$-degree and computes the $c_j$.
          \\\\
          The server saves the information locally and sends back a random number $c$ to the client. This number is used when the client wants to recall the reconstruction phase to get proof of who is trying to access. With this simple number a possible attacker cannot perform multiple requests in order to gather the $E_k$ information.
          \\\\
          Compared to the first scheme the client has to remember just the password, $c$ and $r^\prime$.

          \begin{figure}[!htb]
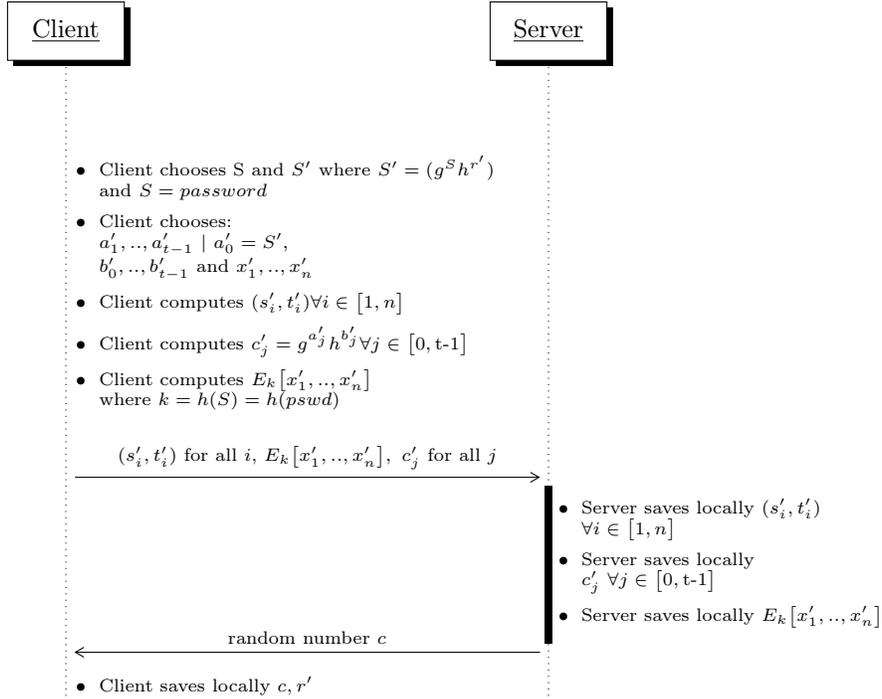

          \tikzset{
              % add this style to all tikzpicture environments
              every picture/.append style={
                % enable scaling of nodes
                transform shape,
                % set scale factor
                scale=0.97
              }
            }
          \centering
          \begin{sequencediagram}
                \newinst{C}{Client}
                \newinst[5]{S}{Server}
                \stepcounter{seqlevel}
              \stepcounter{seqlevel}
              \stepcounter{seqlevel}
              \stepcounter{seqlevel}
              \stepcounter{seqlevel}
              \stepcounter{seqlevel}
              \stepcounter{seqlevel}
              \stepcounter{seqlevel}

              \mess[0,font=\small]{C}{\scriptsize$(s_i^\prime,t_i^\prime)$ for all $i$, $E_k\big[x_1^\prime,..,x_n^\prime\big],\ c_j^\prime$ for all $j$}{S}
                \node (t0) at (mess to) {}; 
                \node (t2) at (mess from) {};
                \node [right=-7mm,align=left,anchor=south west](t3) at (t2){\Large{
                \begin{varwidth}{\linewidth}\scriptsize \begin{itemize}
                    \item Client chooses S and $S^\prime$ where $S^\prime= (g^S h^{r^\prime})$ \\ and $S = password$

                    \item Client chooses: \\$a_1^\prime, .., a_{t-1}^\prime\ |\ a_0^\prime=S^\prime$, \\$b_0^\prime, .. ,b_{t-1}^\prime$ and $x_1^\prime,..,x_n^\prime$
                   
                    \item Client computes $(s_i^\prime,t_i^\prime) \forall i \in \big[1,n\big] $       
                    \item Client computes $c_j^\prime = g^{a_j^\prime} h^{b_j^\prime} \forall j \in \big[0,\text{t-1}\big]$
                    \item Client computes $E_k\big[x_1^\prime,..,x_n^\prime\big]$\\ where $k=h(S)=h(pswd)$
                \end{itemize}\end{varwidth}
              }\\\\\\};
              \stepcounter{seqlevel}
              \stepcounter{seqlevel}
              \stepcounter{seqlevel}
              
              \mess[0,font=\small]{S}{\scriptsize random number $c$}{C}
                \node (t1) at (mess from) {}; 
                \node [right=-7mm,align=left](t4) at (t0){\\\\\\\\\\\\\\\Large{
                \begin{varwidth}{\linewidth}\scriptsize \begin{itemize}

                    \item Server saves locally $(s_i^\prime,t_i^\prime)$\\$ \forall i \in \big[1,n\big]$ 
                    \item Server saves locally \\$c_j^\prime\ \forall j \in \big[0,\text{t-1}\big] $
                    \item Server saves locally $E_k\big[x_1^\prime,..,x_n^\prime\big]$
                  
                \end{itemize}\end{varwidth}
              }};

              \node [right=-7mm,align=left](5) at (mess to){\\\\\\\Large{
                \begin{varwidth}{\linewidth}\scriptsize \begin{itemize}

                    \item Client saves locally $c, r^\prime$
                \end{itemize}\end{varwidth}
              }};
              \draw[line width=3pt] (t0) -- (t1);

          \end{sequencediagram}
          \label{protocol:ExchangeMessagesSharing3_0}
          \caption{Exchange messages of Sharing phase for 3.0 scheme}
        \end{figure}
        
        \paragraph{Reconstruction phase}
          
          \begin{figure}[H]
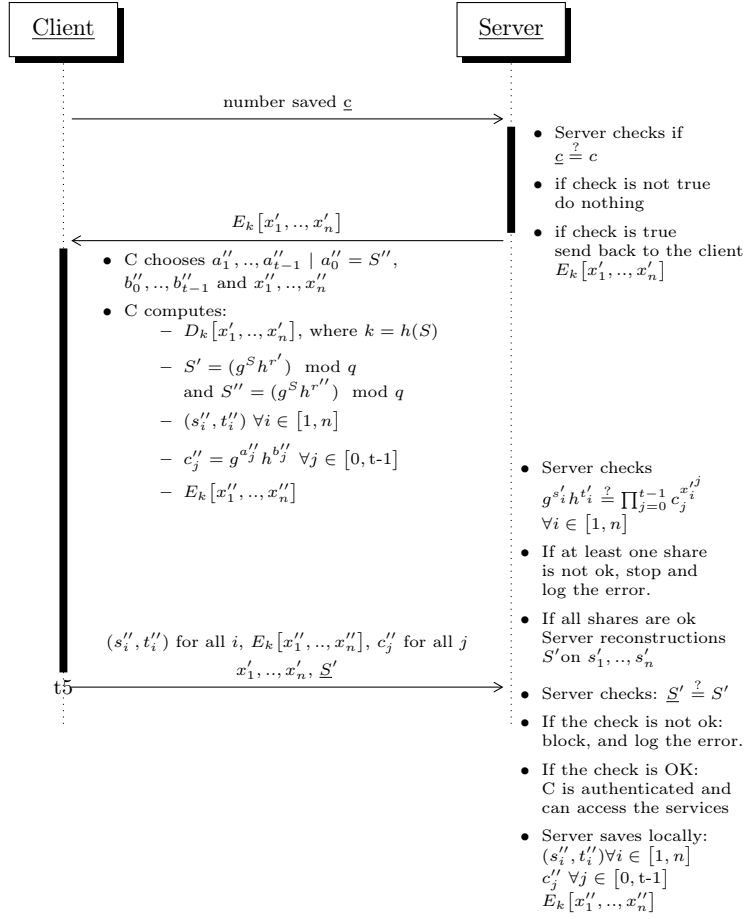

            \tikzset{
              % add this style to all tikzpicture environments
              every picture/.append style={
                % enable scaling of nodes
                transform shape,
                % set scale factor
                scale=0.9
              }
            }
            \centering
            \begin{sequencediagram}
                  \newinst{C}{Client}
                  \newinst[5]{S}{Server}

                  \mess[0,font=\small]{C}{\scriptsize number saved \underline{c}}{S}
                    \node (t0) at (mess to) {};
                    \node (t6) at (mess from) {};

                  \stepcounter{seqlevel}
                \stepcounter{seqlevel}

                  \mess[0,font=\small]{S}{\scriptsize  $E_k\big[x_1^\prime,..,x_n^\prime\big]$}{C}
                    \node (t1) at (mess from) {}; 
                    \node (t3) at (mess to) {}; 
                    %draw=blue!80,
                    \node [right=16mm,align=left,below](t2) at (t0){\Large{
                  \begin{varwidth}{\linewidth}\scriptsize \begin{itemize}
                      \item Server checks if \\$\underline{c} \stackrel{?}{=} c$
                      \item if check is not true \\do nothing
                      \item if check is true \\send back to the client \\$E_k\big[x_1^\prime,..,x_n^\prime\big]$
                  \end{itemize}\end{varwidth}
                }};
              \draw[line width=3pt] (t1) -- (t0);
              
              \stepcounter{seqlevel}
                \stepcounter{seqlevel}
              \stepcounter{seqlevel}
                \stepcounter{seqlevel}
              \stepcounter{seqlevel}
              \stepcounter{seqlevel}
              \stepcounter{seqlevel}
              \stepcounter{seqlevel}
              \stepcounter{seqlevel}
              \stepcounter{seqlevel}

              \mess[0,font=\small]{C}{\scriptsize\shortstack{$(s_i^{\prime\prime},t_i^{\prime\prime})$ for all $i$, $E_k\big[x_1^{\prime\prime},..,x_n^{\prime\prime}\big]$, $c_j^{\prime\prime}$ for all $j$
              \\$x_1^{\prime},..,x_n^{\prime}$, $\underline{S}^\prime$}} {S}

                \node (t5) at (mess from) {t5}; 
                \node (t8) at (mess to) {}; 
                %,draw=blue!80,
                    \node [right=27mm,align=left,below](t4) at (t3){\\\Large{
                  \begin{varwidth}{\linewidth}\scriptsize \begin{itemize}
                    \item C chooses $a_1^{\prime\prime}, .., a_{t-1}^{\prime\prime}\ |\ a_0^{\prime\prime}=S^{\prime\prime},$ \\$b_0^{\prime\prime}, .. ,b_{t-1}^{\prime\prime}$
                      and $x_1^{\prime\prime},..,x_n^{\prime\prime}$
                    
                    \item C computes: \\
                        \begin{varwidth}{\linewidth}\scriptsize \begin{itemize}
                          \item $D_k\big[x_1^\prime,..,x_n^\prime \big]$, where $k=h(S)$
                          \item $S^\prime= (g^S h^{r^\prime})\mod{q}$\\ and $S^{\prime\prime} = (g^S h^{r^{\prime\prime}})\mod{q}$
                        
                        \item $(s_i^{\prime\prime},t_i^{\prime\prime})\ \forall i \in \big[1,n\big]$
                          \item $c_j^{\prime\prime} = g^{a_j^{\prime\prime}} h^{b_j^{\prime\prime}}\ \forall j \in \big[0,\text{t-1}\big]$
                          \item $E_k\big[x_1^{\prime\prime},..,x_n^{\prime\prime}\big]$
                          %\\ where $k=h(S)=h(pswd)$
                      \end{itemize}\end{varwidth}
                  \end{itemize}\end{varwidth}
                }\\};

                \draw[line width=3pt] (t3) -- (t5);

              \node [right=-7mm,align=left](t4) at (t8){\Large{
                  \begin{varwidth}{\linewidth}\scriptsize \begin{itemize}
                    \item Server checks \\ $ g^{s_i^\prime} h^{t_i^\prime} \stackrel{?}{=} \prod_{j=0}^{t-1}{c_j^{x_i^{\prime^j}}}$\\$ \forall i \in \big[1,n\big]$

                    \item If at least one share\\ is not ok, stop and\\ log the error.

                    \item If all shares are ok 
                    \\Server reconstructions\\
                     $S^\prime $on $s_1^\prime,..,s_n^\prime$

                     \item Server checks: $\underline{S}^\prime \stackrel{?}{=} S^\prime$

                     \item If the check is not ok: \\ block, and log the error. %that someone has steal x1,..,xn but not r’

                     \item If the check is OK: \\ C is authenticated and \\can access  the services
                     
                     \item Server saves locally: 
                     \\ $(s_i^{\prime\prime},t_i^{\prime\prime}) \forall i \in \big[1,n\big]$
                     \\ $c_j^{\prime\prime}\ \forall j \in \big[0,\text{t-1}\big]$
                     \\ $E_k\big[x_1^{\prime\prime},..,x_n^{\prime\prime}\big]$
                  \end{itemize}\end{varwidth}
                }};

            \end{sequencediagram}
            \label{protocol:ExchangeMessagesRec3_0}
              \caption{Exchange messages of Reconstruction phase for 3.0 scheme}
          \end{figure}
          In this phase first of all the client will send the number $\underline{c}$ to the server in order to give a primary proof. The server will check the correctness of $\underline{c}$ and in a positive case it will send back $E_k\big[x_1^\prime,\dots,x_n^\prime \big]$. The client, they have obtained $E_k$, will open it because they alone know the key $k$, that is the hash of the password. With $r^\prime$ the client will be able to rebuild $S^\prime$ and compute a $S^{\prime\prime}$ with $r^{\prime\prime}$ (with $r^\prime \ne r^{\prime\prime}$) in order to compute other shares $s_i^{\prime\prime}$ to give to the server as a proof for the next session. Once these operations have been completed the client will send to the Dealer:
          \begin{itemize}
            \item the old data to authenticate himself.
            \item the new data to replace the older one for the next login.
          \end{itemize}  
          The server will check the shares correctness. If everything goes well it will rebuild $S^\prime$ and check if it is the same as that sent by the client. If the two values are equal the server will replace the old information with the new one and the client will authenticate.
    
      \subsubsection{Attack Model check}
        Compared to the attack model check of the \ref{Scheme_Second} scheme, the \ref{Scheme_III} scheme is much simpler. This is due to the fact that only two actors are involved.
        \\\\
        The attacker in this type of scheme can attack only the client or the server. 
        
        \paragraph{Client}
          In this case the attacker can act actively or passively. Let us consider a \textit{passive attack}:
          \begin{itemize}
            \item \textbf{Sharing phase.} In a sharing phase the information about $S$ is not saved locally and is only typed when requested, so there is no way for the passive attacker to obtain it. The only way to read the data is with keylogging methods or read data in RAM, but this lies outside the scope of our study. The only information that a passive attacker can take are $r^\prime$ and $c$. With this information the attacker can obtain $E_k\big[x_1^\prime,..,x_n^\prime\big]$ and try to open it with a dictionary attack or other known methods.

            \item \textbf{Reconstruction phase.} In this case a passive attacker can only take  $r^{\prime\prime}$ because it is the only info saved locally. The other information is computed but it is not saved locally.
          \end{itemize}
          On a passive level the attacker can not take any compromising info apart from $E_k\big[x_1^\prime,..,x_n^\prime\big]$, but this, however, has a temporal validity because every login changes because the coordinates change. Hence, even if the attacker can open the message the information obtained will be useless.
          \\\\
          Let us now consider what happens if the attacker operates actively:
          \begin{itemize}
            \item \textbf{Sharing phase.} In this case the attacker could send inconsistent shares. Since also the commitments $c_j$ are sent the wrong shares will be verified during reconstruction phase. In this way the malicious client is discovered.

            \item \textbf{Reconstruction phase.} In this case too the attacker could not give non-consistent shares about $S^{\prime\prime}$ because they are verified at the next reconstruction phase.
          \end{itemize}
          Hence the attacker can not perform malicious actions because they would be discovered by the server in the reconstruction phase.

        \paragraph{Server}
          Moving to the server, let us consider the \textit{passive attacks}:

          \begin{itemize}
            \item \textbf{Sharing phase.} In the sharing phase the only information that could be taken by the attacker are:
            \begin{itemize}
              \item the shares, which are useless without coordinates.
              \item the $ E_k\big[x_1 ^\prime, .., x_n^\prime \big]$ information but the attacker should be able to decrypt it. In order to do that they need a lot of computing power. However, even if they succeed in opening the encrypted message, at the next login this info will be changed, so the data taken will become useless.
            \end{itemize}

            \item \textbf{Reconstruction phase.} In the reconstruction phase the passive attacker can take new shares. With this new information the attacker can not do anything. Also with the coordinates and the $S^{\prime}$ information the attacker can not do anything because $S^{\prime} = (g^S h^{r^\prime}\mod{p})$ so they can not obtain $S$. The information $S$ is information-theoretic secure thanks to $g$, $h$, $p$ and $r^\prime$. Hence even if $S^{\prime}$ is discovered it is useless data.
          \end{itemize}
          Therefore since the server does not save any important information even if the attacker succeeds in obtaining the server access it can not retrieve any useful data.
          \\\\\\
          Now let us look at how an active attacker can operate:
          \begin{itemize}
            \item \textbf{Sharing phase.} The attacker can not do anything because they only save the info and send $c$.

            \item \textbf{Reconstruction phase.} The attacker can act as in the passive case:
            \begin{itemize}
              \item the new shares but without the coordinates they can not do anything.
              \item the $ E_k\big[x_1^{\prime\prime}, .., x_n^{\prime\prime} \big]$ information.
              \item the old coordinates  $[x_1 ^\prime, .., x_n^\prime \big]$.
              \item $S^\prime$.
            \end{itemize} 
            With all this information the attacker can not do anything. With the new data and with the old ones the malicious server cannot retrieve any useful information.
          \end{itemize}
          Hence even on an active level the scheme is very safe.

      \subsubsection{Errors classification}

        In this section an error classification of the third solution will be presented. In the \ref{Scheme_Second} scheme there are more possibilities to make mistakes than in the \ref{Scheme_III} ones. This happens because there are more parties involved then the risk of cheating increases. If there are tests there can be errors when the test fails. In both schemes the tests are done only in the reconstruction phase, so the sharing phase is free of errors, that is since there are no checks errors cannot be discovered, since they aren't detected.
        \\\\
        The errors can be of two types:
          \begin{itemize}
            \item \textbf{Fatal}: this means that if this error arises the system must not move on.
            \item \textbf{Not-fatal}: this means that if this error arises the system can move on with some precautions.
          \end{itemize}
        In the \ref{Scheme_III} model if something goes wrong the server blocks the operations and it logs the error. This happens because server computes the test and checks that the client is correct. There are three errors that can be found in the third scheme. Each error found corresponds to a certain security level, if the user passes all levels it is authenticated. If an attacker tries to cheat or bypass the security levels they need a lot of computer power and a lot of extra-information.
        \\\\
        In this case we have three checks and then three possible errors: 
          \begin{enumerate}
            \item The first error happens when the client sends $\underline{c}$, in this case if the data does not correspond to the value $c$ sent previously this means that the client is cheating and it is trying to gather $E_k\big[ x_1,..., x_n\big]$. The server will discover the error and will close the connection.

            \item The second error arises when the client sends the information about the coordinates. If the client is lying and trying to send random coordinates, these will be checked thanks to the Pedersen's work. In this way the server can immediately understand if the coordinates sent are correct or not. 
            In case of error it means that the client has not correctly opened the message or an attacker is trying to cheat. In this case the server stops and closes the connection.
            
            \item The last security level guarantees that even if the attacker is able to guess $c$ and opens the encrypted message $E_k\big[ x_1,..., x_n\big]$ they also have to know the secret $S$ and $r^\prime$ in order to compute  $S^\prime = g^S h^{r^\prime}$. The third test is made by checking that the secret rebuilt from the server side with the data sent in the sharing phase and the secret $\underline{S}^\prime$ sent in the last message by the client are equal. The two items of information are compared and if they do not match this means the client has mistaken something or the attacker does not know the $S$ and $r^\prime$ information. In this situation the server does not allow the client's authentication.
          \end{enumerate}
        In all three cases the errors are fatal because they do not allow the operation to continue.

    \subsection{Final considerations}
      In this sections the pros, cons and limitations of the two schemes will be presented with their differences and common limitations. Both schemes are valid solutions to the problem introduced in \ref{NewWayOfThinking} but they manage the information and the operations in a different way.

      \subsubsection{Pros, cons and limitations}\label{subsection:Pros, contras and limits} 
        We will begin with the common limitations. A common limit of the two schemes is what the user must memorize in addition of the password. In the first case the user must save more information than in the second solution. In the first scheme the user has to keep the $r,\ r^\prime,\ k,\ MS,\ S$, and $\big[ x_1,\dots,x_n\big]$ data. Furthermore at every reconstruction phase all the information will change except $S$.
        In the second case, instead, the information that the user has to keep are $r^\prime,\ c$ and $S$ and at every reconstruction phase $r^\prime$ will change.
        \\\\
        Another limit of this scheme arises when the user has more than one device for logging in. In this case, the synchronization of the information between all the devices is needed. For the first study case a constraint is set: a user can have at most one active device where he is logged in. In this way we will not have a synchronization problem between different devices. We can postpone this study to a future work.
        \\\\
        The system must give the user an opportunity to change device on which they log in. The different ways to implement this case are the following:
        \begin{enumerate}
          \item Leave the information saved on the device on which the user has logged in.
          \item Give the opportunity to save the information in a safe place.
          \item Forget the information.
        \end{enumerate}
        The key point of the security is the secret $S$, that is the password. Obviously the secret $S$ \textbf{must not be saved anywhere but kept in mind by the user}. This is because, without the secret $S$ even if the attacker could steal the other information, the latter would be useless for achieving the attacker's goal. Indeed, besides $r,\ r^\prime,\ k,\ MS$, and $\big[ x_1,\dots,x_n\big]$ the attacker must know $S$ in order to compute $S^\prime = g^S h^{r}$ and the same thing is valid for $(g^{S^\prime} h^{r^\prime})$. The same is true the \ref{Scheme_III} scheme. 
        \\\\
        The second solution is the better to adopt, this means that the system give the possibility to the user to save the information in a safe place like an external USB. In this way the constraint that every user can have at least one device where he is logged in is respected but at the same time if the user changes the device, he can bring with him the information that he needs in order to log in. Hence, if the user accesses in an internet cafe, once the session is over, he has to save the information on an external hard disk. The system will delete the additional data from the pc at the end of the session. In this way the user can change the device to log in when he wants.
        \\\\
        Another problem we must examine is how it is possible to handle the operation from the client-side. Since the client has to execute some particular operations in both schemes we need a special client. Since the client needs to be special we should implement some operations with some client-side language. Since it is an online service we will use a browser to access the two phases hence since JavaScript is the language used we should implement these operations in that language. The problem with this language is the facility with which one can make an injection, so an attacker could take possession of a machine and perform malicious actions to get the information he needs. In any case, if $S$ remains hidden, with other information the attacker can not do anything. A solution to solve this problem entirely could be to implement all the client logic in a browser add-on. In this way an attacker can not compromise the source code written in javascript and can not take useful information.
        \\\\
        The \textbf{pros} of the \ref{Scheme_Second} scheme are the following:
        \begin{itemize}
          \item The secret S is never revealed 
          \item Heavy cryptographic operations are done on the server side
        \end{itemize}
        The \textbf{cons} instead are:
        \begin{itemize}
          \item Many interactions between the system's actors.
          \item Storing a lot of information over $S$.
          \item A broadcast channel is required.

        \end{itemize}   
        Regarding the \ref{Scheme_III} solution, the \textbf{pros} are the following:
        \begin{itemize}
          \item The secret S is never revealed.
          \item Few interactions between the system's actors.
          \item Storing of little information over $S$.
          \item A broadcast channel and a set of servers is not required.
        \end{itemize}  
        The \textbf{cons} instead are:
        \begin{itemize}
          \item Heavy cryptographic operations are done on the client side
          \item Having $E_k$ saved in the server side, in the presence of a malicious dealer it can launch a brute force attack against $E_k$ and try to get the coordinates and then get $S^\prime$
        \end{itemize} 
        Finally the two schemes are equivalent in terms of security. Therefore the choice must be made based on each individual case and how the scheme will be used.

      \subsubsection{Static dealer vs dynamic dealer}
        This section is dedicated to the problem of choosing whether it is more convenient to have a static or dynamic dealer. In the previous chapters and sections the network model proposed was based on a fixed dealer. However a dynamic dealer could be implemented.  In the dynamic dealer model, when a client's request comes in, before handling the message the system launches a leader election. The algorithm terminates with an elected dealer. In this way a dynamic dealer model is implemented.
        \\\\
        There are some positive and negative aspects. One of the positive aspects is that all the schemes we have seen so far are still valid because once the system has chosen who should behave as dealer, the solutions seen so far are applicable. In this way the static case is seen as a particular case of the dynamic one. Hence a dynamic model is an additional feature that can be adopted in the system but at the same time is not mandatory for the correct working of the system.
        \\\\
        Let us now consider the benefits:
        \begin{enumerate}
          \item The dealer is not fixed then the attacker does not know who the dealer will be in the next election so they should take control of more than one participant if they want to gather as much information as possible about users. In this way there is an additional layer of security.
          \item In case of a dealer failure, the system continues working. The system could tolerate up to $n-k$ failures only if the remaining $t$-shares are valid.
        \end{enumerate} 
        Instead the negative aspects are:
        \begin{enumerate}
          \item With the dealer election for each phase the number of messages increases and therefore the response time of the system increases too.
          \item The election must be managed safely, there is the risk of fault shareholders. In order to perform the election an extra effort is needed.
        \end{enumerate} 
        Since the election is not preponderant for the functionality of the system it will be considered later in the future work. At this moment we focus on the case of a static dealer because the election is a previous step and it does not influence the other operations, so we can first focus on feasibility with a static dealer and in the forthcoming studies we could be also handle this case.

  %%%%%%%%%%%%%%%%%%%%%%%%%%%%%%%%%%%%%%%%%%%%%
  %   Proof of concept: a working prototype   %
  %%%%%%%%%%%%%%%%%%%%%%%%%%%%%%%%%%%%%%%%%%%%%
  \newpage
  \section{Proof of concept: a working prototype}\label{chap:Proof of concept: a working prototype}
    In this chapter a proof of concept will be presented.  The proof of concept (PoC) is a prototype built in order to prove the \textit{feasibility} and the \textit{correctness} of the studied model. After extensive study one of the three proposed solutions has been implemented. The solution implemented is the second one (see section \ref{Scheme_Second}). In the following chapter the system infrastructure, the network infrastructure, the technologies used and, a performance analysis will be treated. 
    
    \subsection{The proposed authentication system}
      % introduzione del sistema realizzato. Approccio top-down cioè si parte dalle funzionalità del sistema (registrazione-login) fornite all’utente.
      A top-down approach will be used, starting from the functionalities offered by the system to the user, that is the sign up and login, in order to explain the system infrastructure. 
      \subsubsection{Introduction to the system}
        % spiegare gli obbietivi, cioè registrazione ed autenticazione dell'utente.
        
        % spiegazione delle features in generale (come presenteresti il sistema cosi a chi non sa nulla??)
        % obbitettivo, web-app, comunicazione sicura, come viene salvata la password?
        % problema dati aggiuntivi
        The goal of the system is very easy: allow the sign up and login to the users of a given service. The sign up is done through the so far called sharing phase, instead the login is done through the so far called reconstruction phase. The two functions are implemented with two different web pages (it will be seen in detail in section \ref{ImplementedApplication}). The paradigm of the password-based authentication is used with some exceptions. How it is explained in  section \ref{Scheme_Second}, the user has to preserve some additional information apart from the password. Only the union of the password and these data allows the authentication of the user to the service. The user needs a simple browser to access to the two phases because it is a web application. The additional information is saved in the browser exploiting the local storage technology that will be treated in section \ref{Local_storage}. In this way the only info the user has to keep in mind is the password. Since only the union of the two pieces of information allows the authentication also in this case if one of the two data were stolen from the user, the hacker could not retrieve any useful information. Note that the key point of all these data is the password $S$ because without that it is not possible to rebuild the other ones. 
        \\\\
        From the user's point of view the prototype is a simple web-application, but as it is explained in section \ref{sec:system_infrastructure}, behind the web interface there is a cluster of machines that work together in order to reach a common aim.
        \\\\
        Being a web application and using a browser to access to the system HTTPS protocol is used for the communication between client and server. The dealer gives its certificate to the client for the two functions (sign up and login). At the end of the login, the authenticated user will communicate with the service, through the HTTPS protocol but in this case the service certificate will be displayed. The HTTPS protocol ensures the sever authentication of the accessed website and protection of the privacy and integrity of the exchanged data. It protects against MIMA and bidirectional encryption protects against eavesdropping and tampering of the communication. 
        \\\\
        The last thing to note is how the password is saved. The password is never saved in the browser and it is never sent in plaintext. The information used in order to authenticate the user is $g^{password}\ h^r\mod{p}$. In this way the data are information-theoretic secure so it cannot be broken even if the hacker has unlimited computing power. This is an important improvement because, in the worst case when the dealer or at least $t$-shareholders are under attack, the hacker cannot retrieve any useful data.

      \subsubsection{The implemented application}\label{ImplementedApplication}
        Let's see the front-end of the application that is the two phases from the user's point of view:
          \paragraph{Sign up}
            In this phase the client signs up to a service. The user has to provide the following information:
            \begin{itemize}
              \item Username
              \item Email
              \item Password
            \end{itemize}
            The password will be used in order to compute the following additional information:
            \begin{itemize}
              \item $S^\prime = g^{password}\ h^r\mod{q}$
              \item $ MC = E_k \big[g^{S^\prime} h^{r^\prime}\mod{p} \big]$
            \end{itemize}
            The password information will never be saved in the client side but it must be kept in mind by the user.
            \\\\ 
            Once the fields are filled, the information is sent to the dealer. The dealer will proceed with the sharing phase. As it is shown in section \ref{Scheme_Second} is seen, the client sends:
            \begin{itemize}
              \item two messages to the Dealer.
              \item a message to the Service
            \end{itemize}
            All this operated in a transparent manner. At the end of the sign up phase the information saved locally on the browser is the following:
            \begin{itemize}
              \item The random number $r $
              \item The random number $r^\prime$
              \item $MC = E_k \big[g^{S^\prime} h^{r^\prime}\mod{p} \big]$
              \item Key $k$ 
              \item $MS = E_{k^\prime} \big[MC\big]$
              \item The coordinates: $x_1, \dots, x_n$
            \end{itemize}
            This information will be used by the user in the login phase.
            The technologies used, like local storage, how $MC$ is calculated and how the random numbers are computed, will be explained in section \ref{Client_side} . 
            \\\\
            Below the web-interface is shown:
            \begin{figure}[h]
              \centering
              \includegraphics[width=0.8\textwidth]{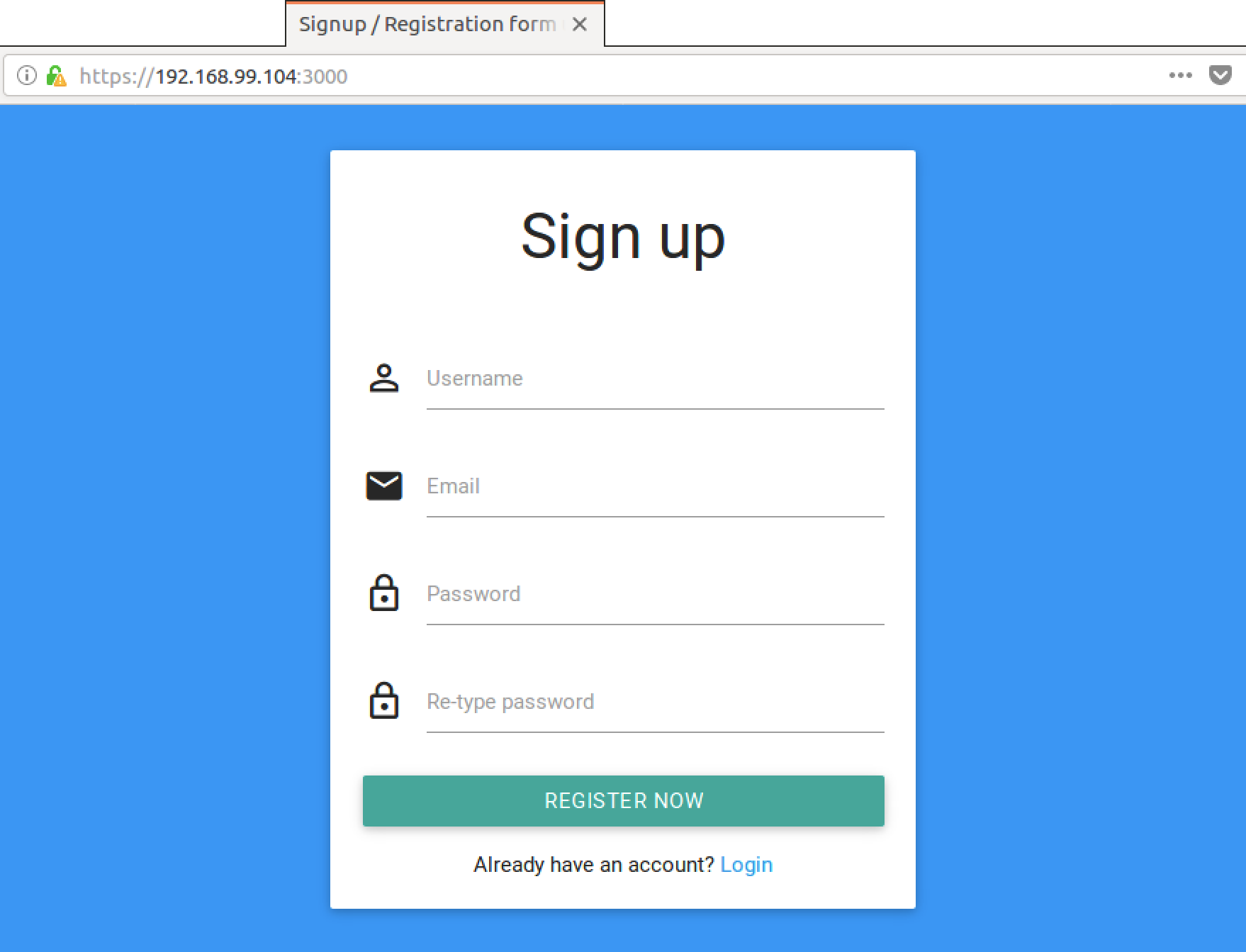}
              \caption{Sign up interface}
              \label{fig:Signup}
            \end{figure}

          \paragraph{Login}
            In this phase the user has to prove their identity in order to be authenticated. The only information to type is username and password because the others is transparent. The are two kinds of additional data, the "old" ones and the "new" ones. The "old" data are stored on the browser at the end of the reconstruction phase or at the end of the previous login. The "new" data are computed in the following manner: 
            \begin{itemize}
              \item $S^{\prime\prime} = g^{password}\ h^{r^{\prime\prime}}\mod{q}$
              \item $ MC^\prime = E_{k^\prime} \big[g^{S^{\prime\prime}} h^{r^{\prime\prime\prime}}\mod{p} \big]$
              \item $E_{MS} = [k,\ g^{S^\prime} h^{r^\prime}]$        
            \end{itemize}
            The latter calculated information will be used in order to replace the old shares if the authentication succeeds. Also in this case the calculation of this info is transparent to the user. These data are sent to the dealer that will proceed with the reconstruction phase. As it is shown in section \ref{Scheme_Second}, the client sends:
            
            \begin{itemize}
              \item a message to the Dealer.
              \item two messages to the Service.
            \end{itemize}
            At the end of the login phase a communication channel with the Service is created. It will open a session towards the user. Furthermore the saved information on the browser is the same as the one of the sharing phase but with different values.
            % foto
            \begin{figure}[h]
              \centering
              \includegraphics[width=.8\textwidth]{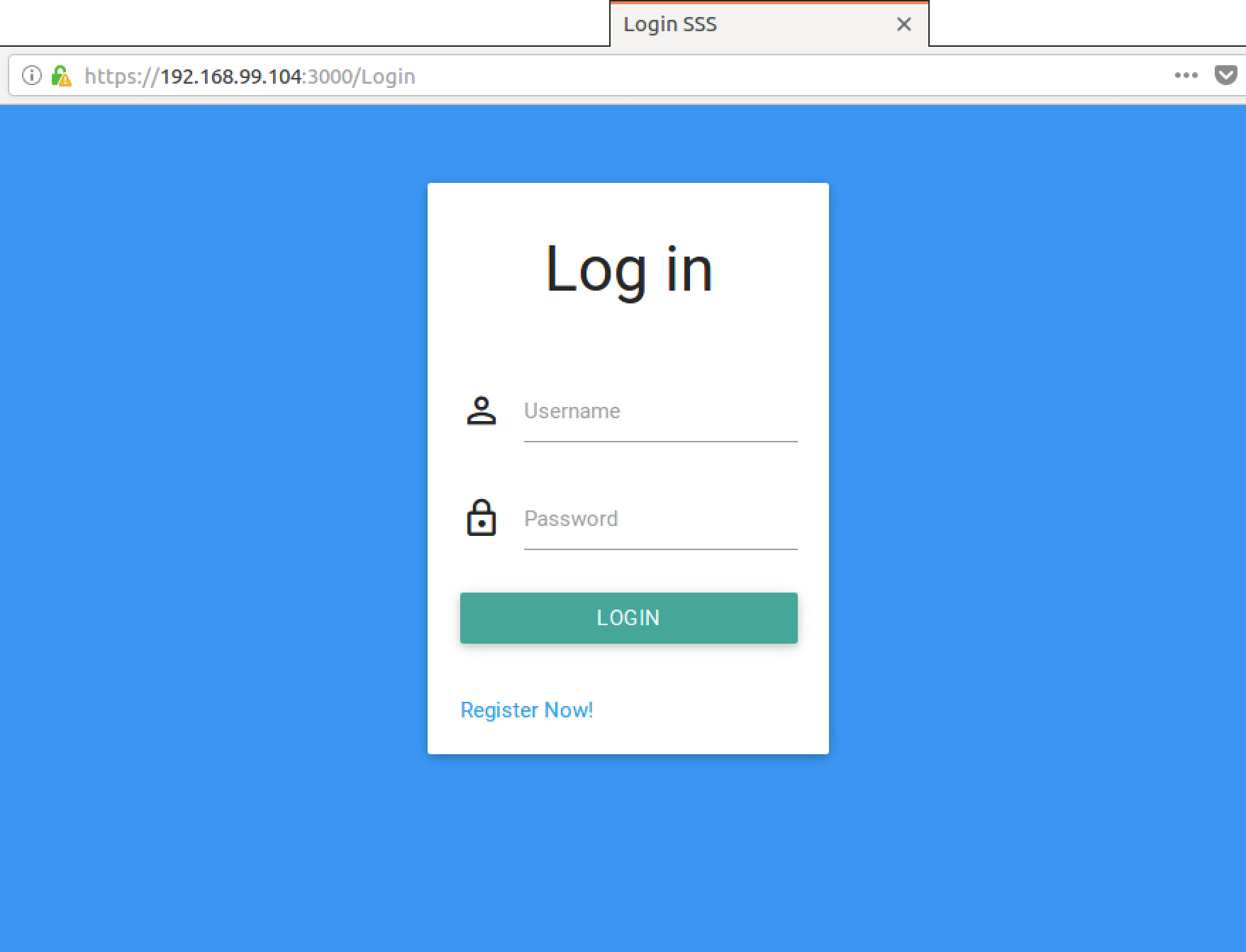}
              \caption{Login interface}
              \label{fig:Login}
            \end{figure}  

    \subsection{The technologies used}
      % spiegazione delle varie tecnologie usate dai due princiapli attori: Client e server sider
      In this section the technologies used in the PoC will be seen. First of all the technologies used in the Client side and then those technologies used on the "server" side, ie the dealer, service and shareholders

        \subsubsection{Client side}\label{Client_side}
          \paragraph{JavaScript and AJAX call}
            % presa da internet
            <<JavaScript is a programming language commonly used in web development... JavaScript is a client-side scripting language, which means the \textbf{source code is processed by the client's web browser} rather than on the web server... JavaScript code can be inserted anywhere within the HTML of a webpage>>\cite{JS_DESC}. Since the application proposed works on the internet, the users communicate with the system through a web page. Hence the programming language used on the client side is JavaScript.
            \\\\
            Since on the client side JavaScript is used, in order to send the messages to the Dealer and Shareholders AJAX call technology is used. Hence, <<the Asynchronous JavaScript And XML (AJAX) is not a programming language. AJAX just uses a combination of:
            \begin{itemize}
              \item A browser built-in XMLHttpRequest object (to request data from a web server)
              \item JavaScript and HTML DOM (to display or use the data)
            \end{itemize}
            AJAX allows web pages to be updated asynchronously by \textbf{exchanging data} with a web server behind the scenes>>\cite{AJAX_DESC}. 

            \paragraph{Advanced Encryption Standard}
            % presa da internet
            Ad it is possible to see in scheme \ref{Scheme_Second} the client must encrypt some messages like:
            \begin{itemize}
              \item $MC =E_{k}\big[g^{S^{\prime}} h^{r^{\prime}} \big]$
                \item $E_{MS}\big[k,\ (g^{S^\prime} h^{r^\prime})\big]$
            \end{itemize}
            In the same way the client must decrypt some messages like: $MS$. 
            Hence cryptographic cipher is needed; the one used in the PoC is AES.\\\\
            <<AES, is a cryptographic cipher that is responsible for a large amount of the information security that you enjoy on a daily basis. Applied by everyone from the NSA to Microsoft to Apple, AES is one of the most important cryptographic algorithms being used in 2018>>\cite{AES_DESC}. In the PoC, keys with 256 bit size are used and the CBC mode. 
            \\\\
            <<The implementation used on the client side is the CryptoJS library in Google Code Archive. CryptoJS is a growing collection of standard and secure cryptographic algorithms implemented in JavaScript using best practices and patterns. They are fast, and they have a consistent and simple interface>> \cite{AES_IMPL}.

          \paragraph{Local Storage}\label{Local_storage}
            % presa da internet
            The Local storage technology is used to store the extra information like:
            \begin{itemize}
              \item $x_1,\dots,x_n$
                \item $r,\ r^\prime$
                \item $MS$
                \item $k$
            \end{itemize}
            in the browser.
            \\\\
            <<HTML5 local storage makes it possible to store values in the browser which can survive the browser session... Local storage is a new specification in HTML5... HTML5 local storage is similar to cookies in that both mechanisms can be used to store data in the browser between HTTP requests. But there is a difference between HTML5 local storage and cookies. Cookies are small pieces of data which a server can store in the browser. The cookie is sent by the browser along with all future HTTP requests to the server that set the cookie. Cookies cannot be bigger than 4KB in total. Instead, HTML5 local storage is set via JavaScript executed in the browser. HTML5 local storage properties are never sent to any server - unless you explicitly copy them out of the local storage and appends them to an AJAX request. HTML5 local storage can store somewhere between 2MB and 10MB data in the browser (per origin - domain name). Exactly how much data is allowed depends on the browser. A limit of 5MB to 10MB is most common. The properties set in the HTML5 local storage can only be read by pages from the same domain as the page that set the properties.>> \cite{LocalStorage_DESC}

          \paragraph{Pseudorandom generator cryptography quality}
            % non so se mettercelo
            % vedi https://developer.mozilla.org/en-US/docs/Web/API/RandomSource/getRandomValues
            Since the client must choose some random numbers, for security reasons it is not possible to use the classical Math.random() function. A pseudo random generator cryptography quality is used. 
            \\\\
            <<Pseudorandom generators (PRG) are used to create random sequences of numbers in deterministic devices. All computer algorithms are strictly deterministic. PRGs allow encryption of many data blocks using data generated from secret keys which have only few bits. Pseudorandom generator has to be \textbf{unpredictable}. There must not be any efficient algorithm that after receiving the previous output bits from PRG would be able to predict the next output bit with probability non-negligibly higher than 0.5>> \cite{PNRG_DESC}.
            \\\\
            The implementation used on the client side is the Crypto.getRandomValues() of MDN \cite{PNRG_IMPLE}.

        \subsubsection{Server side}
          \paragraph{Multi thread server}
            % presa da internet
            The system can handle and communicate with more than one Client at the same time because a multithreaded Server is implemented. Dealer, Shareholders and Service are written in Python using a multithreaded logic.

          \paragraph{Flask}
            % presa da internet
            % mettere foto se ti va di mettere los chema.
            Flask is a small and powerful web framework for Python. It's easy to learn and simple to use, enabling to build a web app in a short amount of time \cite{Flask_DESC}. Applications that use the Flask framework include Pinterest, LinkedIn, and the community web page for Flask itself. \cite{Flask_DESC2}
            \\
            The Flask technology is used to make the web app, that is what the client can see on the browser. Flask is used to link the system, namely the Dealer logic, Service and Shareholders, to the web-app.

          \paragraph{Session}\label{Session}
            % presa da internet
            A session technology is a way to remember who is talking with the server. Since the HTTP protocol is stateless, to remember who is the user that logs in, a session technology is used.
            \\\\
            A session implementation is used on the FLask. <<a session with each client is assigned a Session ID. The Session data is stored on top of cookies and the server signs them cryptographically. For this encryption, a Flask application needs a defined SECRET\_KEY.
            Session object is also a dictionary object containing key-value pairs of session variables and associated values>>\cite{SESSION_DESC}.

          \paragraph{Advanced Encryption Standard}
            % presa da internet
            AES is also used on the server side. It is used for encryption and decryption messages by the Service. Also in this case the size of the keys are 256 bit and the CBD mode is used.
            \\\\
            The implementation used on the server side is the cryptography library. <<Cryptography includes both high level recipes and low level interfaces to common cryptographic algorithms such as symmetric ciphers, message digests, and key derivation functions>>\cite{AES_IMPL_SERVER}.

          \paragraph{Pseudorandom generator cryptography quality}
            % non so se mettercelo
            % vedi https://developer.mozilla.org/en-US/docs/Web/API/RandomSource/getRandomValues
            Pseudo random generator cryptography quality is also used on the server side in order to generate the $k^\prime$ by the Service and other parameters on the Dealer side. 
            \\\\
            The implementation used on the server side is the secret module. <<The secrets module is used for generating cryptographically strong random numbers suitable for managing data such as passwords, account authentication, security tokens, and related secrets. In particularly, secrets should be used in preference to the default pseudo-random number generator in the random module, which is designed for modelling and simulation, not security or cryptography>> \cite{PNRG_IMPLE_SERVER}.

      \newpage
      \subsection{The system infrastructure}\label{sec:system_infrastructure}
        % presentazione dell’infrastruttura. Infrastruttura network. Container vs Virtualisation. Broadcast channel.
        In the previous section the web-interface and how the user interacts with the system have been introduced. In this section the system infrastructure will be presented. The system infrastructure is the back-end of the application. The back-end is built as a cloud computing, there are $n$ machines that work together to reach a common result. The implementation choices, the network cloud computing infrastructure, the technologies used and the final result will be treated in the following reading. Besides, finally, a system performance qualitative analysis and the tests' results will be seen.

        \subsubsection{How to implement a cloud: the Docker solution}
          % foto, definzione, vantaggi/svantaggi, Docker 
          The first problem to resolve is how the cloud can be implemented. It is necessary to know the difference between container and virtualization technology before understanding the various solutions that can be adopted in order to build a cloud. Afterwards the Docker framework with its components will be introduced. Finally, the two possible solutions to build a cloud are listed and it will be explained which one will be used for the PoC implementation.

          \paragraph{Containers against virtualisations}
            Containers and virtual machines reach the same properties but in a different way. The difference is in the implementation: containers virtualize the operating system instead the virtual machines virtualize the hardware. In the following lines a brief description and comparison between the two technologies is introduced. The descriptions are taken from the Docker and Google web sites:
            \begin{itemize}
              \item \textbf{Containers.} <<Containers are an abstraction at the app layer that packages code and dependencies together. Multiple containers can run on the same machine and share the OS kernel with other containers, each running as isolated processes in user space. Containers take up less space than VMs (container images are typically tens of MBs in size), and start almost instantly. A container image is a lightweight, stand-alone, executable package of a piece of software that includes everything needed to run it: code, runtime, system tools, system libraries, settings. Available for both Linux and Windows based apps, containerized software will always run the same, regardless of the environment. Containers isolate software from its surroundings, for example differences between development and staging environments and help reduce conflicts between teams running different software on the same infrastructure>>\cite{Docker_def}.
              \\\\
              <<Containers offer a logical packaging mechanism in which applications can be abstracted from the environment in which they actually run. This decoupling allows container-based applications to be deployed easily and consistently, regardless of whether the target environment is a private data center, the public cloud, or even a developer’s personal laptop. Containerization provides a clean separation of concerns, as developers focus on their application logic and dependencies, while IT operations teams can focus on deployment and management without bothering with application details such as specific software versions and configurations specific to the app>> \cite{Google_Container}.

              \item \textbf{Virtual machines.} 
              <<Virtual machines (VMs) are an abstraction of physical hardware turning one server into many servers. The hypervisor allows multiple VMs to run on a single machine. Each VM includes a full copy of an operating system, one or more apps, necessary binaries and libraries - taking up tens of GBs>>\cite{Docker_def}.
              \\\\
              <<For those coming from virtualized environments, containers are often compared with virtual machines (VMs). Virtual Machine is a guest operating system such as Linux or Windows runs on top of a host operating system with virtualized access to the underlying hardware. Like virtual machines, containers allow you to package your application together with libraries and other dependencies, providing isolated environments for running your software services. As you’ll see below however, the similarities end here as containers offer a far more lightweight unit for developers and IT Ops teams to work with, carrying a myriad of benefits>> \cite{Google_Container}.
            \end{itemize}
            \begin{figure}[!htb]
              \centering
              \begin{minipage}{1\textwidth}
                  \centering
                  \subfloat[Containers]{
                      \includegraphics[width=0.45\textwidth]{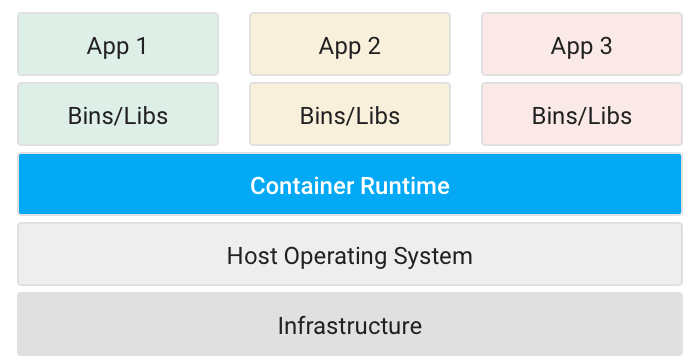}
                      \label{fig:Containers}
                  }
                  \subfloat[Virtualization]{
                      \includegraphics[width=0.45\textwidth]{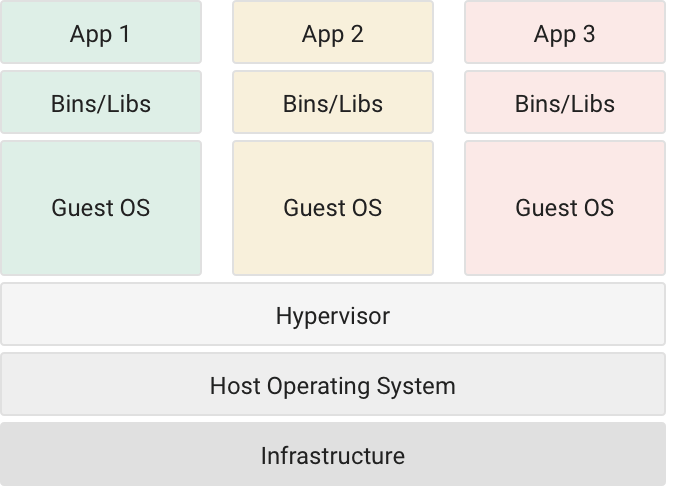}
                      \label{fig:Virtualization}
                  }
                  \caption{Containers against virtualisations \cite{Google_Container}}
                  \label{fig:Containers_VS_VM}
              \end{minipage}
            \end{figure} 
            Therefore the main container's properties are the following:
            \begin{itemize}
              \item <<\textbf{Consistent Environment.} Container technology allows to create isolated custom environments. The customize takes place at the level of operating system, software dependencies, versions of programming language runtimes and other software libraries that the service needs. In this way no matter where the service is deployed. This lead to spend less time during debugging and diagnosing phase when the service is moved from different environments, i.e. moving application from development to deployment environment.

              \item \textbf{Run Anywhere.} Containers can run virtually anywhere. In this way from the development to deployment phase it is not required very effort. They can run are on: Linux, Windows, and Mac operating systems and on each virtual machines.

              \item \textbf{Isolation.} Containers virtualize the hardware resources at the OS-level and each application is logically isolated from the other ones.>>\cite{Google_Container}. 
            \end{itemize}
            In a few words the containers are:
            \begin{itemize}
              \item \textbf{portable} and \textbf{efficient}.
              \item \textbf{cost effective},  \textbf{easy to use} and  \textbf{highly scalable}.
              \item \textbf{fast}, and they use \textbf{less memory} compared to virtualisation.
            \end{itemize}
            % 3- Foto finale presa da google Container-VM   
            \begin{figure}[h]
              \centering
              \includegraphics[width=0.9\textwidth]{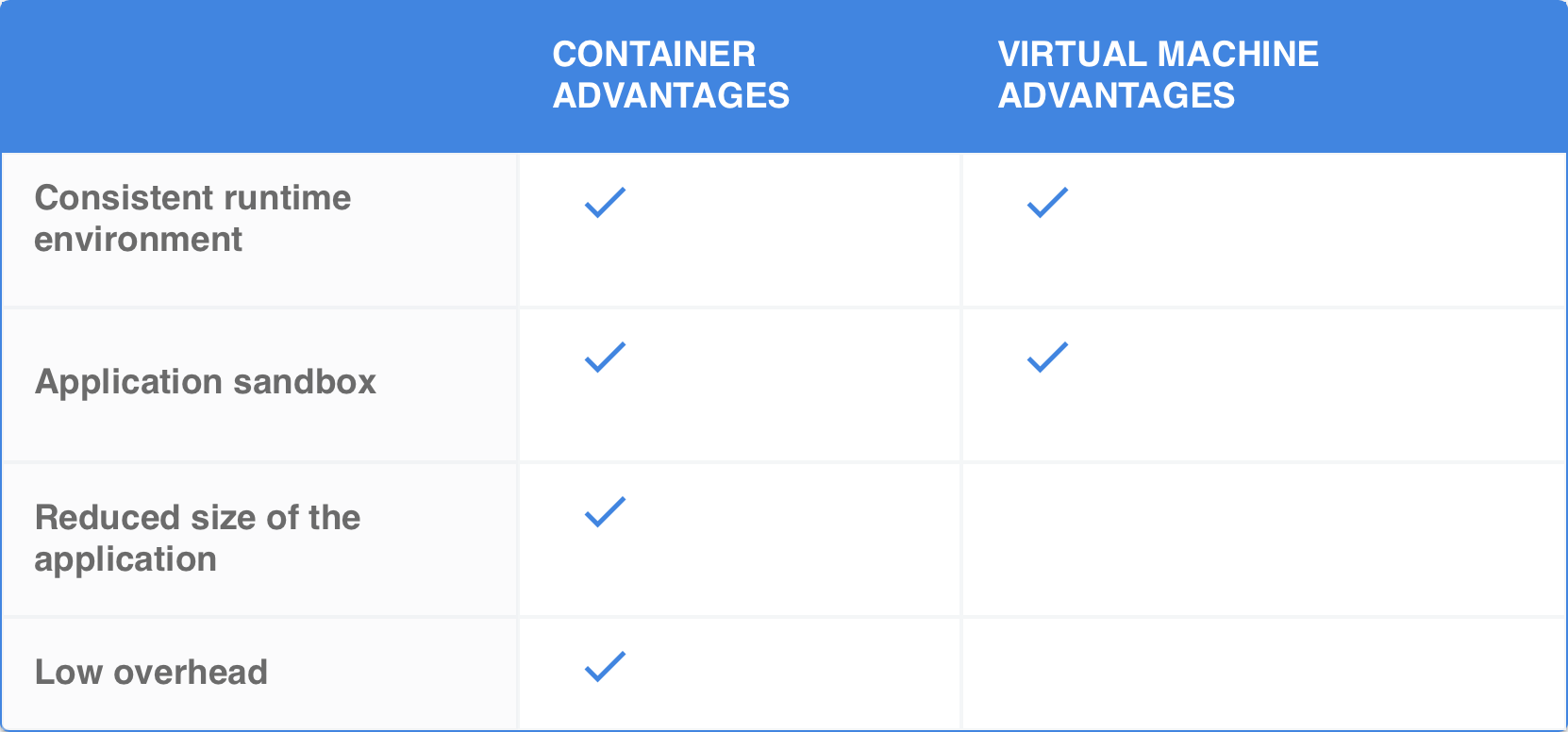}
              \caption{Container vs virtualisation \cite{Google_Container}}
              \label{fig:Container_VM}
            \end{figure}
            In a real business world, the biggest IT company claims about containers: \textit{<<From Gmail to YouTube to Search, everything at Google runs in \textbf{containers}. Containerization allows our development teams to move fast, deploy software efficiently, and operate at an unprecedented scale. Each week, we start over two billion containers. We have learned a lot about running containerized workloads in production over the past decade, and we have shared this knowledge with the community along the way: from the early days of contributing cgroups to the Linux kernel, to taking designs from our internal tools and open sourcing them as the Kubernetes project. We have packaged this expertise into Google Cloud Platform so that developers and businesses of any size can easily tap the latest in container innovation.>> - Google, Inc} \cite{Google_Container}
            \\\\
            This is a big proof of about the future usage of this technology. This directs the design choices of the system to this tool. In the following reading a Docker framework will be seen.

          \paragraph{Docker}
            % 4- Presentazione di Docker CE
            %I parte: Che cosa è Docker?. Perchè è diventato lo standar defacto dei container. Le tre caratteristiche. Crescita nel tempo di docker.
            "Docker is the company driving the container movement and the only container platform provider to address every application across the hybrid cloud. Today’s businesses are under pressure to digitally transform but are constrained by existing applications and infrastructure while rationalizing an increasingly diverse portfolio of clouds, datacenters and application architectures. Docker enables true independence between applications and infrastructure and developers and IT ops to unlock their potential and creates a model for better collaboration and innovation."\cite{Docker_def}.
            \\\\
            There are many container formats available. Docker is the most popular, open-source container format because it is robust platform, code registry, and it has an integrated management tools. The most important features are:
            \begin{itemize}
              \item <<\textbf{LIGHTWEIGHT.} Docker containers running on a single machine share that machine's operating system kernel; they start instantly and use less compute and RAM. Images are constructed from filesystem layers and share common files. This minimizes disk usage and image downloads are much faster.

              \item \textbf{STANDARD.} Docker containers are based on open standards and run on all major Linux distributions, Microsoft Windows, and on any infrastructure including VMs, bare-metal and in the cloud.

              \item \textbf{SECURE.}  Docker containers isolate applications from one another and from the underlying infrastructure. Docker provides the strongest default isolation to limit app issues to a single container instead of the entire machine>>\cite{Docker_def}.
          
            \end{itemize}
            The growth of Docker has been exponential, it was launched in 2013 and it brought software containers to the masses.
            By June 2015 Docker donated the specification and runtime code now known as runC, to the Open Container Initiative (OCI) to help establish \textit{standardization}.
            \begin{figure}[h]
              \centering
              \includegraphics[width=0.9\textwidth]{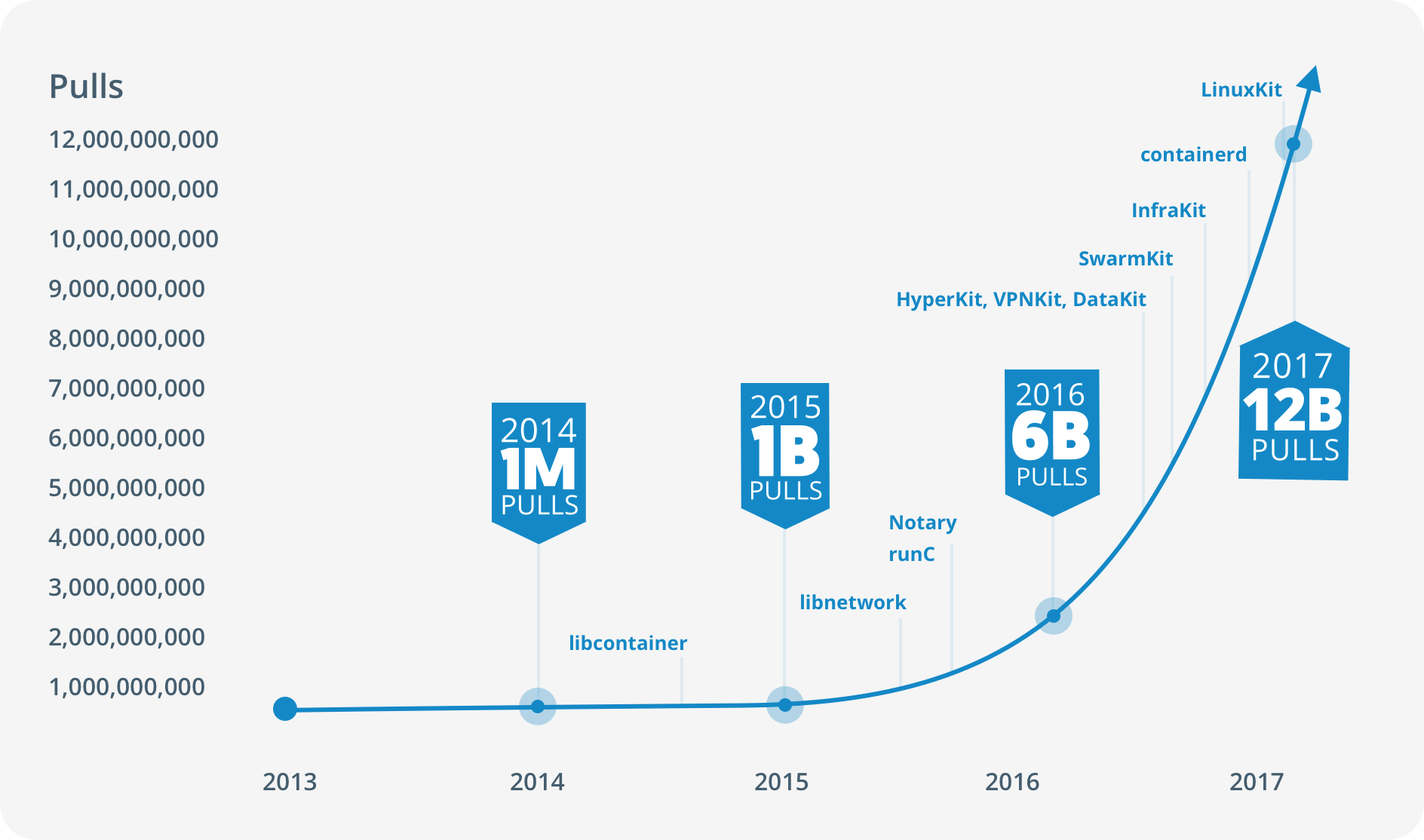}
              \caption{The growth of Docker \cite{Docker_def}}
              \label{fig:Docker_improvement}
            \end{figure}

          %%%%%%%%%%%%%%%%%%%%%%%%%%%%%%%%%%%%%%%%%%%%%%%%%%%%%%%%%%%%%%%%%%%%%%%%%%%%%%%%%%%%%%
          % II parte: i tool di docker (Machine, File, Compose ecc) con le caratteristiche   %
          %   Spiegare i vari pezzi di Docker: Docker Machine, Compose, Swarm, Stack       %
          %%%%%%%%%%%%%%%%%%%%%%%%%%%%%%%%%%%%%%%%%%%%%%%%%%%%%%%%%%%%%%%%%%%%%%%%%%%%%%%%%%%%%%
          \newpage
          The main tools and concepts of Docker world are the following, the test was taken from the official docker documentation (see the official docs for a deeply reading \cite{Docker_Docs}):
          \begin{itemize}
            \item \textbf{Dockerfile.} <<A Dockerfile is a text document that contains all the commands a user could call on the command line to assemble an image. Using docker build users can create an automated build that executes several command-line instructions in succession. The main concept is the following: the developer write the \textbf{Dockerfile}, \textbf{built} it to make the image and in the end the image can \textbf{run} in a containers. Therefore, the main benefit is that with one built image it is possible to run many containers>>\cite{Docker_Docs}.

            \begin{figure}[h]
                \centering
                \includegraphics[width=0.9\textwidth]{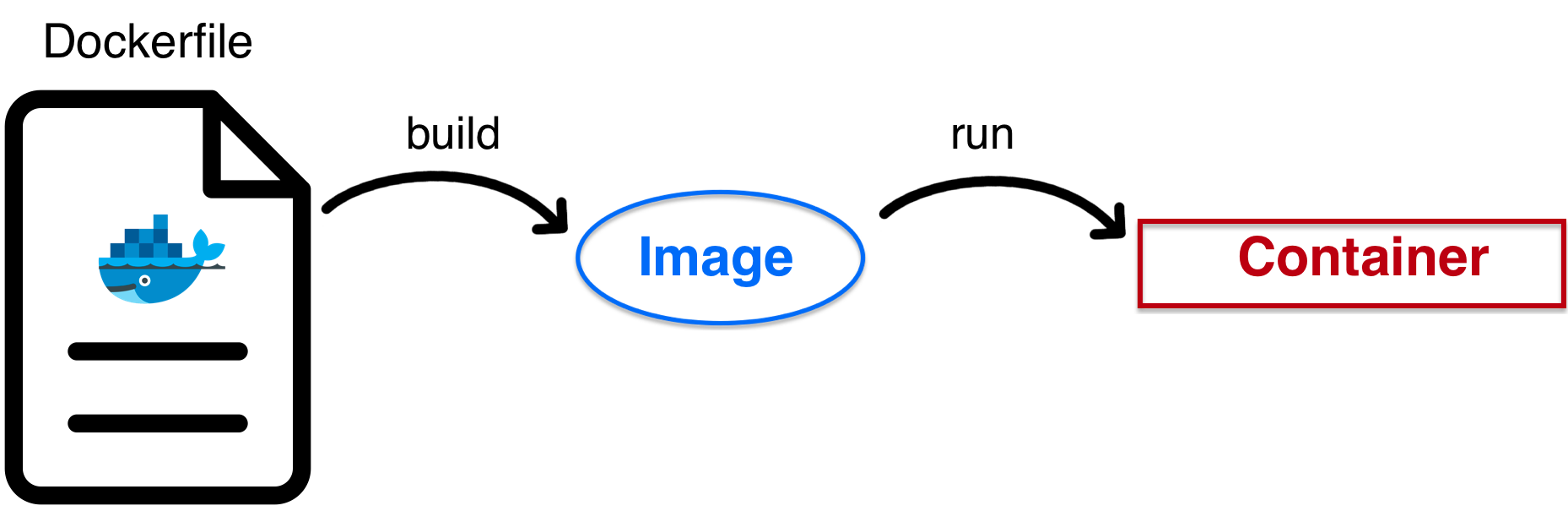}
                \caption{How to use Dockerfile}
                \label{fig:DockerFile}
            \end{figure}

            \item \textbf{Docker Compose.} <<Compose is a tool for defining and running \textbf{multi-container} Docker applications. With Compose, YAML file is used to configure the application’s services. Then, with a single command, it is possible to create and start all the services from the configuration. Compose works in all environments: production, staging, development, testing, as well as CI workflows>>\cite{Docker_Docs}.

            \item \textbf{Docker Machine.} <<Docker Machine is a tool that lets you install Docker Engine on \textbf{virtual hosts}. Docker Machine is used to create Docker hosts on local Mac or Windows box, on company network, in datacenter, or on cloud providers like Azure, AWS, or Digital Ocean>> \cite{Docker_Docs}.
            \begin{figure}[h]
                \centering
                \includegraphics[width=0.9\textwidth]{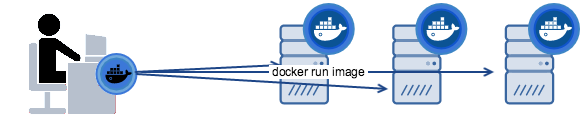}
                \caption{Docker Machine \cite{Docker_Docs}}
                \label{fig:DockerMachine}
              \end{figure}

            \item \textbf{Docker Service.} <<In a distributed application, different pieces of the app are called “services.” Services are really just “containers in production.” A service only runs one image, but it codifies the way that image runs—what ports it should use, how many replicas of the container should run so the service has the capacity it needs, and so on. Scaling a service changes the number of container instances running that piece of software, assigning more computing resources to the service in the process>> \cite{Docker_Docs}.

            \textsuperscript{testo in apice}
          
            \item \textbf{Docker Swarm.}<<A swarm is a \textbf{group of machines} that are running Docker and joined into a cluster. The machines in a swarm can be physical or virtual. After joining a swarm, they are referred to as nodes. Docker Swarm implements the cluster management and orchestration features embedded in the Docker Engine. \textit{Swarm managers} are the only machines in a swarm that can execute the user's commands, or authorize other machines to join the swarm as \textit{workers}. Workers are just there to provide capacity and do not have the authority to tell any other machine what it can and cannot do.
            \\\\
            % 7- Containers and Virtual Machines Together
            Containers and VMs used together provide a great deal of flexibility in deploying and managing apps>>\cite{Docker_Docs}.

            \begin{figure}[h]
              \centering
              \includegraphics[width=0.9\textwidth]{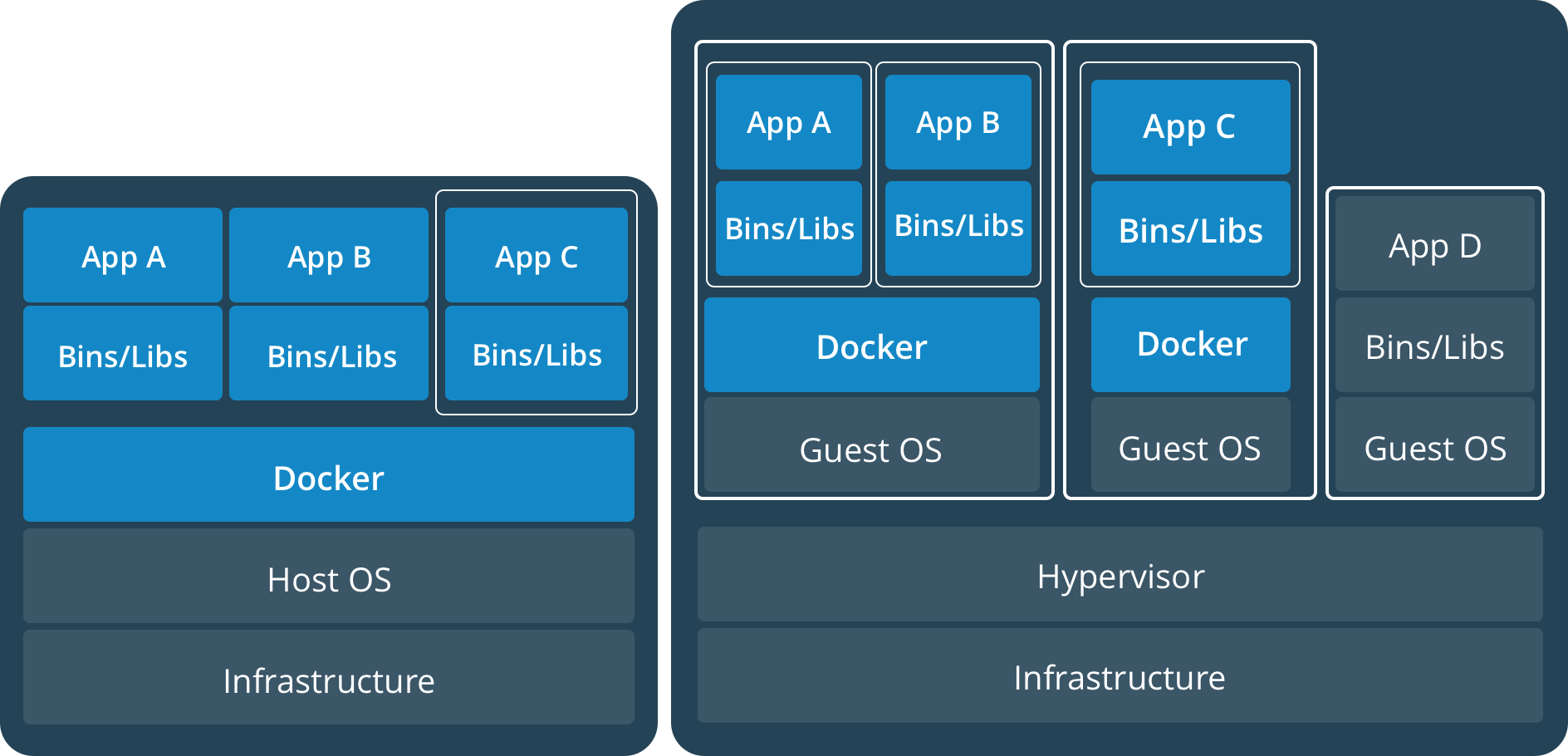}
              \caption{Docker and virtualisation \cite{Docker_def}}
              \label{fig:VM_vs_Docker}
            \end{figure}

            Feature highlights:
            \begin{itemize}
              \item<<\textbf{Cluster management integrated with Docker Engine:} Use the Docker Engine CLI to create a swarm of Docker Engines where you can deploy application services. You don’t need additional orchestration software to create or manage a swarm.

              \item \textbf{Decentralized design:} Instead of handling differentiation between node roles at deployment time, the Docker Engine handles any specialization at runtime. You can deploy both kinds of nodes, managers and workers, using the Docker Engine. This means you can build an entire swarm from a single disk image.

              \item \textbf{Declarative service model:} Docker Engine uses a declarative approach to let you define the desired state of the various services in your application stack. For example, you might describe an application comprised of a web front end service with message queueing services and a database backend.

              \item \textbf{Scaling:} For each service, you can declare the number of tasks you want to run. When you scale up or down, the swarm manager automatically adapts by adding or removing tasks to maintain the desired state.

              \item \textbf{Desired state reconciliation:} The swarm manager node constantly monitors the cluster state and reconciles any differences between the actual state and your expressed desired state. For example, if you set up a service to run 10 replicas of a container, and a worker machine hosting two of those replicas crashes, the manager creates two new replicas to replace the replicas that crashed. The swarm manager assigns the new replicas to workers that are running and available.

              \item \textbf{Multi-host networking:} You can specify an overlay network for your services. The swarm manager automatically assigns addresses to the containers on the overlay network when it initializes or updates the application.

              \item \textbf{Service discovery:} Swarm manager nodes assign each service in the swarm a unique DNS name and load balances running containers. You can query every container running in the swarm through a DNS server embedded in the swarm.

              \item \textbf{Load balancing:} You can expose the ports for services to an external load balancer. Internally, the swarm lets you specify how to distribute service containers between nodes.

              \item \textbf{Secure by default:} Each node in the swarm enforces TLS mutual authentication and encryption to secure communications between itself and all other nodes. You have the option to use self-signed root certificates or certificates from a custom root CA.

              \item \textbf{Rolling updates:} At rollout time you can apply service updates to nodes incrementally. The swarm manager lets you control the delay between service deployment to different sets of nodes. If anything goes wrong, you can roll-back a task to a previous version of the service>>\cite{Docker_Docs}.

            \end{itemize}
            
            \begin{figure}[h]
              \centering
              \includegraphics[width=0.7\textwidth]{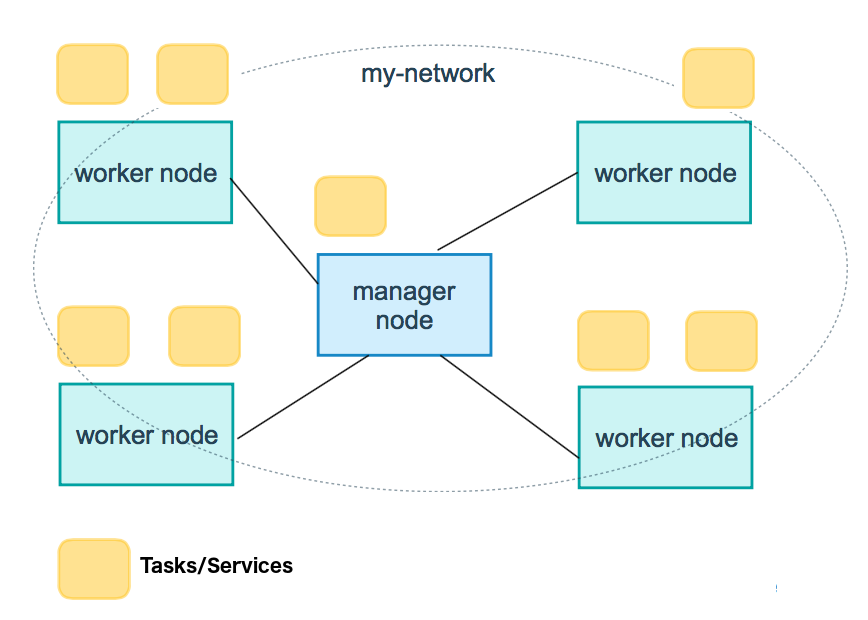}
              \caption{Docker Swarm logically view \cite{Docker_Docs}}
              \label{fig:DockerSwarm}
            \end{figure}

            \item \textbf{Docker Stack.} <<In the previously point, how to set up a swarm was learned, which is a cluster of machines running Docker, and deployed an application to it, with containers running in concert on multiple machines.
            \\\\
            With the Docker Stack the top of the hierarchy of distributed applications is reached. A stack is a group of interrelated services that share dependencies, and can be orchestrated and scaled together. A single stack is capable of defining and coordinating the functionality of an entire application (though very complex applications may want to use multiple stacks).
            \\\
            Docker Compose is used for single service stack running on a single host, whit Docker stack it is possible to make multiple services relate to each other, and run them on multiple machines. 
            \\\\
            Indeed, a stack is a collection of services that make up an application in a specific environment. A stack file is a file in YAML format, similar to a docker-compose.yml file, that defines one or more services. Stacks are a convenient way to automatically deploy multiple services that are linked to each other, without needing to define each one separately. Stack files define environment variables, deployment tags, the number of containers, and related environment-specific configuration. Because of this, you should use a separate stack file for development, staging, production, and other environments>>\cite{Docker_Docs}.

             % 6- Caratteristiche di docker: service found, tls, load balacence, che dato il docker file poi si monta facile!

          \end{itemize}

          \paragraph{How to implement a cloud}
            Basically there are two solutions:
            \begin{itemize}
              \item \textbf{OpenStack.} \guillemotleft OpenStack is a cloud operating system that controls large pools of compute, storage, and networking resources throughout a datacenter, all managed through a dashboard that gives administrators control while empowering their users to provision resources through a web interface>> \cite{Openstack_def}.

              \item \textbf{Docker and its tools}.
            \end{itemize}
            In order to implement the PoC, Docker and its tools have been chosen because it is an innovative and powerful technology.

          \subsubsection{The infrastructure realization}
            % come è in generale: abbiamo n serve fra cui delaer, shareholer, logger e service. Un cline che gira su un browser. I server giradno con python server multih. Per realizzare la web app si usa Flask. La logiga poi è implementata in python
            The PoC system infrastructure is made with many software layers because all run on one physical machine. To realize the PoC a virtualization technology is used. Above the physical machine there is a virtualize machine with Ubuntu operating system. In the end, thanks to Docker Machine it is possible to create three nodes in order to build a swarm that acts like a cloud. In each node some containers run.
            \\\\
            The following list presents layers the different layers with their main features. Only the first layer is the physical one, the others are virtualised:
            \begin{itemize}
              \item MacBook Pro 2017:
                \begin{itemize}
                  \item CPU: 2.9GHz Dual-core Intel Core i5
                  \item Memory: 16GB 2133MHz LPDDR3 SDRAM
                  \item Operating System: macOs Sierra
                \end{itemize}

              \item Ubuntu:
                \begin{itemize}
                  \item CPU: 2.9GHz Dual-core Intel Core i5
                  \item Memory: 12GB 2133MHz LPDDR3 SDRAM
                  \item Operating System: Ubuntu Desktop 16.04.3
                \end{itemize}

              \item Swarm-Manager, Swarm-Worker-0, Swarm-Worker-1:
                \begin{itemize}
                  \item CPU: 1 (of 2.9GHz Dual-core Intel Core i5)
                  \item Memory: 1024Mb 2133MHz LPDDR3 SDRAM
                  \item Operating System: boot2docker \footnote{<<Boot2Docker is a lightweight Linux distribution made specifically to run Docker containers. It runs completely from RAM, is a 45MB download and boots quickly>> \cite{boot2docker_def}.}
                \end{itemize}

              \item A container:
                \begin{itemize}
                  \item CPU: 1 (2.9GHz Dual-core Intel Core i5)
                  \item Memory: 995Mb 2133MHz LPDDR3 SDRAM
                  \item Operating System: Ubuntu Server 17.10.1 
                \end{itemize}
            \end{itemize}
            \begin{figure}[h]
              \centering
              \includegraphics[width=0.95\textwidth]{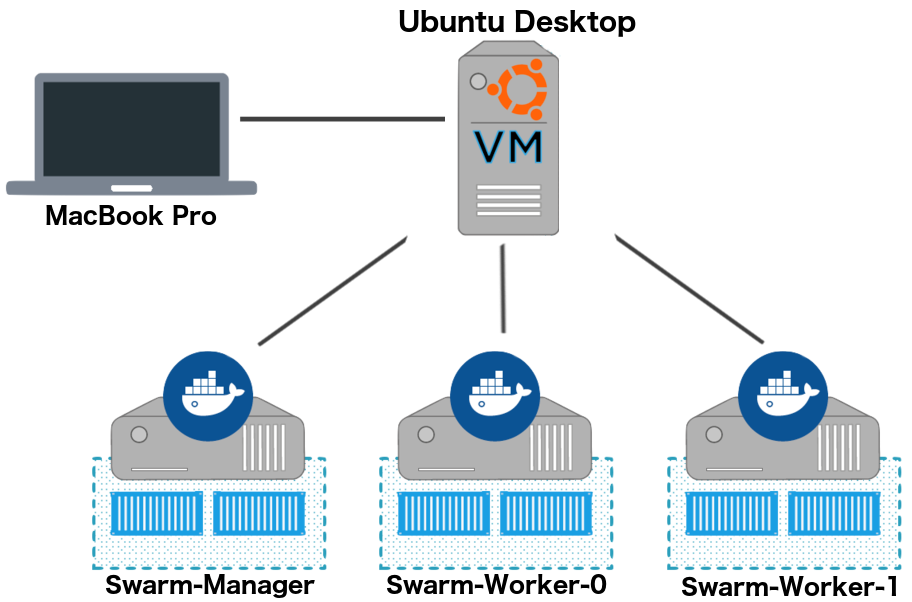}
              \caption{System Infrastructure}
              \label{fig:SystemInfrastucture}
            \end{figure}

            \newpage
            Logically there are the following containers:
            \begin{itemize}
              \item 1 Dealer
              \item 1 Service
              \item 3/5/7/10 Shareholders
              \item 1 Logger
            \end{itemize}
            In figure \ref{fig:Containers} it is possible to see where each container runs. The shareholders are called "Jarvis-i" and the Dealer is called "Master-Jarvis". In the next section how these components communicate each other is explained. 

            \begin{figure}[h]
              \centering
              \includegraphics[width=0.9\textwidth]{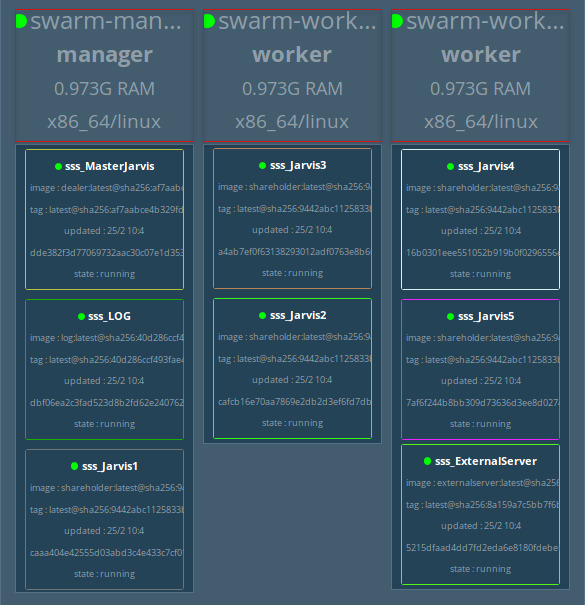}
              \caption{Containers}
              \label{fig:Containers}
            \end{figure}

            \newpage
            \subsubsection{Network infrastructure}
              % spiegazine della net infra con logger!! (Msg & error)
              In this section the network infrastructure is treated. Since there are two levels: \textit{nodes} and \textit{containers} there are two points of view for the network: 

              \begin{itemize}
                \item \textbf{Nodes. } There are three machines that are executed with VirtualBox \footnote{<<Oracle VM VirtualBox (formerly Sun VirtualBox, Sun xVM VirtualBox and Innotek VirtualBox) is a free and open-source hypervisor for x86 computers currently being developed by Oracle Corporation>> \cite{Virtuabox}.}. VirtualBox has different network adapters, the one used to connect the three machines is the Host Only Networking \footnote{<<Host-only networking is a networking mode. It can be thought of as a hybrid between the bridged and internal networking modes: as with bridged networking, the virtual machines can talk to each other and the host as if they were connected through a physical Ethernet switch. Host-only networking is particularly useful for preconfigured virtual appliances, where multiple virtual machines are shipped together and designed to cooperate>> \cite{network_hostonly}.}. The network topology is the following:
                \begin{figure}[h]
                  \centering
                  \includegraphics[width=0.7\textwidth]{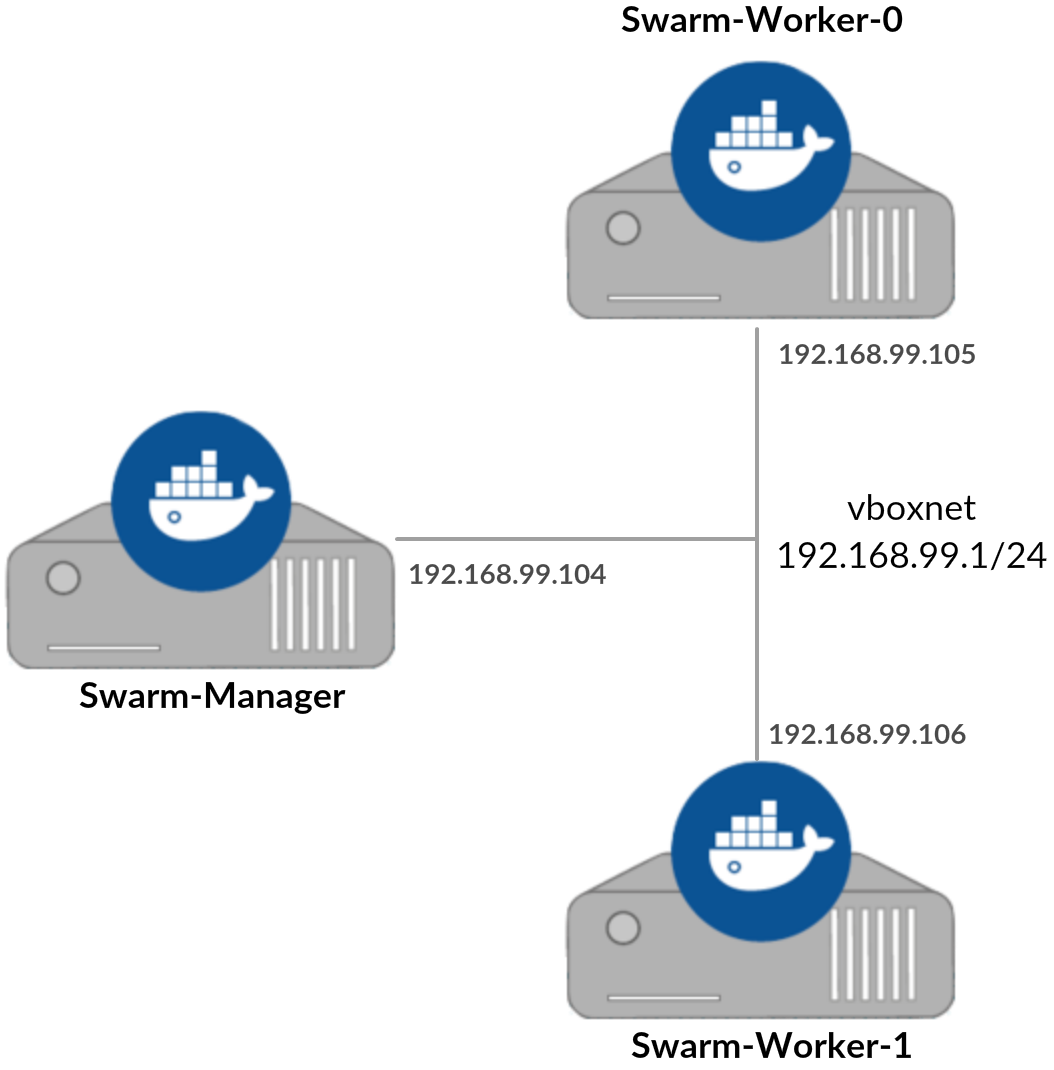}
                  \caption{Network infrastructure: node view}
                  \label{fig:Node_Network_Infra}
                \end{figure}

                \item \textbf{Containers. } Since containers run on the three nodes, they are connected with an \textbf{overlay network} \footnote{<<An overlay network is a computer network that is built on top of another network>>\cite{OverlayNetwork}.}. In order to implement the overlay network a Virtual Extensible LAN (VXLAN) technology is used. <<Virtual Extensible LAN (VXLAN) is a proposed encapsulation protocol for running an overlay network on existing Layer 3 infrastructure. An overlay network is a virtual network that is built on top of existing network Layer 2 and Layer 3 technologies to support elastic compute architectures. VXLAN will make it easier for network engineers to scale out a cloud computing environment while logically isolating cloud apps and tenants>>\cite{VXLAN_desc}. This technology encapsulates the OSI layer 2 inside the OSI layer 4 with UDP datagram. VXLAN is a standard for the overlay networks \cite{VXLAN_IETF}. With VXLAN it is possible to send UDP packets whit inside there are encapsulated packets of level 2. The format of a packer is the following:
                \begin{figure}[h]
                  \centering
                  \includegraphics[width=0.9\textwidth]{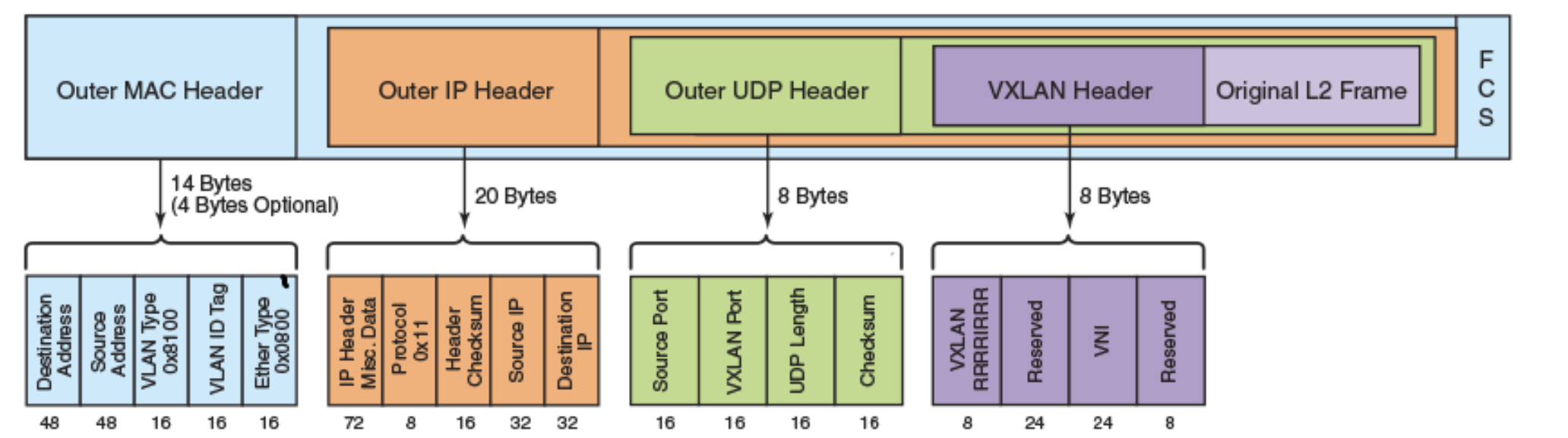}
                  \caption{VXLAN Segments and Packet Format}
                  \label{fig:VXLAN Segments and Packet Format}
                \end{figure}
                \\\\
                In the following image there is an example of how the containers can communicate with each other through VXLAN technology:
                \begin{figure}[h]
                  \centering
                  \includegraphics[width=0.9\textwidth]{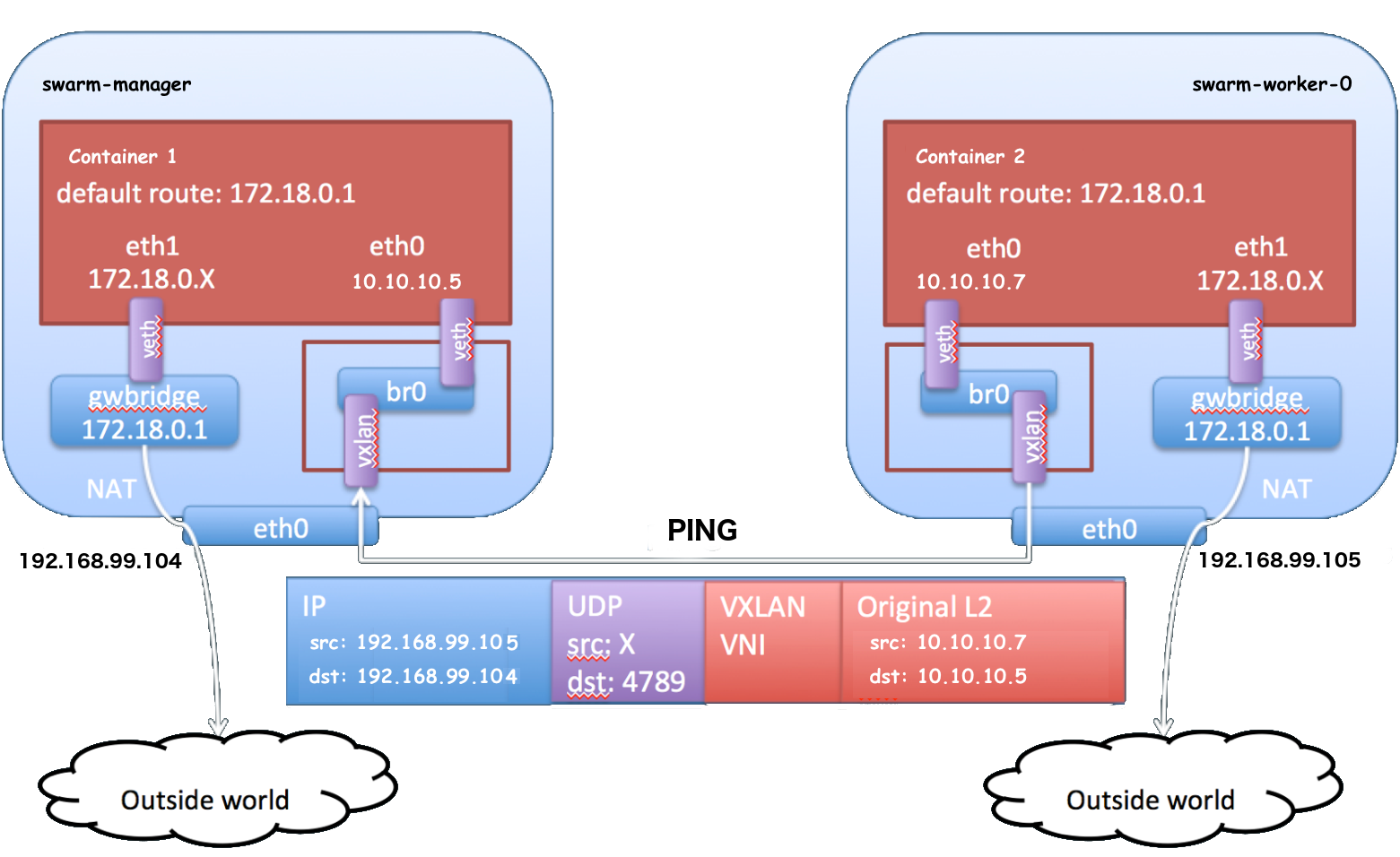}
                  \caption{VXLAN overlay architecture}
                  \label{fig:vlax_overlay_architecture}
                \end{figure}
                \\
                Finally, the network infrastructure of the containers are represented in \ref{fig:Network infrastructure: containers view}. There are two overlay networks:

                \begin{itemize}
                  \item \textbf{\texttt{Backend network.}} It is the network where the Shareholders and Dealer communicate in an isolated channel. It acts like a LAN network so the broadcast channel is implemented. Since the default overlay network driver offered by Docker does not provide the broadcast primitive, an external plug-in, implemented by WeaveNet, is used \cite{WeaveNet_Docker} to provides the broadcast primitive.

                  \item \textbf{\texttt{Dealer\char`_Service}} It is the network where Dealer and Service can communicate.
                \end{itemize}
                The cloud is accessible by the Client from the Dealer or Service. Thanks to Docker Swarm and Docker Stack it is possible to \textit{expose} some ports on the outside of the cloud. In this way, since the three nodes are in the 192.168.99.0 network, it is possible to load the Login, Sign up and then the personal page in the Service through the two different ports. This can be reached thanks to load balancing features of Docker Swarm. It allows to reach the port through the three different IPs (192.168.99.104, 192.168.99.105 or 192.168.99.106). Therefore the Login page can be loaded through the three different links:
                \begin{itemize}
                  \item https://192.168.99.\textbf{104}:3000/Login 
                  \item https://192.168.99.\textbf{105}:3000/Login 
                  \item https://192.168.99.\textbf{106}:3000/Login 
                \end{itemize}
                Finally, the container network infrastructure is the following:
                \begin{figure}[h]
                  \centering
                  \includegraphics[width=1.\textwidth]{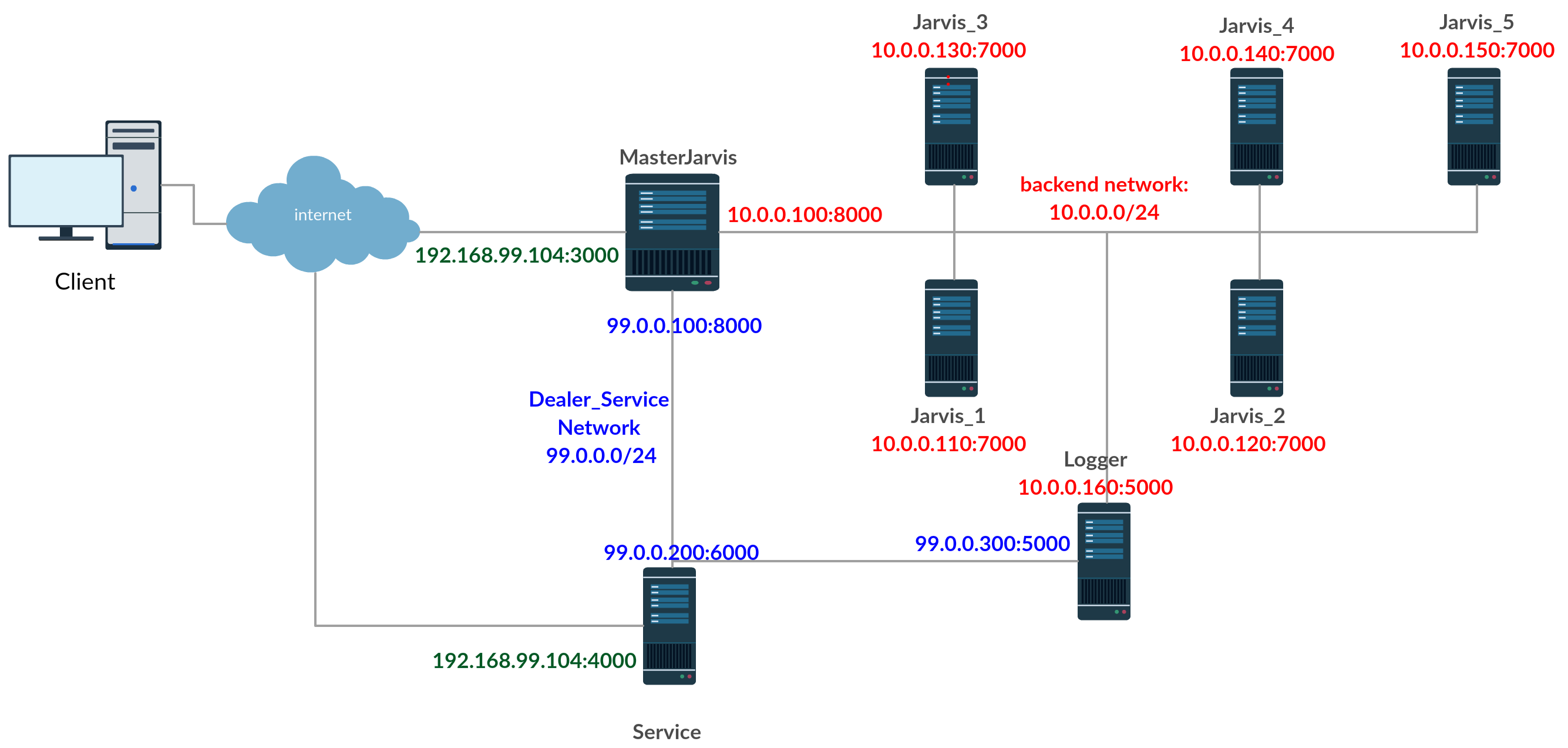}
                  \caption{Network infrastructure: containers view}
                  \label{fig:Network infrastructure: containers view}
                \end{figure}  
              \end{itemize}

          \subsection{A proof of working}
            % griglia dei vari messaggi, vedi file che avevi fatto
            In this section a proof of working is shown. In order to reach the purpose a Logger is put in the network. The Logger receives the messages from the other parts and stores them. The Logger also works  for the message error. If something goes wrong, each actor sends an alert to the Logger, in this way the administrator of the system will be able to check if something wrong has happened through the report made by the Logger. This concept will be deepened in chapter \ref{chap:Security analysis}.
            \\\\
            The result of the logger is reported in the following images. The messages exchanged between each player are the same as the once proposed in scheme \ref{Scheme_Second}. Here are the two phases: Sharing and Reconstruction.
            \\\\
            \begin{figure}[!htb]
              \centering
              \includegraphics[width=1.\textwidth]{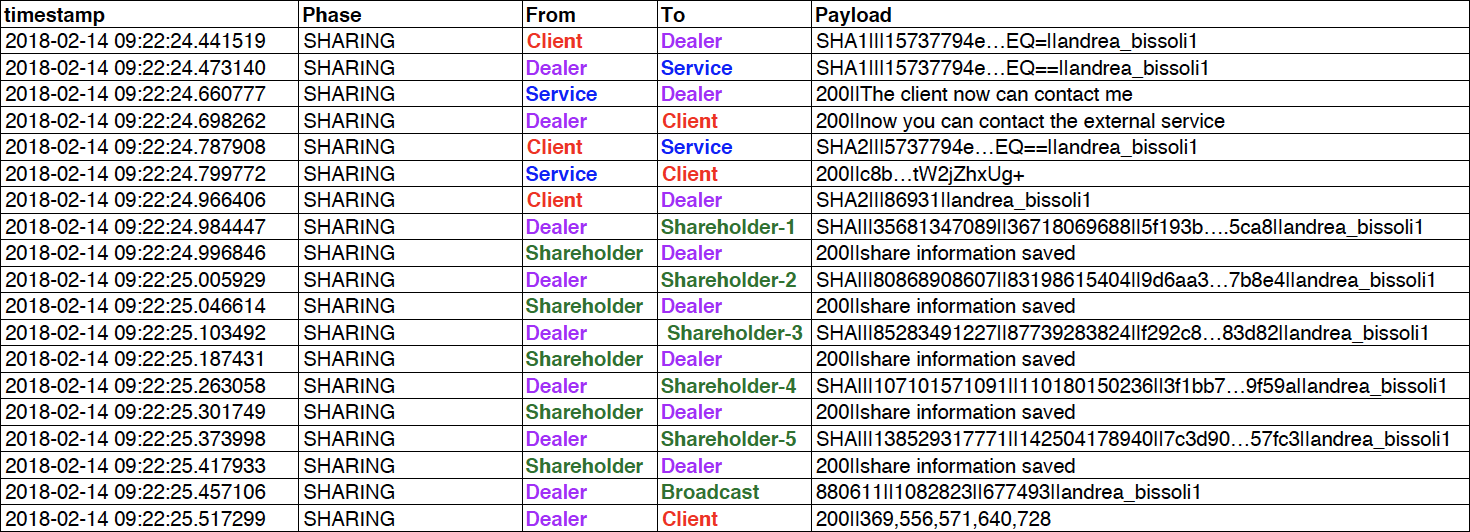}
              \caption{Log of Sharing phase}
              \label{fig:Log of Sharing phase}
            \end{figure}
            
            \begin{figure}[!htb]
              \centering
              \includegraphics[width=1.\textwidth]{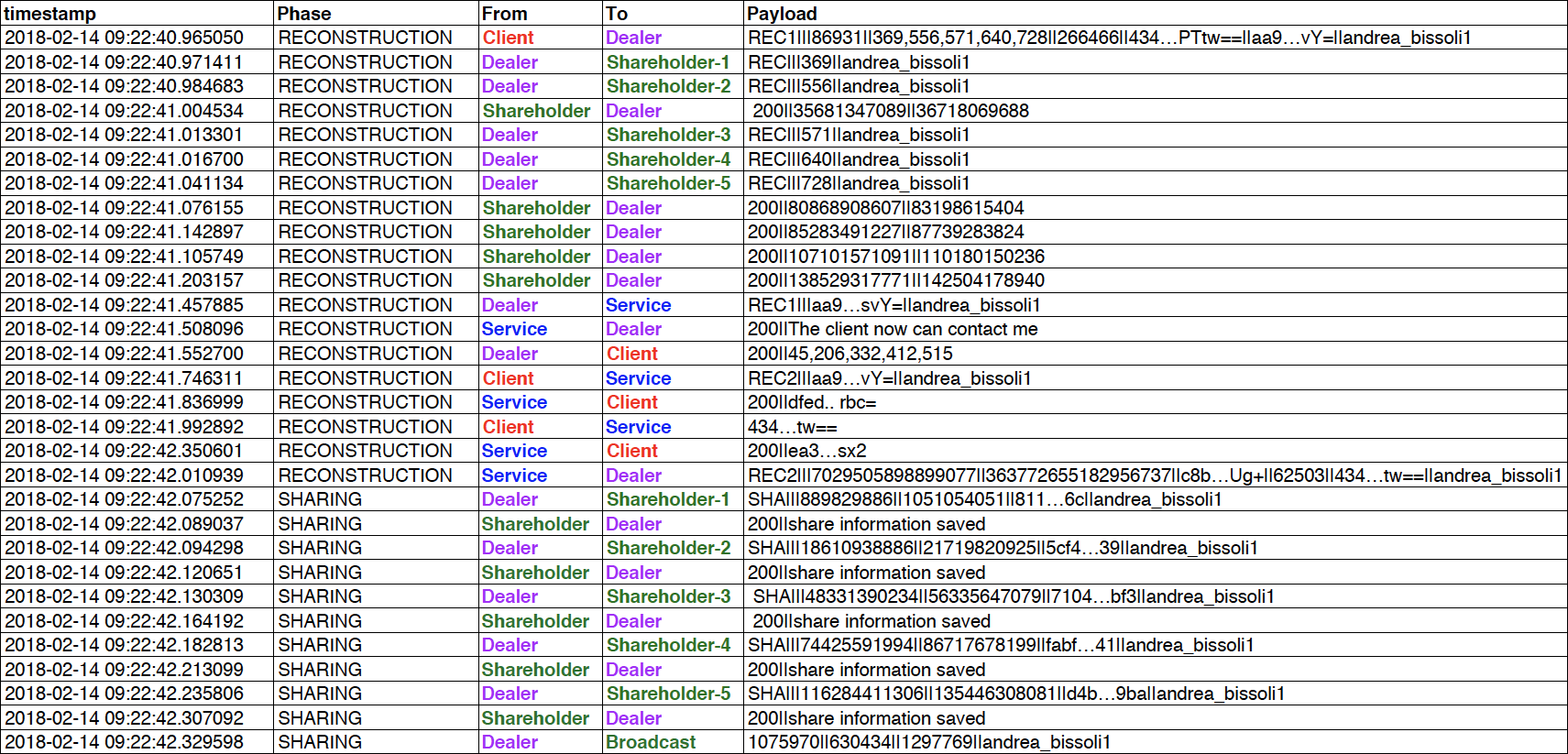}
              \caption{Log of Reconstruction phase}
              \label{fig:Log of Reconstruction phase}
            \end{figure}

          \newpage  
          \subsection{Performance qualitative analysis}
            In this section the performed tests will be shown. First of all the setting of environment for the testing is treated, then the obtained results and finally the conclusion.

            \subsubsection{The environment for the testing}
              % come cambia l'env? uso di un clien python perchè JS non permette multiu-th. Dire che tutto gira in una macchina.
              In order to perform the tests it is not possible to use the client on the browser because the test is made to stress the whole system. In order to test the system it is necessary to have many requests at the same time. To simulate this situation it is necessary to have a multithread client. JavaScript does not admit multithreading, so for the test, a python client is used. The python client can launch many different requests to the system at the same time. We have to observe that the tests are made in only one physical machine, so the response time with 100 concurrent requests is very high. Then, the important aspect that comes out from these tests is the relationship between each phase and what happens if some parameters change.
              \\\\
              The parameters of the tests are the following:
              \begin{itemize}
                \item $p$: 50/250/700 digit (that is 166/830/2325 bit)
                \item $t$: 2/3/5/10 
                \item $n$: 3/5/7/10 
              \end{itemize}
              Hence, in the tests, the parameter about how large $p$, how many shareholders and the minimum number of shareholders needed to rebuild the secret is changed. For each test there are up to 100 concurrent requests through steps of 5 or 10. For each request time is taken, and in the end the average is calculated. For example, if the test is running with 50 request, all the fifty requests are launched in different threads. When the thread has finished its task time is taken. When all fifty threads have finished the average is computed. In the next section it is possible to see the results.

            \subsubsection{The performed tests}
              % spiegazione dei vari parameri cambiati per i test, visione dei test con FOTO,s 
              % Per ogni gruppo si fa un commento.
              The performed tests are divided into \textbf{four groups}:
              \begin{enumerate}
                \item t=2, n=3
                \item t=3, n=5
                \item t=5, n=7
                \item t=10, n=10
              \end{enumerate}
              For each group the two phases have been tested and $p$ has been changed with 50, 250 and 700 digit value. Hence 24 tests have been made. The tests are put together to have comparisons. 
              \\\\
              In the first case the $p$-value is hold fixed, and $t$ and $n$ change. In figure \ref{fig:Sharing phase with p: 50, 250 and 700 digit} and \ref{fig:Reconstruction phase with p: 50, 250 and 700 digit} it is possible to see with the same size of $p$, the \textbf{four main groups}; having many requests, the delay per client increases if $t$ and $n$ increase. This is true for all the plots. Hence, it is possible to claim that increasing $t$ and $n$, with the same $p$-value, the delay per client increase. As it is expected, the Reconstruction phase is slower than the Sharing phase.
              \begin{figure}[!htb]
                \centering
                \begin{minipage}{1\textwidth}
                    \centering
                    \subfloat[Sharing - 166 bit]{
                        \includegraphics[width=0.47\textwidth]{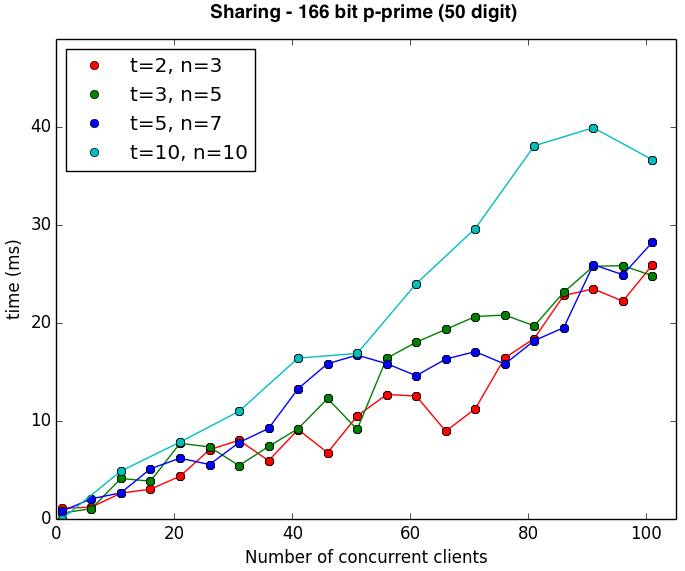}
                        \label{fig:first_sub}
                    }
                    \subfloat[Sharing - 830 bit]{
                        \includegraphics[width=0.47\textwidth]{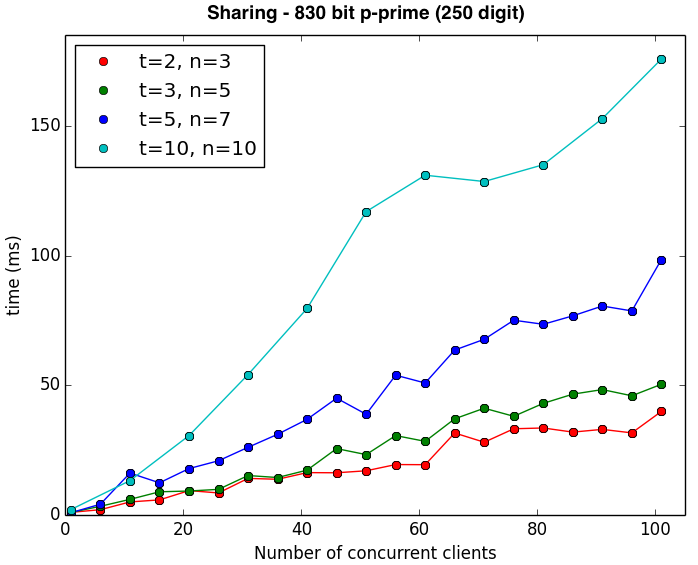}
                        \label{fig:second_sub}
                    }
                    \\
                    \subfloat[Sharing - 2325 bit]{
                        \includegraphics[width=0.5\textwidth]{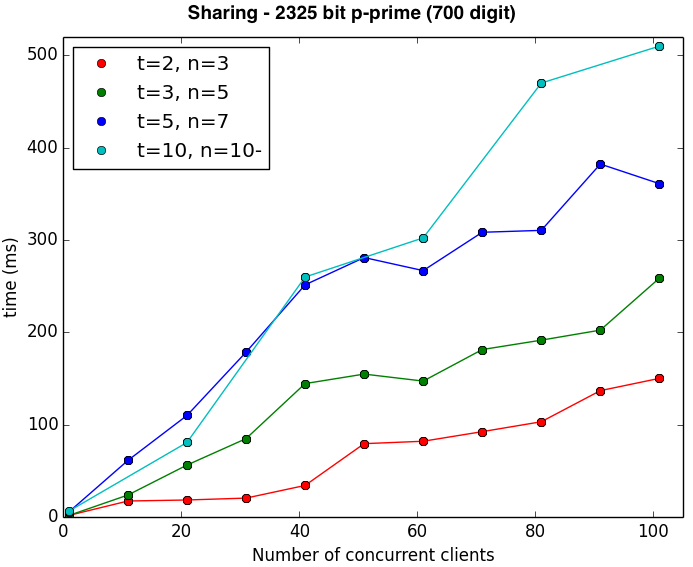}
                        \label{fig:third_sub}
                    }
                    \caption{Sharing phase with fixed $p$}
                    \label{fig:Sharing phase with p: 50, 250 and 700 digit}
                \end{minipage}
              \end{figure}

              \begin{figure}[!htb]
                \centering
                \begin{minipage}{1\textwidth}
                    \centering
                    \subfloat[Reconstruction - 166 bit]{
                        \includegraphics[width=0.47\textwidth]{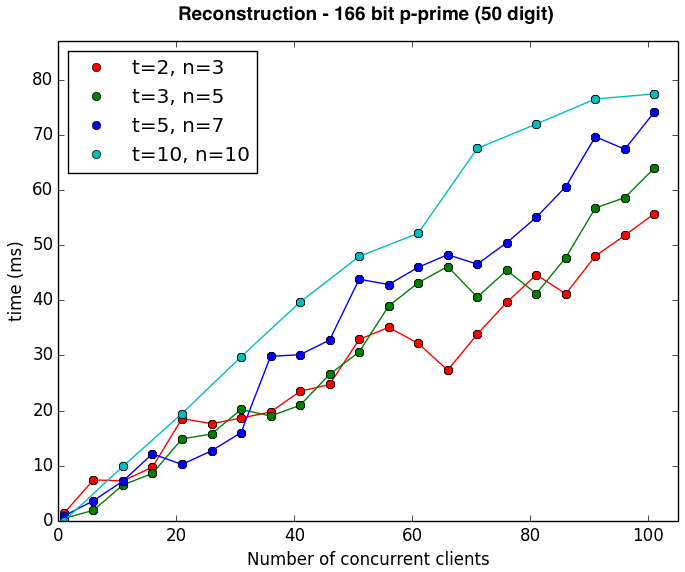}
                        \label{fig:first_sub}
                    }
                    \subfloat[Reconstruction - 830 bit]{
                        \includegraphics[width=0.47\textwidth]{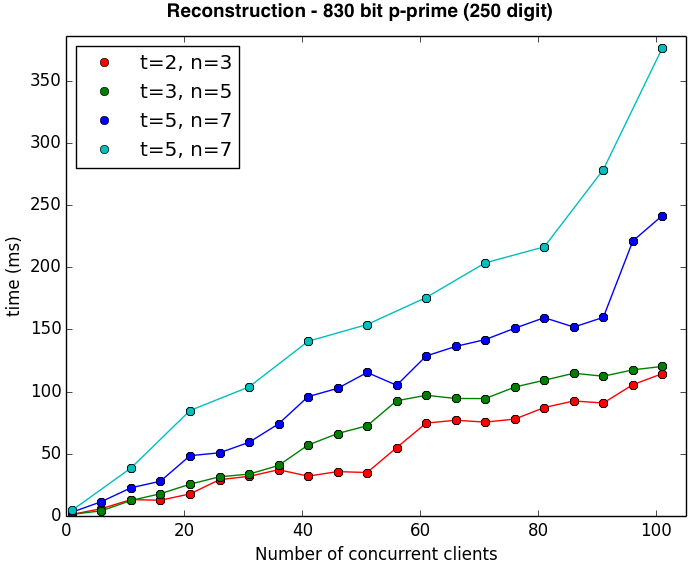}
                        \label{fig:second_sub}
                    }
                    \\
                    \subfloat[Reconstruction - 2325 bit]{
                        \includegraphics[width=0.5\textwidth]{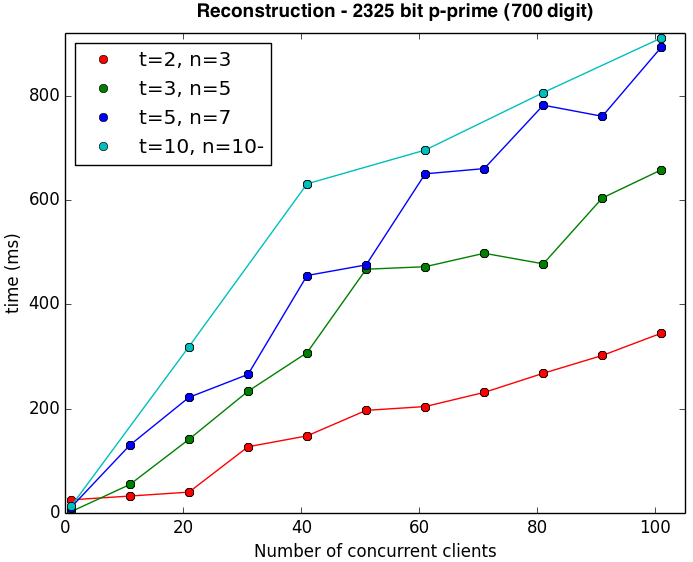}
                        \label{fig:third_sub}
                    }
                    \caption{Reconstruction phase with fixed $p$}
                    \label{fig:Reconstruction phase with p: 50, 250 and 700 digit}
                \end{minipage}
              \end{figure}

              \null\newpage
              In the second case the $p$-value changes and $t$ and $n$ are held fixed. In figures 4.20 and 4.21 it is possible to see the results. What stands out from these groups of tests is that if the $p$-value increases the delay per client increase. However, the gap between 50 and 250 digit compared to 700 digit is not the same. Since the bigger $p$ is, the safer the system is, it is better to take $p$ with 250 digit. The problem with great values of $p$ is the time computations complexity, for this reason the delay increases. 

              \begin{figure}[H]
                \centering
                \includegraphics[width=0.9\textwidth]{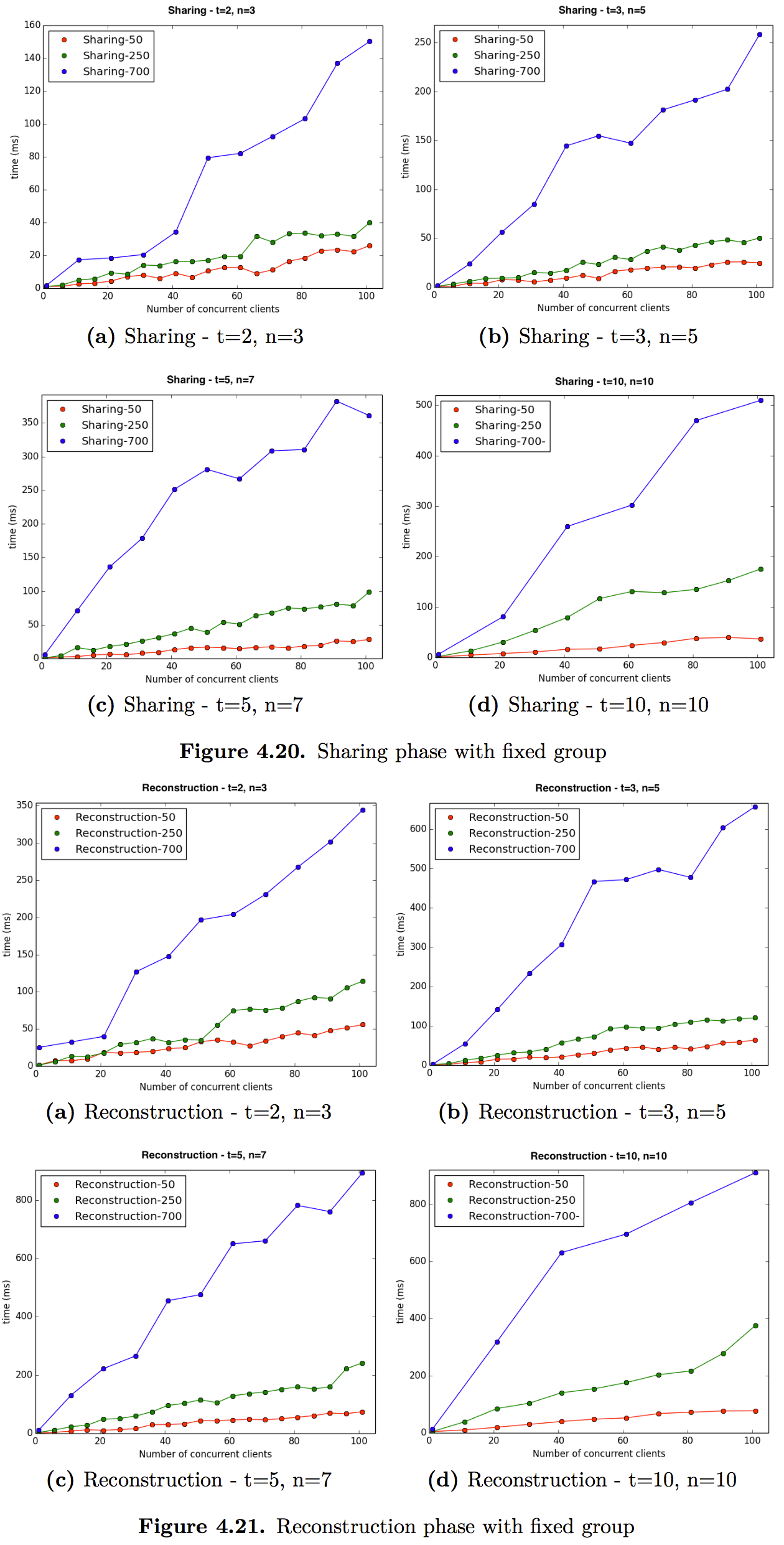}
                \label{fig:Log of Reconstruction phase}
              \end{figure}

            \subsubsection{Conclusion}
            % dopo i tests, possiamo fare una misurazione delle performance. Specificare che abbiamo usato una sola macchina, container ecc!
              The Reconstruction phase is always slower than the Sharing phase, increasing the number of bit of $p$-prime, the complexity of the calculations augments and so also the execution time and the response time of the system. In the same way increasing the number of Shareholders, the number of messages increase and as a result the response time of the system.
              \\\\
              The tests were done in one physical machine, so with high parameters like $p$ with 700 digit and with 10 Shareholders the average response time of the system is unsatisfactory. The results are useful in order to understand the \textbf{proportions} between each phase and among the different values of $p,\ t$ and $n$.

  %%%%%%%%%%%%%%%%%%%%%%%%%%%%%%%%%%%%%%%%%%%%%
  %     Security analysis.              %
  %%%%%%%%%%%%%%%%%%%%%%%%%%%%%%%%%%%%%%%%%%%%%

  \section{Security analysis}\label{chap:Security analysis}
    In this chapter we will see how scheme \ref{Scheme_Second}, implemented in the PoC reacts to various types of attack will be treated. Each case will be studied. When something goes wrong, the player sends the error message to the Logger. The Logger stores the received information to make a report for the system administrator. In this way if someone is under attack, the admin can check its correctness and restore its status. 
    \\\\
    Finally, an analysis of how many faults the system can manage before it no longer works will be made.
    \subsection{Message error analysis}
      All the errors that the system can handle are reported. For each phase there is a classification of the issues and for each of them an identification code is given. The code is valid for all actors of the system.
      \\\\
      Let's see the two phases in detail:
      \begin{itemize}
        \item \textbf{Sharing. } In the Sharing phase the main errors that may arise due to the incorrect behaviour of some actors are:
        \begin{itemize}
          \item the Client wants to register but is already registered. (COD100)
          \item the Dealer sends the information of a user who is already registered in the Service. (COD150)
          \item the Client is anticipating the call to the Service. (COD170)
          \item MC sent by the client to the Service is different from the one sent by the Dealer.(COD400)
          \item the Client is anticipating the call to the Dealer.
        \end{itemize}

        In the below figure there are all the listed errors, the description and what event can trigger them.

          \begin{figure}[H]
            \centering
            \includegraphics[width=1\textwidth]{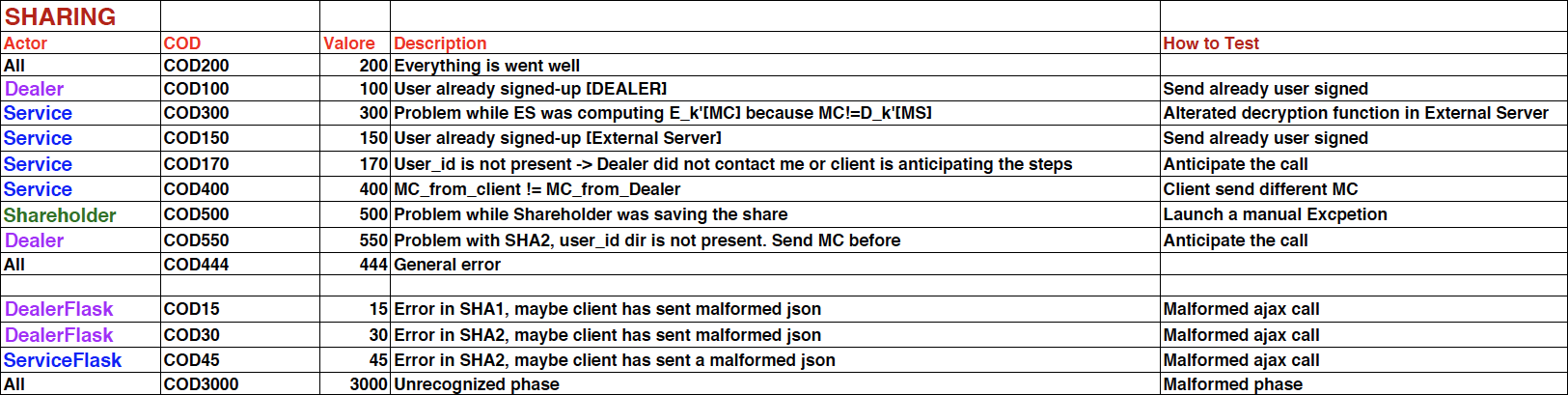}
            \label{fig:Codes of Error in Sharing phase}
            \caption{Codes of error in the Sharing phase}
          \end{figure}

        \item \textbf{Reconstruction. } In the Reconstruction phase the main errors that may arise due to the incorrect behaviour of some actors are:

          \begin{itemize}
            \item the Client wants to log in but is not registered. (COD600)
            \item the Dealer sends the wrong information of $x_i$ to the Shareholder. (COD700)
            \item the Dealer sends non-consistent shares during Sharing phase. (COD750)
            \item the Dealer has failed to collect enough shares. In this case there may be two causes: either the client has sent the wrong coordinates or some Shareholders are compromised. (COD800)
            \item the Dealer succeeded in collecting enough shares, but some are compromised. The secret can be rebuilt (correct shares $\ge t$) but this means that some Shareholder is compromised. (COD830)
            
            \item the Dealer succeeded in collecting enough shares, but many shares are compromised. The secret cannot be rebuilt (correct shares $\le t$). Many Shareholders are compromised.(COD850)
            
            \item The client gave the right coordinates but the secret is not the same as the rebuilt one ($S^\prime \ne \overline{S^\prime}$). (COD860)
            \item The last Dealer checks cannot be successful (COD2000, COD2400, COD2600) if the Service or the Client lie.
          \end{itemize}

          \begin{figure}[H]
            \centering
            \includegraphics[width=1\textwidth]{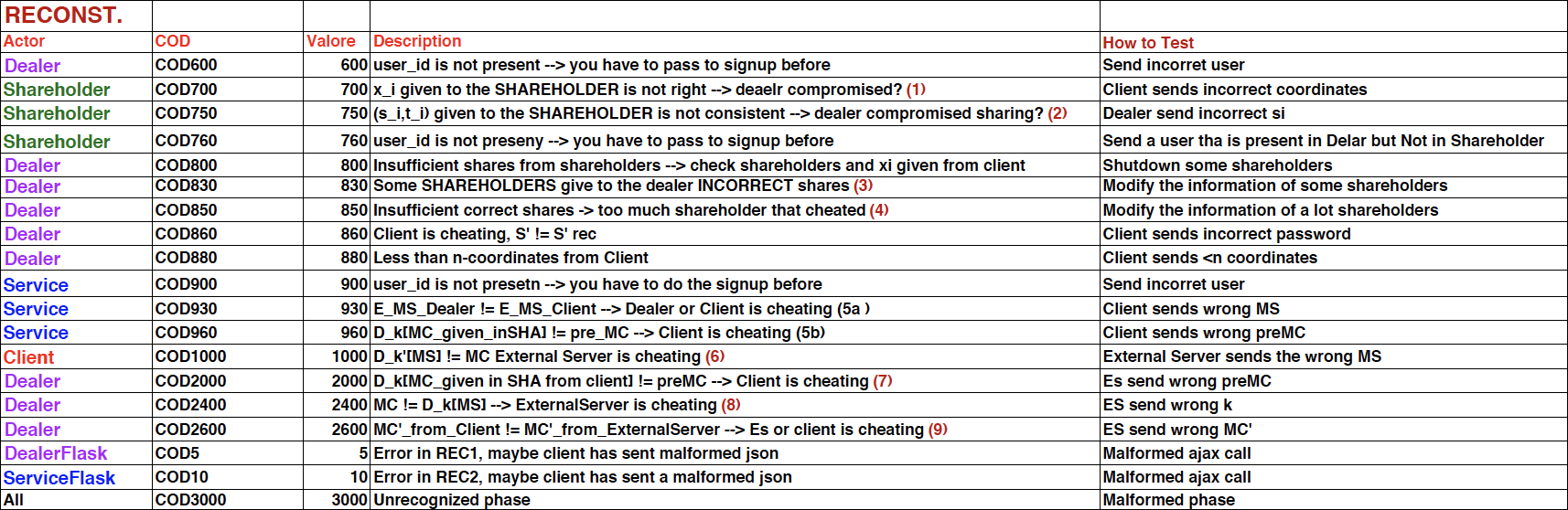}
            \caption{Codes of error in the Reconstruction phase}
            \label{fig:Codes of Error in Reconstruction phase}
          \end{figure}

      \end{itemize}
      Remember that every time a player finds an error, they notify it to the Logger. In this way the system admin can help to fix the issues. Indeed, the system admin, thanks to the Logger, can identify which machine is under attack and act accordingly. As future work, a smarter management of the errors can be thought. Besides sending the report to the Logger, the system can self-balance and disqualify players who are considered compromised by most of the participants. This improvement is possible if we have a dynamic Dealer.

    \subsection{A fake Client}
      There are two types of attacks:
      \begin{itemize}
        \item A hacker tries to lie about their identity and they try to impersonate another user. To perform the attack they need much information that is difficult to obtain. The data are: $S,\ r,\ [x_1,.., x_n],\ r^{\prime \prime},\ k,\ r^\prime,\ MS$.

        \item A hacker takes control of the user device where the data are stored. The data stored are: $r,\ [x_1,.., x_n],\ r^{\prime \prime},\ k,\ r^\prime,\ MS$, but without $S$ it is impossible to log in. Then, only with the union of the whole information the hacker would be able to log in.
      \end{itemize}

    \subsection{A compromised Dealer}
      If the Dealer is under attack, depending on the hacker's intentions, the system reacts differently.

      \begin{itemize}
        \item If the hacker's goal is to authorize a fake user, additional data are needed, such as $MS,\ k,\ g^{S^\prime} h^{r^\prime}$, to submit them to the Service. The Dealer fails because to have this information the attacker must have hacked that specific user.
        
        \item The Dealer manages the data $S^\prime$. If their goal is to get the password $S$, the hacker will never succeed. This is because, being $S^\prime = g^S h^r$ information-theoretic secure, they cannot retrieve $S$ also with unbounded computational power.

        \item If the goal of the attacked Dealer is to create misunderstanding in the protocol it can succeed but the hacker would be discovered by the other participants. The participants are able to notify to the Logger if someone tries to lie. Then the admin can investigate and find the causes of the problem and fix them. In the dynamic Dealer case, if the majority of the participants agree, they will disqualify that Dealer until it is re-established as safe.
      \end{itemize}

    \subsection{Compromised Shareholders}
      If one or more Shareholders are under attack, depending on the hacker's intentions, the system reacts differently.

      \begin{itemize}
        \item In the worst case the hacker takes $n$-Shareholders under their control. Since the coordinates are not stored on the Shareholder side, the attacker cannot rebuild the secret. Even if they succeed in obtaining at least $t$-coordinates, and it happens during the login, the secret $S^\prime$ becomes no longer valid because it is replaced with the new one $S^{\prime\prime}$. Furthermore,  being $S^\prime = g^S h^r$ information-theoretic secure, they cannot retrieve $S$ also with unbounded computational power. 

        \item If the Shareholders lie and give no consistent shares to the Dealer, they can detect them and can notify the errors to the Logger. In this way the system admin can help to fix the issues.
      \end{itemize}

    \subsection{A compromised Service}
      A Service does not handle sensitive information such as a password. The information is:
      \begin{itemize}
        \item $MC = E_k\big[g^{S^\prime} h^{r^\prime}\big]$
        \item $k$
        \item $g^{S^\prime} h^{r^\prime}$, where $S^\prime = g^S h^r$
      \end{itemize}
      Hence, even if the Service is under attack there is no leakage of the passwords.

    \subsection{Fault tolerant analysis}
      This section deals with quantitative analysis of the faults that the system can support. At present the system tolerates up to $k$ failures, where the following equation is respected:
                  \[n-k \ge t\] 
      Shareholder failures. Hence, $k$ must be:
              \[k \le n-t\] 
      This is happens because to reconstruct the secret at least $t$ correct shares are needed. Failure means *:
      \begin{itemize}
        \item the Shareholders go down.
        \item the Shareholders are under the attack and the hacker tries to perform misunderstanding.
      \end{itemize}
      Regarding the Dealer, in the static case that is implemented in the PoC, if they go down the system goes down too. In the dynamic Dealer case, if the Dealer fails*, they could be disqualified from the election until they will be recovered. In the dynamic Dealer case the equation $n-k \ge t$ is true for all type of failures of each actor, that is for Shareholders failures and Dealer failures.

  %%%%%%%%%%%%%%%%%%%%%%%%%%%%%%%%%%%%%%%%%%%%%
  %     Authentication as a Services      %
  %%%%%%%%%%%%%%%%%%%%%%%%%%%%%%%%%%%%%%%%%%%%%

  \section{Authentication as a services}\label{chap:Authentication as a service}
    \subsection{The idea}
      Scheme \ref{Scheme_Second} could be adopted and used to build an Authentication As a Service. Authentication as a Service is an online system which provides the users with the possibility of authenticating into others' online services. There are two points of view: the user and the service:
      \begin{itemize}
        \item \textbf{User. } The basic idea is to have the same password for a lot of services and still have a higher security level. The key point is that the password is the source to generate other passwords for each External Service.
        \item \textbf{External Service. } The External Service has no more the responsibility of the passwords. Hence, if it is attacked, it has nothing of important to steal.
      \end{itemize}
      
      \begin{figure}[H]
        \centering
        \includegraphics[width=1\textwidth]{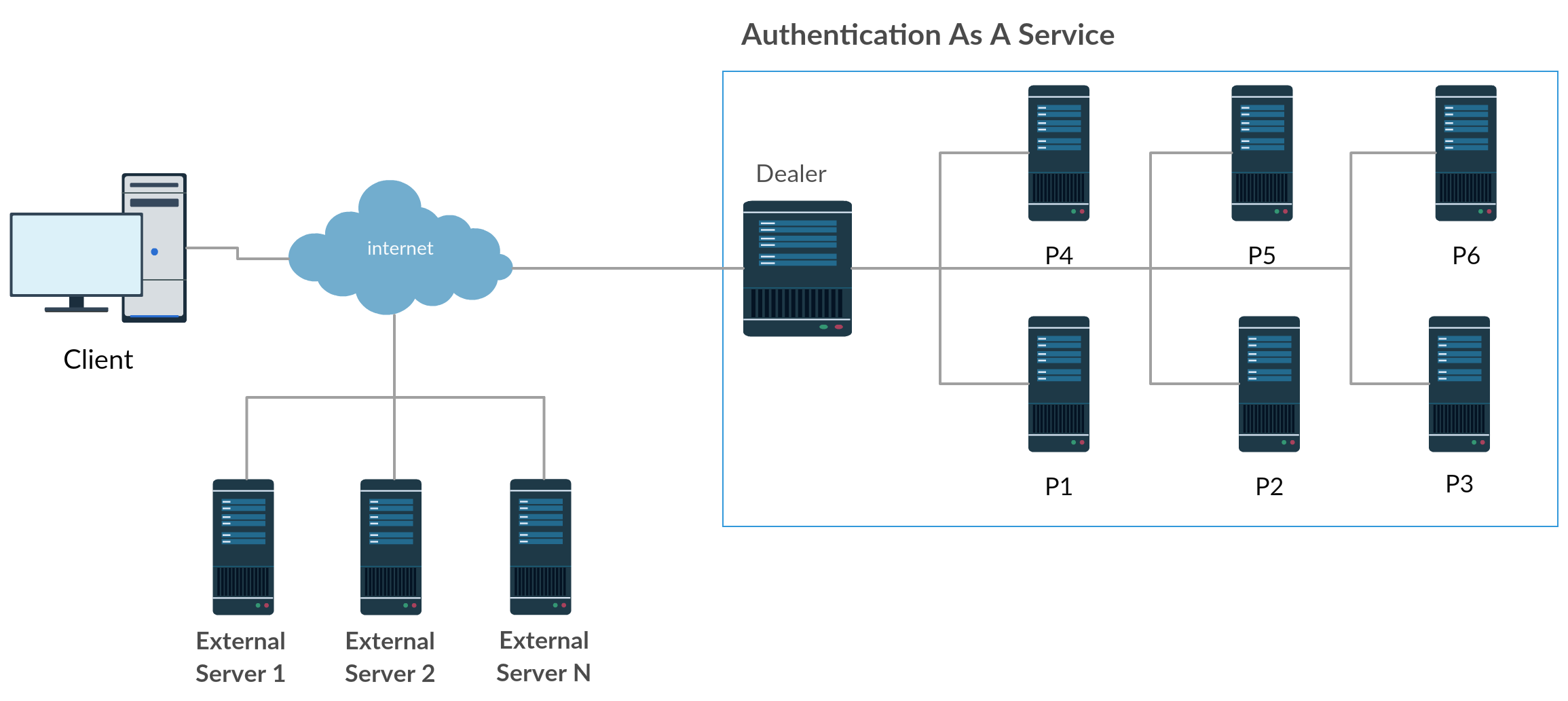}
        \caption{Authentication As A Service}
        \label{fig:AAAS}
      \end{figure}
\begin{figure}[H]
        \centering
        \includegraphics[width=0.6\textwidth]{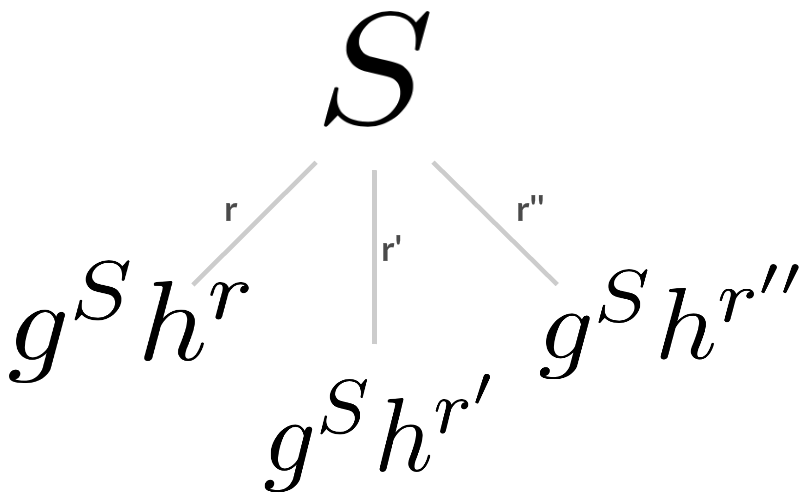}
        \caption{How handle different passwords for each External Service}
        \label{fig:AAAS_how_TO_managePassword}
      \end{figure}

      Until now, in scheme \ref{Scheme_Second} there have been four actors: Client, Dealer, Shareholders and Service. From now on, in order to implement an Authentication as a Service it is possible to consider the role of the Service in a different way. The Service, in the network model \autoref{fig:DealerFixedNetworkModelExternalServe2} had two network interfaces: one towards the Dealer and the other towards the Client. Instead, in this situation, the External Services can be seen as separate entity. Hence, the tree main actors of the Authentication as a Service are:
      \begin{itemize}
        \item Client
        \item System (Dealer + Shareholders)
        \item External Servers/Services
      \end{itemize}
      In this scheme, when the client wants to log in or sign up in a web service they must use the System. The password that the user must memorize is only one but with the same information and with additional element it is possible to create different other passwords. In figure \ref{fig:AAAS_how_TO_managePassword} it is possible to see a simple scheme about the adopted basic idea. Thus, with the additional data $r$ and the password $S$ the shared information will change depending on which External Service the client is using.
      In this way the External Services can delegate the burden of authentication to the system for various reasons. The reasons could be that the External Services do not want to take responsibility for handling sensitive information such as a user's password because they do not have the resources to put a secure system online. 
      %This is because the system can be used as an authentication-as-a-service. It means that external services can delegate the burden of authentication to our system for various reasons. The reasons could be that the external services do not want to take responsibility for handling sensible information such as a user's password because they do not have the resources in order to put online a secure system. Obviously the service relies on our system but it can happen that it is not 100\% secure because secure systems do not exist. For this reason, the $MC$ and $MS$ mechanism are adopted so that even if the dealer is under attack this is detected by the external service. The \textbf{main issue} to care about  is that the dealer does not allow access to unauthorized users.

    \subsection{How it works}
      Looking at the \ref{Scheme_Second} scheme the actions of each actor do not change. The only thing that will change is the network model, as we can see in Figure \ref{fig:AAAS}. Thus, the main problem is to understand how it is possible to make the different External Services and the Authentication as a service System coexist.
      \\\\
      As it is possible to see in scheme \ref{Scheme_Second}, the Service has its own logic. If the service provider wants to use the authentication as a service, it must be compliant with that logic. If the service provider follows the protocol, it can use the System and authenticate its users.  
      \\\\
      Logically speaking, the service provider has to implement these actions:
      \begin{itemize}
        \item \textbf{Sharing phase. } 
          \begin{itemize}
            \item Locally save $MC$ for each user.
            \item Check if $\overline{MC}$ received from the Dealer is equal to $MC$ received from the Client.
            \item Compute and store $MS$, that is: $MS=E_k^\prime\big[MC \big]$ and keep secret $k^\prime$ secret.
            \item If some error arise, notify it to the Logger or the system administrator.
          \end{itemize}
        \item \textbf{Reconstruction phase} 
          \begin{itemize}
            \item Locally save $E_{MS}$
            \item Check if $\overline{E_{MS}}\big[k,g^{S^\prime} h^{r^\prime} \big]$ received from the Dealer is equal to $E_{MS}\big[k,\ g^{S^\prime} h^{r^\prime} \big]$  received from the Client.
            \item Decrypt $E_{MS}\big[k,g^{S^\prime} h^{r^\prime} \big]$.
            
            \item Send back $k^\prime$ to the Client.

            \item Locally save $MC^\prime$ for each user. 

            \item Compute and store $MS^prime$, that is: $MS=E_k^{\prime\prime\prime}\big[MC^\prime \big]$ and keep secret $k^{\prime\prime\prime}$.

            \item Send the whole data to the Dealer: $k,\ k^\prime,\ MS,\ g^{S^\prime} h^{r^\prime},\ \overline{MC^\prime}$.
            
            \item If some error arise, notify it to the Logger or the system administrator.
        
          \end{itemize}
      \end{itemize}

  %%%%%%%%%%%%%%%%%%%%%%%%%%%%%%%%%%%%%%%%%%%%%
  %     Conclusions and future working    %
  %%%%%%%%%%%%%%%%%%%%%%%%%%%%%%%%%%%%%%%%%%%%%

  \section{Conclusions and future work}
    The proposed study in this research allows a user to have a secure authentication thanks to the combination of cloud tools, Shamir and Pedersen's work. The study ensures that the system can offer important guarantees from the point of view of security, respectively:
    \begin{itemize}
      \item Integrity
      \item Confidentiality
      \item Authenticity
      \item Authentication
      \item Password is never saved in plaintext but it is stored in a secure information-theoretically way.
    \end{itemize}
    In the proposed system the user must keep only the password in mind , while the other additional information is saved into the browser. Only the union of this two data will allow the authentication to the user. By modify Shamir's work in order not to make the coordinates public, and with the addition of  Pedersen's work, the built system ensures a certain degree of security supported by mathematical demonstrations. Since the back-end is a cloud computing, there are $n$ machines that work together to reach a common aim. The hacker must take under their control at least $t$ machines in $n$ and steal the coordinates to rebuild the secret $S^\prime$. The core of the matter is how $S^\prime$ is computed, in fact the password is inside $S^\prime$ but it is information-theoretically secure, hence it will be unfeasible to open $S^\prime$ and retrieve the password.
    \\\\
    The cloud computing infrastructure is realized using Docker and its container technology. Docker is a powerful framework, reliable and the first to have popularized this type of technology. With the containerisation it is possible to set up a complex system with a few commands. Besides, Docker allows many primitives to exploit and build a nice system. To evaluate the performances achieved by the system, particular tests were carried out that allowed the qualitative assessment of the two phases and to understand which best values to adopt for the parameters $p,\ t,\ n,$ and $t$ in order to have the safest authentication operations.
    \\\\
    To conclude, the studied protocol can be used to implement an authentication as a service. In this way a user could have different passwords for different services but only one piece of information to memorize. 
    \\\\
    Moreover, in this work different schemes, depending on the assumptions made, are proposed. For each scheme the attack model and the risk analysis is studied. We have achieved two final solutions but only one is implemented in the PoC. For the PoC a thorough security analysis evaluating the effective robustness of the system against the different types of attack that could occur and that could threaten the user's privacy is studied. In conclusion, the system has proved to be robust and safe and can provide the users with important guarantees of safety.
    \\\\
    In conclusion, we would like to suggest several improvements to take into account, for those who are well-intentioned to continue the work addressed in this research:
    \begin{itemize}
      \item Implement the Dealer election before the other two phases. Hence the possibility of having a dynamic Dealer. This would have a negative impact on the system response but on the other hand it would have many security benefits.

      \item At the beginning of the discussion, the presence of only one device per user was placed as a constraint. The user would have an improvement if they had the possibility of using more devices. In order to implement this aspect a synchronization logic must be implemented to keep the additional information updated on each device.

      \item At this moment if an error occurs, that is, if through the controls the actors detect something wrong, they notify the fact to the Logger. The Logger then provides a report to be read by the system admin. An improvement could be to implement an automation smart error manager where the players could disqualify other actors if the majority agrees. In this way the system would be self-balancing. 

      \item When the system is used as an authentication-as-a-service, APIs must be provided to developers so that they can interface and use the system.

      \item To do tests on real machines to obtain an estimate of the quantitative response of the system.
    \end{itemize}

  \newpage
  \printbibliography
  %\listoffigures
\end{document}